\documentclass[preprint,prd,nofootinbib,showpacs]{revtex4}
\usepackage{graphicx}% Include figure files
\usepackage{epsfig}
\usepackage{bm}% bold math
\begin{document}   

\title{Effective theory approach to new physics in $b \to u$ and $b \to c$ leptonic and semileptonic decays.}
\author{Rupak~Dutta${}$}
\email{rupak@iith.ac.in}   
\author{Anupama~Bhol${}$}   
\email{ph10p004@iith.ac.in}
\author{Anjan~K.~Giri${}$}
\email{giria@iith.ac.in}
\affiliation{
${}$Indian Institute of Technology Hyderabad, Hyderabad 502205, India
\\
}

\begin{abstract}
Recent measurements of exclusive $B^{-} \to \tau^{-}\,\nu$ and $B^0 \to \pi^{+}\,l^{-}\,\bar{\nu}_l$ decays via the $b \to u\,l\,\nu$ transition process differ from the standard model
expectation and, if they persist in future $B$ experiments, will be a definite hint of the physics beyond the standard model. Similar hints 
of new physics have been observed in $b \to c$ semileptonic transition processes as well. BABAR measures the ratio of branching fractions of 
$B \to (D,\,D^{\ast})\,\tau\,\nu$ to the corresponding $B \to (D,\,D^{\ast})\,l\nu$, where $l$ represents either an electron or a muon, and finds $3.4\sigma$ discrepancy with the standard model expectation.
In this context, we consider a most general effective Lagrangian for the $b \to u\,l\,\nu$ and $b \to c\,l\,\nu$ transition processes in the presence of 
new physics and perform a combined analysis
of all the $b \to u$ and $b \to c$ semi-(leptonic) data to explore various new physics operators and their couplings.
We consider various new physics scenarios and give predictions for the $B_c \to \tau\nu$ and $B \to \pi\tau\nu$ decay
branching fractions. We also study the effect of these new physics parameters on the ratio of the branching ratios of $B \to \pi\tau\nu$ to the corresponding
$B \to \pi\,l\,\nu$ decays.   
\end{abstract}

 % 	leptonic + semileptonic decays

\maketitle

\section{Introduction}
Although, the standard model~(SM) of particle physics can explain almost all the existing data to a very
good precision, there are some unknowns which are beyond the scope of the SM. The latest discovery of a Higgs-like
particle by CMS~\cite{CMS} and ATLAS~\cite{atlas} further confirms the validity of the SM as a low energy effective theory.
There are two ways to look for
evidence of new physics~(NP): direct detection and indirect detection. The Large Hadron Collider~(LHC), which is running successfully at CERN, in principle, has the ability to detect new particles
that are not within the SM, while, on the other hand the LHCb experiment has the ability to perform indirect searches of NP effects, and since any NP will affect the SM observables, 
any discrepancy between measurements and the SM expectation will be an indirect evidence of NP beyond the SM.

Recent measurements of $b \to u\,\tau\,\nu$ and $b \to c\,\tau\,\nu$ leptonic and semileptonic $B$ decays differ from SM 
expectation. The measured branching ratio of $(11.4\pm 2.2)\times 10^{-5}$~\cite{belle, babar, pdg} for the 
leptonic $B^{-} \to \tau^{-}\,\nu$ decay mode is larger than the SM expectation~\cite{Charles, ckm, Bona}. However, the 
measured branching ratio of $(14.6\pm 0.7)\times 10^{-5}$~\cite{babar-pilnu, belle-pilnu, HFAG} for the exclusive semileptonic $B^0 \to \pi^{+}\,l\,\nu$ decays is consistent with 
the SM prediction. The SM calculation, however, depends on the hadronic quantities such as $B$ meson decay constant and $B \to \pi$ transition form
factors and the Cabibbo-Kobayashi-Maskawa~(CKM) element $|V_{ub}|$. The ratio of branching fractions defined by
\begin{eqnarray}
&&R_{\pi}^l = \frac{\tau_{B^0}}{\tau_{B^-}}\,\frac{\mathcal B(B^{-} \to \tau^{-}\,\nu)}{\mathcal B(B^{0} \to \pi^{+}\,l^{-}\,\nu)}
\end{eqnarray}
is independent of the CKM matrix elements and is measured to be $(0.73\pm0.15)$~\cite{Fajfer:2012jt}, and there is still more than $2\sigma$ discrepancy with the SM
expectation. More recently, BABAR~\cite{babar-dstnu} measured the ratio of branching fractions of $B \to (D,\,D^{\ast})\,\tau\,\nu$ to the corresponding $B \to (D,\,D^{\ast})\,l\nu$
and found $3.4\sigma$ discrepancy with the SM expectation~\cite{Fajfer}.
The measured ratios are
\begin{eqnarray}
\label{rdds}
&& R_D=\frac{\mathcal B(\bar B \to D\tau^- \bar{\nu}_\tau)}{\mathcal B(\bar B \to D\, l^- \bar{\nu}_l)}=0.440\pm0.058\pm0.042 \,, \nonumber\\
&& R_{D^{\ast}}=\frac{\mathcal B(\bar B \to D^{\ast}\tau^- \bar{\nu}_\tau)}{\mathcal B(\bar B \to D^{\ast}\,l^- \bar{\nu}_l)}=0.332\pm0.024\pm0.018 \,,
\end{eqnarray}
where the first error is statistical and the second one is systematic. For definiteness, we consider $B^- \to l^{-}\,\bar{\nu}_l$, $\bar{B}^0 \to \pi^{+}\,l^{-}\,\bar{\nu}_l$, 
$B^{-} \to D^0\,l^{-}\,\bar{\nu}_l$, and $B^{-} \to D^{\ast\,0}\,l^{-}\,\bar{\nu}_l$ throughout this paper. However, for brevity, we denote all these decay modes as $B \to l\,\nu$, $B \to \pi\,l\,\nu$,
$B \to D\,l\,\nu$, and $B \to D^{\ast}\,l\,\nu$, respectively. 

Due to the large mass of the tau lepton, decay processes with a tau lepton in
the final state are more sensitive to some new physics effects than processes with first two generation leptons. These NP, in principle, can enhance the decay rate for these helicity-suppressed
decay modes quite significantly from the SM prediction. In Ref.~\cite{Fajfer}, a thorough investigation of the lowest dimensional effective operators that leads to modifications in the $B \to D^{\ast}\tau\nu$
decay amplitudes has been done. Possible NP effects on various observables have been explored.
Among all the leptonic and semileptonic decays, decays with a tau lepton in the final state can be an excellent probe of new physics as these are sensitive        
to non-SM contributions arising from the violation of lepton flavor universality~(LFU). 
A model-independent analysis to identify the new physics models has been explored in Ref.~\cite{Fajfer:2012jt}. They also 
look at the possibility of a scalar leptoquark or a vector leptoquark, which 
can contribute to these decay processes at the tree level and obtain a bound of $m \ge 280\,{\rm GeV}$ on the mass of the scalar electroweak triplet leptoquark.
Model with composite quarks and leptons also modify these $b \to u$ and $b \to c$ semileptonic measurements~\cite{Fajfer:2012jt}.
The enhanced production of a tau lepton in leptonic and semileptonic decays can be explained by NP contribution with different models among which the minimal supersymmetric standard model~(MSSM) 
is well motivated and is a charming candidate of NP whose Higgs sector contains the two Higgs doublet model~(2HDMs). 
There are four types of 2HDMS such as type-I, type-II, lepton specific, and flipped~\cite{Branco}. New particles such as charged Higgs bosons whose coupling is proportional to the masses of particles 
in the interaction can have significant effect on decay processes having a tau lepton in the final state.
In Ref.~\cite{Hou}, the author uses the 2HDM model of type-II for purely leptonic $B$ decays that are
sensitive to charged Higgs boson at the tree level. This model, however, cannot explain 
all the $b \to c$ semileptonic measurements simultaneously~\cite{babar-dstnu}. A lot of studies have been done using the 2HDM of type II and type III models~\cite{2HDM}. However,
none of the above 2HDMS can accommodate all the existing data on $b \to u$ and $b \to c$ semi-(leptonic) decays. Recently,
a detailed study of a 2HDM of type III with MSSM-like Higgs potential and flavor-violation in the up sector in Ref.~\cite{Crivellin} has demonstrated
that this model can explain the deviation from the SM in $R_{\pi}^l$, $R_D$, and $R_{D^{\ast}}$ simultaneously and predict enhancement in the $B \to \tau\nu$,  
$ B \to D\tau\nu $, and the $ B \to D^{\ast}\,\tau\nu$ decay branching ratios. Also, in Refs.~\cite{datta,datta1}, the authors have used a model independent way
to analyse the $B \to D\tau\nu$ and $B \to D^{\ast}\tau\nu$ data by considering an effective theory for the $b \to c\,\tau\,\nu$ processes in the presence of NP and obtain bounds on each NP parameter.
They consider two different NP scenarios and see the effect of various NP couplings on different observables. This analysis, however, does not include the $B \to \tau\nu$ data. A similar analysis has been performed in Ref.~\cite{fazio} considering a tensor operator in the effective weak Hamiltonian. Also, in Ref.~\cite{Crivellin1}, the author investigates the effects of an effective right handed charged currents on the determination of $V_{ub}$ and $V_{cb}$ from inclusive and exclusive $B$ decays.
Moreover, the aligned two Higgs doublet model~(A2HDM)~\cite{Celis} and, more recently a non-universal left-right model~\cite{He} have been explored in order to explain the discrepancies 
between the measurements and the SM prediction.

The recent measurements suggest the possibility of having new physics in the third generation leptons only. However, more experimental studies are needed to confirm the presence of NP. 
A thorough investigation of these decays will enable us to have significant constraints on NP scenarios. In this report, we use the most general effective Lagrangian for the $b \to q$ semi-(leptonic) transition 
decays and do a combined analysis of $b \to u$ and $b \to c$ semi-(leptonic) decay processes where we use constraints from all the existing data related to these decays.
It differs considerably from earlier treatments. First, we have 
introduced the right-handed neutrinos and their interactions for our analysis. Second, we have performed a combined analysis of all the $b \to u$ and $b\to c$ data. 
We illustrate four different scenarios of the new physics and the effects of each NP coupling on various observables are shown. 
We predict the branching ratio of $B_c \to \tau\nu$ and $B \to \pi\tau\nu$ decay processes in all four different scenarios. We also consider the ratio of branching ratio $R_{\pi}$ of 
$B \to \pi\tau\nu$ to the corresponding $B \to \pi\,l\,\nu$ decay mode for our analysis.

The paper is organized as follows. In Sec.~\ref{th}, we start with a brief description of the effective Lagrangian for the $b \to (u,\,c)\,l\,\nu$ processes and then present all
the relevant formulae of the decay rates for various decay modes in the presence of various NP couplings. We then define several observables in $B \to \pi\tau\nu$, $B \to D\tau\nu$, and $B \to D^{\ast}\tau\nu$ decays.
The numerical prediction for various NP couplings and the effects of each NP coupling on various observables are presented in Sec.~\ref{rnd}. We also discuss the effects of these NP couplings
on $\mathcal B(B_c \to \tau\nu)$, $\mathcal B(B \to \pi\tau\nu)$, and the ratio $R_{\pi}$ for various NP scenarios in this section. We conclude with a summary of our results in Sec.~\ref{con}.
We report the details of the kinematics and various form factors in the Appendix.
 
\section{Effective Lagrangian and decay amplitude}
\label{th}
The most general effective Lagrangian for $b \to q^\prime\,l\,\nu$ in presence of NP, where $q^\prime = u,\, c$, can be written as~\cite{Bhattacharya, Cirigliano}
\begin{eqnarray}
\mathcal L_{\rm eff} &=&
-\frac{g^2}{2\,M_W^2}\,V_{q^\prime b}\,\Bigg\{(1 + V_L)\,\bar{l}_L\,\gamma_{\mu}\,\nu_L\,\bar{q^\prime}_L\,\gamma^{\mu}\,b_L +
V_R\,\bar{l}_L\,\gamma_{\mu}\,\nu_L\,\bar{q^\prime}_R\,\gamma^{\mu}\,b_R \nonumber \\
&&+
\widetilde{V}_L\,\bar{l}_R\,\gamma_{\mu}\,\nu_R\,\bar{q^\prime}_L\,\gamma^{\mu}\,b_L +
\widetilde{V}_R\,\bar{l}_R\,\gamma_{\mu}\,\nu_R\,\bar{q^\prime}_R\,\gamma^{\mu}\,b_R \nonumber \\
&&+
S_L\,\bar{l}_R\,\nu_L\,\bar{q^\prime}_R\,b_L +
S_R\,\bar{l}_R\,\nu_L\,\bar{q^\prime}_L\,b_R \nonumber \\
&&+
\widetilde{S}_L\,\bar{l}_L\,\nu_R\,\bar{q^\prime}_R\,b_L +
\widetilde{S}_R\,\bar{l}_L\,\nu_R\,\bar{q^\prime}_L\,b_R \nonumber \\
&&+ 
T_L\,\bar{l}_R\,\sigma_{\mu\nu}\,\nu_L\,\bar{q^\prime}_R\,\sigma^{\mu\nu}\,b_L +
\widetilde{T}_L\,\bar{l}_L\,\sigma_{\mu\nu}\,\nu_R\,\bar{q^\prime}_L\,\sigma^{\mu\nu}\,b_R\Bigg\} + {\rm h.c.}\,,
\end{eqnarray}
where g is the weak coupling constant which can be related to the Fermi constant by the relation $ g^2/\,8 \,M_W^2=G_F/\sqrt 2$
and $ V_{q^\prime b}$ is the CKM Matrix elements. The new physics couplings denoted by $V_{L,R}$, $S_{L,R}$, and $ T_{L}$ involve
left-handed neutrinos, whereas, the NP couplings denoted by $\widetilde{V}_{L,R}$, $\widetilde{S}_{L,R}$, and $\widetilde{T}_{L}$ involve right-handed neutrinos.
We assume the NP couplings to be real for our analysis.
Again, the projection operators are $P_L = \,(1 - \gamma_5)/2$ and $P_R = \,(1 + \gamma_5)/2$. We neglect 
the new physics effects coming from the tensor couplings $T_{L}$ and $\widetilde{T}_{L}$ for our analysis. With this simplification, we obtain 
\begin{eqnarray}
\label{leff}
\mathcal L_{\rm eff} &=&
-\frac{G_F}{\sqrt{2}}\,V_{q^\prime b}\,\Bigg\{G_V\,\bar{l}\,\gamma_{\mu}\,(1 - \gamma_5)\,\nu_l\,\bar{q^\prime}\,\gamma^{\mu}\,b -
G_A\,\bar{l}\,\gamma_{\mu}\,(1 - \gamma_5)\,\nu_l\,\bar{q^\prime}\,\gamma^{\mu}\,\gamma_5\,b \nonumber \\
&&+
G_S\,\bar{l}\,(1 - \gamma_5)\,\nu_l\,\bar{q^\prime}\,b - G_P\,\bar{l}\,(1 - \gamma_5)\,\nu_l\,\bar{q^\prime}\,\gamma_5\,b \nonumber \\
&&+
\widetilde{G}_V\,\bar{l}\,\gamma_{\mu}\,(1 + \gamma_5)\,\nu_l\,\bar{q^\prime}\,\gamma^{\mu}\,b -
\widetilde{G}_A\,\bar{l}\,\gamma_{\mu}\,(1 + \gamma_5)\,\nu_l\,\bar{q^\prime}\,\gamma^{\mu}\,\gamma_5\,b \nonumber \\
&&+
\widetilde{G}_S\,\bar{l}\,(1 + \gamma_5)\,\nu_l\,\bar{q^\prime}\,b - \widetilde{G}_P\,\bar{l}\,(1 + \gamma_5)\,\nu_l\,\bar{q^\prime}\,\gamma_5\,b \Bigg\} + {\rm h.c.}\,,
\end{eqnarray}
where 
\begin{eqnarray} 
&&G_V = 1 + V_L + V_R\,,\qquad\qquad
G_A = 1 + V_L - V_R\,, \nonumber \\
&&G_S = S_L + S_R\,,\qquad\qquad
G_P = S_L - S_R\,, \nonumber \\
&&\widetilde{G}_V = \widetilde{V}_L + \widetilde{V}_R\,,\qquad\qquad
\widetilde{G}_A = \widetilde{V}_L - \widetilde{V}_R\,, \nonumber \\
&&\widetilde{G}_S = \widetilde{S}_L + \widetilde{S}_R\,,\qquad\qquad
\widetilde{G}_P = \widetilde{S}_L - \widetilde{S}_R\,.
\end{eqnarray}
In the SM, $G_V = G_A = 1$ and all other NP couplings are zero.

The expressions for $B \to l\nu$, $B \to P\,l\nu$, and $B \to V\,l\,\nu$ decay amplitude depends on nonperturbative hadronic matrix elements that can be expressed in terms of 
$B_q$ meson decay constants and $B \to (P,\,V)$ transition form factors, where $P$ denotes a pseudoscalar meson and $V$ denotes a
vector meson, respectively . The $B$ meson decay constant and $B \to (P,\,V)$ transition form factors are defined as
\begin{eqnarray}
\langle 0 | \bar{q^{\prime}}\,\gamma_{\mu}\,\gamma_5\,b | B(p) \rangle &=& -i\,f_{B_{q^{\prime}}}\,p_{\mu}\,, \nonumber \\
\langle P(p^{\prime}) | \bar{q^{\prime}}\,\gamma_{\mu}\,b | B(p) \rangle &=& F_{+}(q^2)\,\Big[(p + p^{\prime})_{\mu} - \frac{m_B^2 - m_P^2}{q^2}\,q_{\mu}\Big] + F_0(q^2)\,\frac{m_B^2 - m_P^2}{q^2}\,q_{\mu} \,,  \nonumber\\
\langle V(p^{\prime},\epsilon^*) | \bar{q^{\prime}}\,\gamma_{\mu}\,b | B(p) \rangle &=& \frac{2\,i\,V(q^2)}{m_B+m_V}\, \varepsilon_{\mu\nu\rho\sigma}\,\epsilon^{*\nu}\,{p^\prime}^\rho\, p^\sigma\,, \nonumber\\
\langle V(p^{\prime},\epsilon^*) | \bar{q^{\prime}}\,\gamma_{\mu}\,\gamma_5\,b | B(p) \rangle &=& 2\,m_V\,A_0(q^2)\,\frac{\epsilon^*.\,q}{q^2}\,q_\mu
\,+\,(m_B+m_V)\,A_1(q^2)\,\Big[\,\epsilon{^*_\mu}\,-\,\frac{\epsilon^*.\,q}{q^2}\,q_\mu\,\Big] \nonumber\\
&&-\,A_2(q^2)\,\frac{\epsilon^*.\,q}{(m_B\,+\,m_V)}
\Big[(p+p^\prime)_\mu\,-\,\frac{ m_B^2\,-\,m_V^2}{q^2}\,q_\mu\,\Big]\,,
\end{eqnarray}
where $q = p - p^{\prime}$ is the momentum transfer.
Again, from Lorentz invariance and parity, we obtain
\begin{eqnarray}
&&\langle 0 | \bar{q^{\prime}}\,\gamma_{\mu}\,b | B(p) \rangle = 0\,,\nonumber \\
&&\langle P(p^{\prime}) | \bar{q^{\prime}}\,\gamma_{\mu}\,\gamma_5\,b | B(p) \rangle = 0 \,,\nonumber\\
&&\langle V(p^{\prime},\epsilon^*) | \bar{q^{\prime}}\,b | B(p) \rangle = 0\,.
\end{eqnarray}
We use the equation of motion to find the scalar and pseudoscalar matrix elements. That is
\begin{eqnarray}
&&\langle 0 | \bar{q^{\prime}}\,\gamma_5\,b | B(p) \rangle = i\,\frac{m_B^2}{m_b(\mu) + m_{q^{\prime}}(\mu)}\,f_{B_{q^{\prime}}}\,, \nonumber \\
&&\langle P(p^{\prime}) |\bar{q^{\prime}}\,b | B(p) \rangle = \frac{m_B^2 - m_P^2}{m_b(\mu) - m_{q^{\prime}}(\mu)}\,F_0(q^2)\,, \nonumber\\
&&\langle V(p^{\prime},\epsilon^*) | \bar{q^{\prime}}\,\gamma_5\,b | B(p) \rangle = -\,\frac{\,2\,m_V\,A_0(q^2)\,}{m_b(\mu) + m_{q^{\prime}}(\mu)} \epsilon^*.\,q\,,
\end{eqnarray}
where, for the $B \to \pi$ form factors, we use the formulae and the input values reported in Ref.~\cite{Khodjamirian}. Similarly, we follow Refs.~\cite{Falk, Caprini, Sakaki} and 
employ heavy quark effective theory~(HQET) to estimate 
the $B \to D$ and $B \to D^{\ast}$ form factors. All the relevant formulae and various input parameters pertinent to our analysis are presented in Appendix.~\ref{ffpi} and in Appendix.~\ref{ffdd}. 

Using the effective Lagrangian of Eq.~(\ref{leff}) in the presence of NP, the partial decay width of $B \to l\nu$ can be expressed as
\begin{eqnarray}
\Gamma(B \to l\nu) &=&
\frac{G_F^2\,|V_{ub}|^2}{8\,\pi}\,f_B^2\,m_l^2\,m_B\,\Big(1 - \frac{m_l^2}{m_B^2}\Big)^2\,
\Bigg\{\Big[G_A - \frac{m_B^2}{m_l\,(m_b(\mu) + m_u(\mu))}\,G_P\Big]^2 \nonumber \\
&&+ 
\Big[\widetilde{G}_A - \frac{m_B^2}{m_l\,(m_b(\mu) + m_u(\mu))}\,\widetilde{G}_P\Big]^2\Bigg\}\,,
\end{eqnarray}
where, in the SM, we have $G_A = 1$ and $G_P = \widetilde{G}_A = \widetilde{G}_P = 0$, so that
\begin{eqnarray}
\Gamma(B \to l\nu)_{\rm SM} &=&
\frac{G_F^2\,|V_{ub}|^2}{8\,\pi}\,f_B^2\,m_l^2\,m_B\,\Big(1 - \frac{m_l^2}{m_B^2}\Big)^2\,.
\end{eqnarray}
It is important to note that the right-handed neutrino couplings denoted by $\widetilde{V}_{L,R}$ and $\widetilde{S}_{L,R}$ appear in the decay width quadratically, whereas, the left-handed neutrino couplings denoted by $V_{L,R}$ and $S_{L,R}$ appear linearly in the decay rates. The linear dependence, arising 
due to the interference between the SM couplings and the NP couplings, is suppressed for the right-handed neutrino couplings as it is proportional to a small factor $m_{\nu}$
and hence is neglected. We now proceed to discuss the $B \to P\,l\,\nu$ and $B \to V\,l\,\nu$ decays.

We follow the helicity methods of Refs.~\cite{Korner,Kadeer} for the $B \to P\,l\,\nu$ and $ B \to V\,l\,\nu $ semileptonic decays.
The differential decay distribution can be written as
\begin{eqnarray}
\frac{d\Gamma}{dq^2\,d\cos\theta_l} &=&
\frac{G_F^2\,|V_{q^{\prime}b}|^2\,|\overrightarrow{p}_{(P,\,V)}|}{2^9\,\pi^3\,m_B^2}\,\Big(1 - \frac{m_l^2}{q^2}\Big)\,L_{\mu\nu}\,H^{\mu\nu}\,,
\end{eqnarray}
where $L_{\mu\nu}$ and $H_{\mu\nu}$ are the usual leptonic and hadronic tensors, respectively. Here, $\theta_l$ is the angle between the $P~(V)$ meson and the lepton three momentum 
vector in the $q^2$ rest frame. The three momentum vector $|\overrightarrow{p}_{(P,\,V)}|$ is defined as $|\overrightarrow{p}_{(P,\,V)}| = \sqrt{\lambda(m_B^2,\,m_{P(V)}^2,\,q^2)}/2\,m_B$,
where $\lambda(a,\,b,\,c) = a^2 + b^2 + c^2 - 2\,(a\,b + b\,c + c\,a)$.  
The resulting differential decay distribution for $B \to P \,l\,\nu$ in terms of the helicity amplitudes $H_0$, $H_t$, and $H_S$ is
\begin{eqnarray}
\label{pilnu1}
\frac{d\Gamma}{dq^2\,d\cos\theta_l} 
&=& 2\,N\,|\overrightarrow{p}_P|\,\Bigg\{H_0^2\,\sin^2\theta_l\,\Big(G_V^2+\widetilde{G}_V^2\Big) + \frac{m_l^2}{q^2}\Big[H_0\,G_V\,\cos\theta_l - \Big(H_t\,G_V + \frac{\sqrt{q^2}}{m_l}\,H_S\,G_S \Big)\Big]^2 \nonumber\\
&& + \frac{m_l^2}{q^2} \Big[H_0\,\widetilde{G}_V\,\cos\theta_l  - \Big(H_t\,\widetilde{G}_V + \frac{\sqrt{q^2}}{m_l}\,H_S\,\widetilde{G}_S\Big)\Big]^2\Bigg\}\,,
\end{eqnarray}
where 
\begin{eqnarray}
&&N = \frac{G_F^2\,|V_{q^\prime b}|^2\,q^2}{256\,\pi^3\,m_B^2}\,\Big(1 - \frac{m_l^2}{q^2}\Big)^2\,,\nonumber \\
&&H_0 = \frac{2\,m_B\,|\overrightarrow{p}_P|}{\sqrt{q^2}}\,F_{+}(q^2)\,, \nonumber \\
&&H_t = \frac{m_B^2 - m_P^2}{\sqrt{q^2}}\,F_0(q^2)\,, \nonumber \\
&&H_S=\frac{m_B^2 - m_P^2}{m_b(\mu) - m_{q^\prime}(\mu)}\,F_0(q^2)\,.
\end{eqnarray}
The details of the helicity amplitudes calculation are given in Appendix.~\ref{kha}.
We refer to Refs.~\cite{Korner,Kadeer} for all omitted details. We determine the differential decay rate $d\Gamma/dq^2$ by performing the $\cos\theta_l$ integration,
i.e,
\begin{eqnarray}
\label{pilnu}
\frac{d\Gamma^P}{dq^2} &=&
\frac{8\,N\,|\overrightarrow{p}_P|\,}{3}\Bigg\{\,H_0^2\,\Big(G_V^2 + \widetilde{G}_V^2\Big)\,\Big(1 + \frac{\,m_l^2}{2\,q^2}\Big) \nonumber\\
&& + \frac{3\,m_l^2}{2\,q^2}\,\Big[ \Big(H_t\,G_V + \frac{\sqrt{q^2}}{m_l}\,H_S\,G_S \Big)^2 + \Big(H_t\,\widetilde{G}_V + \frac{\sqrt{q^2}}{m_l}\,H_S\,\widetilde{G}_S\Big)^2\Big] \Bigg\}\,,
\end{eqnarray}
where, in the SM, $G_V = 1$ and all other couplings are zero. One obtains
\begin{eqnarray}
\Big(\frac{d\Gamma^P}{dq^2}\Big)_{\rm SM} &=&
\frac{8\,N\,|\overrightarrow{p}_P|}{3}\,\Bigg\{H_0^2\Big(1 + \frac{m_l^2}{2\,q^2}\Big) + \frac{3\,m_l^2}{2\,q^2}\,H_t^2\Bigg\}\,.
\end{eqnarray}
Our formulae for the differential branching ratio in the presence of NP couplings in Eq.~(\ref{pilnu1}) and Eq.~(\ref{pilnu}) differ slightly from those given in Ref.~\cite{datta}.
The term containing $G_S$ and $\widetilde{G}_S$ is positive in Eq.~(\ref{pilnu1}) and Eq.~(\ref{pilnu}), whereas, it is negative in Ref.~\cite{datta}.
Although, the SM formula is same, the numerical differences may not be negligible once the NP couplings $S_{L,\,R}$ and $\widetilde{S}_{L,\,R}$ are introduced. 
It is worth mentioning that, for $l = e,\,\mu$, the term containing $m_l^2/q^2$ can be safely ignored.
However, same is not true for the $B \to P\tau\nu$ decay
mode as the mass of $\tau$ lepton is quite large and one cannot neglect the $m_{\tau}^2/q^2$ term from the decay amplitude. We assume that the NP affects the third generation lepton only.

Similarly, the differential decay distribution for $B \to V \,l\,\nu$ in terms of the helicity amplitudes $\mathcal{A}_0$, $\mathcal{A}_\parallel$, $\mathcal{A}_\perp $, $\mathcal{A}_P$, and
$\mathcal{A}_t$ is
\begin{eqnarray}
\label{vlnu}
\frac{d\Gamma}{dq^2\,d\cos\theta_l} &=&
N\,|\overrightarrow{p}_V|\,\Bigg\{2\,\mathcal{A}_0^2\,\sin^2\theta_l\Big(G_A^2+\widetilde{G}_A^2\Big) 
 + \Big(1+\cos^2\theta_l\Big)\Big[  \mathcal{A}_\parallel^2\Big(G_A^2+\widetilde{G}_A^2\Big) + \mathcal{A}_\perp^2\,\Big(G_V^2+\widetilde{G}_V^2\Big)\Big]\nonumber\\
&& - 4\,\mathcal{A}_\parallel \mathcal{A}_\perp \cos \theta_l\Big(G_A\,G_V - \widetilde{G}_A\,\widetilde{G}_V\Big) 
+ \,\frac{m_l^2}{q^2}\sin^2\theta_l\,\Big[\mathcal A_\parallel^2\Big(G_A^2+\widetilde{G}_A^2\Big) + \mathcal A_\perp^2\Big(G_V^2+\widetilde{G}_V^2\Big)\Big]  \nonumber\\
&&+\,
\frac{2m_l^2}{q^2}\Big[\Big\{ \mathcal{A}_0\,G_A\cos \theta_l - \Big(\mathcal{A}_t\,G_A + \frac{\sqrt{q^2}}{m_l}\,\mathcal{A}_P\,G_P\Big)\Big\}^2 \nonumber \\
&&+ 
\Big\{ \mathcal{A}_0\,\widetilde{G}_A\cos \theta_l - \Big(\mathcal{A}_t\,\widetilde{G}_A 
+ \frac{\sqrt{q^2}}{m_l}\,\mathcal{A}_P\,\widetilde{G}_P\Big)\Big\}^2 \Big]  \Bigg\}\,, 
\end{eqnarray}
where
\begin{eqnarray}
&&\mathcal{A}_0=\frac{1}{2\,m_V\,\sqrt{q^2}}\Big[\Big(\,m_B^2-m_V^2-q^2\Big)(m_B+m_V)A_1(q^2)\,-\,\frac{4M_B^2|\vec p_V|^2}{m_B+m_V}A_2(q^2)\Big]\,, \nonumber\\
&&\mathcal{A}_\parallel=\frac{2(m_B+m_V)A_1(q^2)}{\sqrt 2}\,,\nonumber\\
&&\mathcal{A}_\perp=-\frac{4m_BV(q^2)|\vec p_V|}{\sqrt{2}(m_B+m_V)}\,,\nonumber\\
&&\mathcal{A}_t=\frac{2m_B|\vec p_V|A_0(q^2)}{\sqrt {q^2}}\,,\nonumber\\
&&\mathcal{A}_P=-\frac{2m_B|\vec p_V|A_0(q^2)}{(m_b(\mu)+m_c(\mu))}\,.
\end{eqnarray}
We perform the $\cos \theta_l $ integration and obtain the differential decay rate  $d\Gamma/dq^2$, that is
\begin{eqnarray}
\label{vlnu1}
\frac{d\Gamma^V}{dq^2} &=&
\frac{8\,N\,|\overrightarrow{p}_V|}{3}\,\Bigg\{ \mathcal{A}_{AV}^2 + \, \frac{ m_l^2}{2\,q^2}\Big[ \mathcal{A}_{AV}^2 + 3\mathcal{A}_{tP}^2 \Big] 
+
\widetilde{\mathcal{A}}_{AV}^2 + \, \frac{m_l^2}{2\,q^2}\Big[ \widetilde{\mathcal{A}}_{AV}^2 + 3\mathcal{\widetilde{A}}_{tP}^2 \Big] \Bigg\}\,,
\end{eqnarray}
where
\begin{eqnarray}
&&\mathcal{A}_{AV}^2 = \mathcal{A}_0^2\,G_A^2 + \mathcal{A}_\parallel^2\,G_A^2 + \mathcal{A}_\perp^2\,G_V^2 \,, \nonumber\\
&&\widetilde{\mathcal{A}}_{AV}^2=\mathcal{A}_0^2\,\widetilde{G}_A^2 + \mathcal{A}_\parallel^2\,\widetilde{G}_A^2 + \mathcal{A}_\perp^2\,\widetilde{G}_V^2\,, \nonumber\\
&&\mathcal{A}_{tP}=\mathcal{A}_t\,G_A + \frac{\sqrt{q^2}}{m_l}\,\mathcal{A}_P\,G_P \,,\nonumber\\
&&\mathcal{\widetilde{A}}_{tP}=\mathcal{A}_t\,\widetilde{G}_A + \frac{\sqrt{q^2}}{m_l}\,\mathcal{A}_P\,\widetilde{G}_P \,.
\end{eqnarray}

In the SM, $G_V = G_A = 1$ and all other NP couplings are zero. We obtain
\begin{eqnarray}
\Big(\frac{d\Gamma^V}{dq^2}\Big)_{\rm SM} &=&
\frac{8\,N\,|\overrightarrow{p}_V|}{3}\,\Bigg\{(\mathcal A_0^2 + \mathcal A_{||}^2 + \mathcal A_{\perp}^2)\Big(1 + \frac{m_l^2}{2\,q^2}\Big) + \frac{3\,m_l^2}{2\,q^2}\,\mathcal A_t^2\Bigg\}\,.
\end{eqnarray}

We want to mention that
our formulae for the $B \to V\,l\,\nu$ differential decay width in Eq.~(\ref{vlnu}) and Eq.~(\ref{vlnu1}) differ slightly from those reported in Ref.~\cite{datta}. Our formulae, however,
agree with those reported in Ref.~\cite{Fajfer}. In Eq.~(\ref{vlnu}), we have $(1 + \cos^2\,\theta_l)$ instead of $(1 + \cos\theta_l)^2$ reported in Ref.~\cite{datta}. Again, note that our
definition of $G_P = S_L - S_R$, different from that of $g_P = S_R - S_L$~\cite{datta}, leads to a sign discrepancy in $\mathcal{A}_{tP}~(\mathcal {\widetilde{A}}_{tP})$. Depending on the NP couplings $G_P$ and $\widetilde{G}_P$, 
the numerical estimates might differ from Ref.~\cite{datta}.

We define some physical observables such as differential branching ratio~DBR$(q^2)$, the ratio of branching fractions $ R(q^2)$, and the forward-backward asymmetry $A_{FB}(q^2)$.
\begin{eqnarray}
&& DBR(q^2) = \Big(\frac{d\Gamma}{dq^2}\Big)/\Gamma_{tot} \,, \qquad\qquad
R(q^2)=\frac{DBR(q^2)\Big(B \to (P,\,V)\,\tau\,\nu\Big)}{DBR(q^2)\Big(B \to (P,\,V)\,l\,\nu\Big)}\,, 
\nonumber\\
&&[A_{FB}]_{(P,\,V)}(q^2)=\frac{\Big(\int_{-1}^{0}-\int_{0}^{1}\Big)d\cos \theta_l\frac{d\Gamma^{(P,\,V)}}{dq^2\,d\cos\theta_l}}{\frac{d\Gamma^{(P,\,V)}}{dq^2}}\,.
\end{eqnarray}
For $B \to P\,l\,\nu$ decay mode, the forward-backward asymmetry in the presence of NP is
\begin{eqnarray}
\label{eq:afbplnu}
A_{FB}^P(q^2) &=& \frac{3\,m_l^2}{2\,q^2}\frac{H_0\,G_V\,\Big[\Big(H_t\,G_V + \frac{\sqrt{q^2}}{m_l}\,H_S\,G_S \Big) + \Big(H_t\,\widetilde{G}_V + \frac{\sqrt{q^2}}{m_l}\,
H_S\,\widetilde{G}_S\Big)\,\Big]}{H_0^2\,(G_V^2+\widetilde{G}_V^2)(1+\frac{m_l^2}{2\,q^2})+\frac{3\,m_l^2}{2\,q^2}\,\Big[\Big(H_t\,G_V + 
\frac{\sqrt{q^2}}{m_l}\,H_S\,G_S \Big)^2 + \Big(H_t\,\widetilde{G}_V + \frac{\sqrt{q^2}}{m_l}\,H_S\,\widetilde{G}_S\Big)^2\,\Big]}\,,\nonumber \\
\end{eqnarray}
where, in the SM, $G_V = 1$ and all other couplings are zero. We obtain
\begin{eqnarray}
\label{afbsm:plnu}
\Big(A_{FB}^P\Big)_{\rm SM}(q^2) &=& \frac{3\,m_l^2}{2\,q^2}\frac{H_0\,H_t}{H_0^2\,\Big(1 + \frac{m_l^2}{2\,q^2}\Big) + \frac{3\,m_l^2}{2\,q^2}\,H_t^2}\,.
\end{eqnarray}
Similarly, for $B \to V\,l\,\nu$ decay mode, in the presence of NP
\begin{eqnarray}
\label{eq:afbvlnu}
A_{FB}^V(q^2) &=& 
\frac{3}{2}\frac{\mathcal{A}_\parallel\,\mathcal{A}_\perp\,\Big(G_A\,G_V-\widetilde{G}_A\widetilde{G}_V\Big)+\frac{m_l^2}{q^2}\mathcal{A}_0\,
G_A\Big[\mathcal{A}_t\,G_A - \frac{\sqrt{q^2}}{m_l}\,\mathcal{A}_P\,G_P + \mathcal{A}_t\,\widetilde{G}_A - \frac{\sqrt{q^2}}{m_l}\,\mathcal{A}_P\,\widetilde{G}_P \Big]} 
{\mathcal{A}_{AV}^2 +  \frac{m_l^2}{2\,q^2}\Big[ \mathcal{A}_{AV}^2 + 3\mathcal{A}_{tP}^2 \Big] 
+ \widetilde{\mathcal{A}}_{AV}^2 + \frac{m_l^2}{2\,q^2}\Big[\widetilde{\mathcal{A}}_{AV}^2 + 3\mathcal{\widetilde{A}}_{tP}^2 \Big]}\,.\nonumber \\
\end{eqnarray} 
In the SM, $G_A = G_V = 1$ while all other NP couplings are zero. Thus we obtain
\begin{eqnarray}
\label{afbsm:vlnu}
\Big(A_{FB}^V\Big)_{\rm SM}(q^2) &=& \frac{3}{2}\,\frac{\mathcal{A}_{\parallel}\,\mathcal{A}_{\perp} + \frac{m_l^2}{q^2}\mathcal{A}_0\,\mathcal{A}_t}{
\Bigg\{(\mathcal A_0^2 + \mathcal A_{||}^2 + \mathcal A_{\perp}^2)\Big(1 + \frac{m_l^2}{2\,q^2}\Big) + \frac{3\,m_l^2}{2\,q^2}\,\mathcal A_t^2\Bigg\}}\,.
\end{eqnarray}
We see that, in the SM, for the light leptons $l = e,\,\mu$, the forward-backward asymmetry is vanishingly small due to the $m_l^2/q^2$ term for the $B \to P\,l\,\nu$ decay modes. 
However, for $B \to V\,l\,\nu$, the first term will contribute and we will get a nonzero value for the forward-backward asymmetry. Any non-zero value of the $A_{FB}$ parameter
for the $B \to P\,l\,\nu$ decay modes will be a hint of NP in all generation leptons. We, however, ignore the NP effects in the case of $l = e,\,\mu$. We strictly assume that only third 
generation leptons get modified due to NP couplings.

We wish to determine various NP effects in a model independent way. The theoretical uncertainties in the calculation of the decay branching fractions come from various input parameters.
first, there are uncertainties associated with well-known input
parameters such as quark masses, meson masses, and lifetime of the mesons. We ignore these uncertainties as these are not important for our analysis.
Second, there are uncertainties that are associated with not so well-known hadronic input parameters such as form factors, decay constants, and the CKM elements. In order to realize the effect
of the above-mentioned uncertainties on various observables, we use a random number generator and perform a random scan of all the allowed hadronic as well as the CKM elements. In our random scan of
the theoretical parameter space, we vary
all the hadronic inputs such as $B \to (P,\,V)$ form factors, $f_{B_q}$ decay constants, and CKM elements $|V_{qb}|$ within $3\sigma$ from their central values.
In order to determine the allowed NP parameter space, we impose the experimental constraints coming from the measured ratio of branching fractions $R_{\pi}^l$, $R_D$, and $R_{D^{\ast}}$ simultaneously.
This is to ensure that the resulting NP parameter space can simultaneously accommodate all the existing data on $b \to u$ and $b \to c$ leptonic and semileptonic decays.
We impose the experimental constraints in such a way that we ignore those theoretical models that are not compatible within $3\sigma$ of the experimental constraints for the $3\sigma$ random scan.

\section{Results and discussion}
\label{rnd}
For definiteness, we summarize the input parameters for our numerical analysis. We use the following inputs from Ref.~\cite{pdg}.
\begin{eqnarray}
\label{inputs}
&&m_b = 4.18\,{\rm GeV}\,,\qquad\qquad
m_c=1.275\,{\rm GeV}\,,\qquad\qquad
m_{\pi} = 0.13957\,{\rm GeV}\,,\nonumber \\
&&m_{B^{-}} = 5.27925\,{\rm GeV}\,,\qquad\qquad
m_{B^{0}} = 5.27955\,{\rm GeV}\,,\qquad\qquad
m_{B_{c}} = 6.277\,{\rm GeV}\,,\nonumber \\
&&m_{D^0} = 1.86486\,{\rm GeV}\,,\qquad\qquad
m_{D^{\ast\,0}} = 2.00698\,{\rm GeV}\,,\qquad\qquad
\tau_{B^0} = 1.519\times 10^{-12}\,{\rm Sec}\,,\nonumber \\
&&\tau_{B^-} = 1.641\times 10^{-12}\,{\rm Sec}\,,\qquad\qquad
\tau_{B_c} = 0.453\times 10^{-12}\,{\rm Sec}\,,
\end{eqnarray}
where $m_b \equiv m_b(m_b)$ and $m_c \equiv m_c(m_c)$ denote the running $b$ and $c$ quark masses in $\overline{\rm MS}$ scheme. We employ a renormalization scale $\mu = m_b$ for which
the strong coupling constant $\alpha_s(m_b) = 0.224$. Using the two-loop expression for the running quark mass~\cite{Buchalla:1995vs}, we find $m_c(m_b) = 0.91\,{\rm GeV}$. Thus, the coefficients $V_{L,\,R}$, $\widetilde{V}_{L,\,R}$, $S_{L,\,R}$, and $\widetilde{S}_{L,\,R}$ are defined at the scale $\mu = m_b$. The error associated with the quark
masses, meson masses, and the mean lifetime of mesons is not important and we ignore them in our analysis. In Table~\ref{tab1} and Table~\ref{tab2}, we present the most important theoretical and experimental inputs with their uncertainties that are used for our random scan.
%%%%%%%%%%%%%%%%%%%%%%%%%%%%%%%%%%
\begin{table}[htdp]
\begin{center}
\begin{tabular}{|c|c|c|c|}
\hline
\multicolumn {2}{|c|}{CKM Elements:} & \multicolumn {2}{|c|}{Meson Decay constants~(in GeV):} \\[0.2cm]
\hline
$ |V_{ub}| $ (Exclusive) & $(3.23 \pm 0.31 ) \times 10^{-3} $~\cite{pdg} & $ f_B $ & $ 0.1906 \pm 0.0047 $~\cite{Bazavov:2011aa, Na:2012kp, latticeavg} \\[0.2cm]
$ |V_{cb}| $ (Average) & $ (40.9 \pm 1.1 ) \times 10^{-3} $~\cite{pdg} &$ f_{B_c} $ & $ 0.395 \pm 0.015 $~\cite{Choudhury} \\[0.2cm]
\hline
\multicolumn{2}{|c|}{Inputs for $(B \to \pi)$ Form Factors:} & \multicolumn{2}{|c|}{Inputs for $(B \to D^{\ast})$ Form Factors:} \\[0.2cm]
\hline
$ F_{+}(0)=F_{0}(0)$ & $ 0.281\pm0.028$ ~\cite{Khodjamirian} & $ h_{A_1}(1)|V_{cb}| $ & $ (34.6\pm1.02)\times 10^{-3} $ ~\cite{Dungel:2010uk} \\[0.2cm]
$ b_1 $ & $ -1.62\pm 0.70 $ ~\cite{Khodjamirian} & $ \rho_1^2 $ & $ 1.214\pm 0.035$ ~\cite{Dungel:2010uk} \\[0.2cm]
$ b_1^0 $ & $ -3.98\pm 0.97 $ ~\cite{Khodjamirian} & $ R_1(1) $ & $ 1.401\pm 0.038$ ~\cite{Dungel:2010uk} \\[0.2cm]
\cline{1-2}
\cline{1-2}
\multicolumn{2}{|c|}{Inputs for $(B \to D)$ Form Factors:} & $ R_2(1) $ & $ 0.864\pm 0.025$ ~\cite{Dungel:2010uk} \\[0.2cm]
\cline{1-2}
$ V_1(1)|V_{cb}| $ & $ (43.0\pm2.36)\times 10^{-3} $ ~\cite{Aubert:2009ac} & $ R_0(1) $ & $ 1.14\pm 0.114$ ~\cite{Fajfer} \\[0.2cm]
$ \rho_1^2 $ & $ 1.20\pm 0.098$ ~\cite{Aubert:2009ac} & $ $  & $ $\\[0.2cm]
\hline
\end{tabular}
\end{center}
\caption{Theory input parameters}
\label{tab1}
\end{table}
%%%%%%%%%%%%%%%%%%%%%%%%%%%%%%%%%%
%%%%%%%%%%%%%%%%%%%%%%%%%%%%%%%%%%%
\begin{table}[htdp]
\begin{center}
\begin{tabular}{|c|c|}
\hline
\multicolumn{2}{|c|}{Ratio of branching ratios:} \\[0.2cm]
\hline
$ R_{\pi}^l $ & $0.73 \pm 0.15$ ~\cite{Fajfer:2012jt} \\[0.2cm]
$ R_D $ & $0.440 \pm 0.072 $ ~\cite{babar-dstnu} \\[0.2cm]
$ R_{D^{\ast}} $ & $0.332 \pm 0.030 $ ~\cite{babar-dstnu} \\[0.2cm]
\hline
\end{tabular}
\end{center}
\caption{Experimental input parameters}
\label{tab2}
\end{table}
%%%%%%%%%%%%%%%%%%%%%%%%%%%%%%%%%%%

We wish to study the effects of each new physics parameter on various observables and the $B_c \to \tau\nu$ and $B^0 \to \pi\tau\nu$ decays in a model independent way. 
We also 
consider the ratio of branching fractions of $B^0 \to \pi\tau\nu$ to $B^0 \to \pi\,l\nu$ decays, defined as
\begin{eqnarray}
&&R_{\pi} = \frac{\mathcal B(B \to \pi\tau\nu)}{\mathcal B(B \to \pi\,l\,\nu)}\,,
\end{eqnarray}
which, in the SM, only depends
on the ratio of form factors $F_0(q^2)/F_{+}(q^2)$. The decay mode $B \to \pi\tau\nu$ is particularly important because it originates from the same flavor changing
interaction as the $B \to \tau\nu$ decay mode and hence can be used as an indicator for NP operators. Similarly, the $B_c \to \tau\nu$ is important as it is mediated via
$b \to c$ transition decays, same as $B \to D\,\tau\,\nu$ and $B \to D^{\ast}\,\tau\,\nu$ decays, and, in principle, can help in identifying the nature of NP in $b \to c$ processes.
The SM prediction for the branching ratios and ratio of branching ratios is reported in Table.~\ref{tab3},
\begin{table}[htdp]
\begin{center}
\begin{tabular}{|c|c|c|}
\hline
$ $ & Central value & $1\sigma$ range\\
\hline
$\mathcal B(B \to \tau\nu)$ & $6.70\times 10^{-5}$ & $(5.22,\,8.45)\times 10^{-5}$ \\[0.2cm]
$\mathcal B(B_c \to \tau\nu)$ & $1.63\times 10^{-2}$ & $(1.43,\,1.85)\times 10^{-2}$ \\[0.2cm]
$\mathcal B(B \to \pi\,l\,\nu)$ & $12.77\times 10^{-5} $ & $(7.39, \, 21.28)\times 10^{-5}$  \\[0.2cm]
$\mathcal B(B \to \pi\,\tau\,\nu)$ & $8.91\times 10^{-5}$ & $(4.93, \, 15.40)\times 10^{-5}$ \\[0.2cm]
$\mathcal B(B \to D\,l\,\nu)$ & $2.32\times 10^{-2}$ & $(1.89,\,2.81)\times 10^{-2}$ \\[0.2cm]
$\mathcal B(B \to D\,\tau\,\nu)$ & $0.72\times 10^{-2}$ & $(0.62,\,0.84)\times 10^{-2}$ \\[0.2cm]
$\mathcal B(B \to D^{\ast}\,l\,\nu)$ & $4.93\times 10^{-2}$ & $(4.51,\,5.39)\times 10^{-2}$ \\[0.2cm]
$\mathcal B(B \to D^{\ast}\,\tau\,\nu)$ & $1.25\times 10^{-2}$ & $(1.14,\,1.37)\times 10^{-2}$ \\[0.2cm]
$R_{\pi}^l$  & $0.486$ & $(0.328,\,0.733)$ \\[0.2cm]
$R_{\pi}$  & $0.698$ & $(0.654,\,0.764)$ \\[0.2cm]
$R_D$  & $0.313$ & $(0.300,\,0.327)$ \\[0.2cm]
$R_D^{\ast}$  & $0.253$ & $(0.245,\,0.261)$ \\[0.2cm]
\hline
\end{tabular}
\end{center}
\caption{Branching ratio and ratio of branching ratios within the SM.}
\label{tab3}
\end{table}
where, for the central values
we have used the central values of all the input parameters from Eq.~(\ref{inputs}) and from Table.~\ref{tab1}. We vary all the theory inputs such as
$B_q$ meson decay constants, $B \to (P,V)$ transition form factors and the CKM matrix elements $|V_{qb}|$ within $1\sigma$ of their central values and obtain
the $1\sigma$ allowed ranges in all the different observables in Table.~\ref{tab3}.
The uncertainties associated with the input parameters for the calculation of the form factors, reported in Appendix~\ref{ffpi} and Appendix~\ref{ffdd}, 
are added in quadrature and tabulated in Table~\ref{tab1}. 

We now proceed to describe four
different scenarios of new physics and the effect of these NP parameters. We consider all the NP parameters to be real for our analysis.
We assume that only the third generation leptons get corrections from the NP couplings in the $b\to(u,\,c)\,l\nu $ processes and for $l=e^-,\mu^-$ cases the NP is absent. We use $3\sigma$ experimental constraint coming from the
ratio of branching ratios $R_{\pi}^l$, $R_D$, and $R_D^{\ast}$ to find the allowed ranges of all the NP couplngs. We then show how different observables
behave with various NP couplings under four different NP scenarios that we consider for our analysis. We also give predictions for the branching ratios
of $B_c \to \tau\nu$ and $B \to \pi\tau\nu$ decays and the ratio $R_{\pi}$ for all the different NP scenarios. 
%%%%%%%%
%%%%%%%%%%%%%%%%%%%
\subsection{Scenario A}
\begin{figure}[htbp]
\begin{center}
\includegraphics[width=8cm,height=5cm]{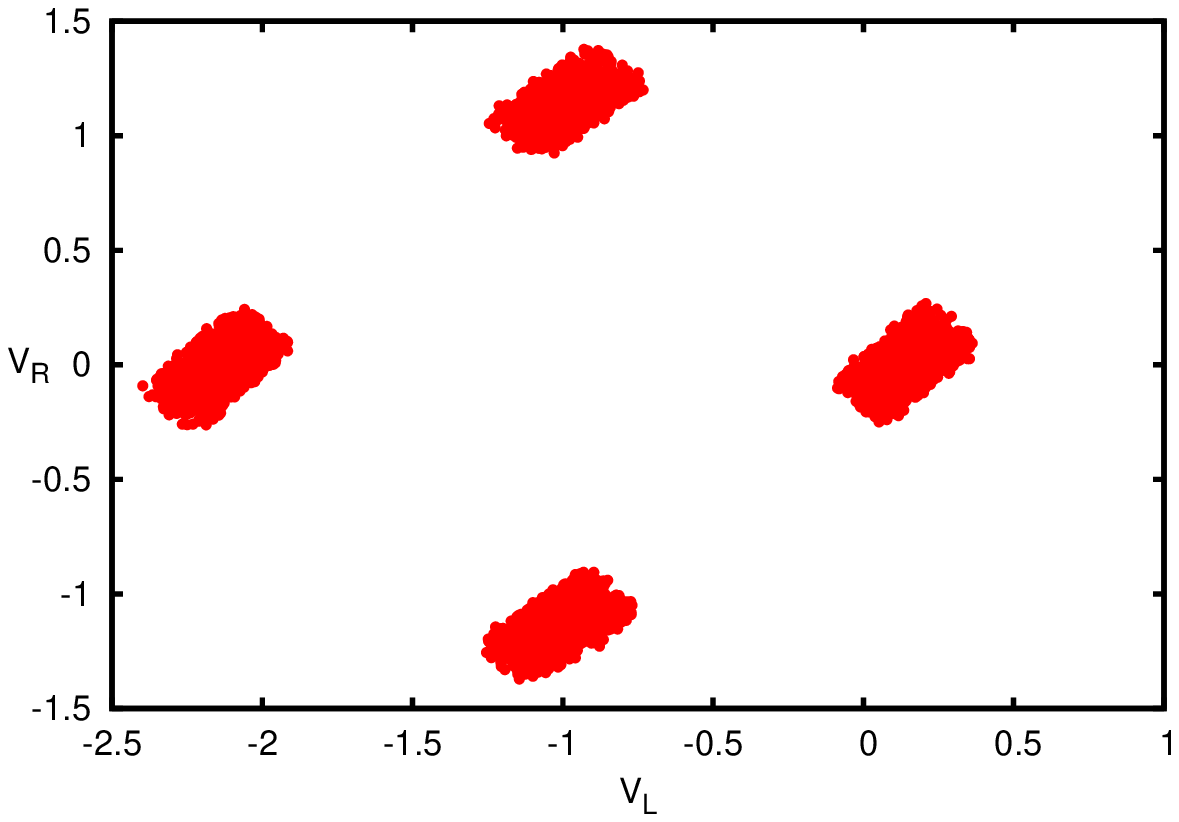}
\includegraphics[width=8cm,height=5cm]{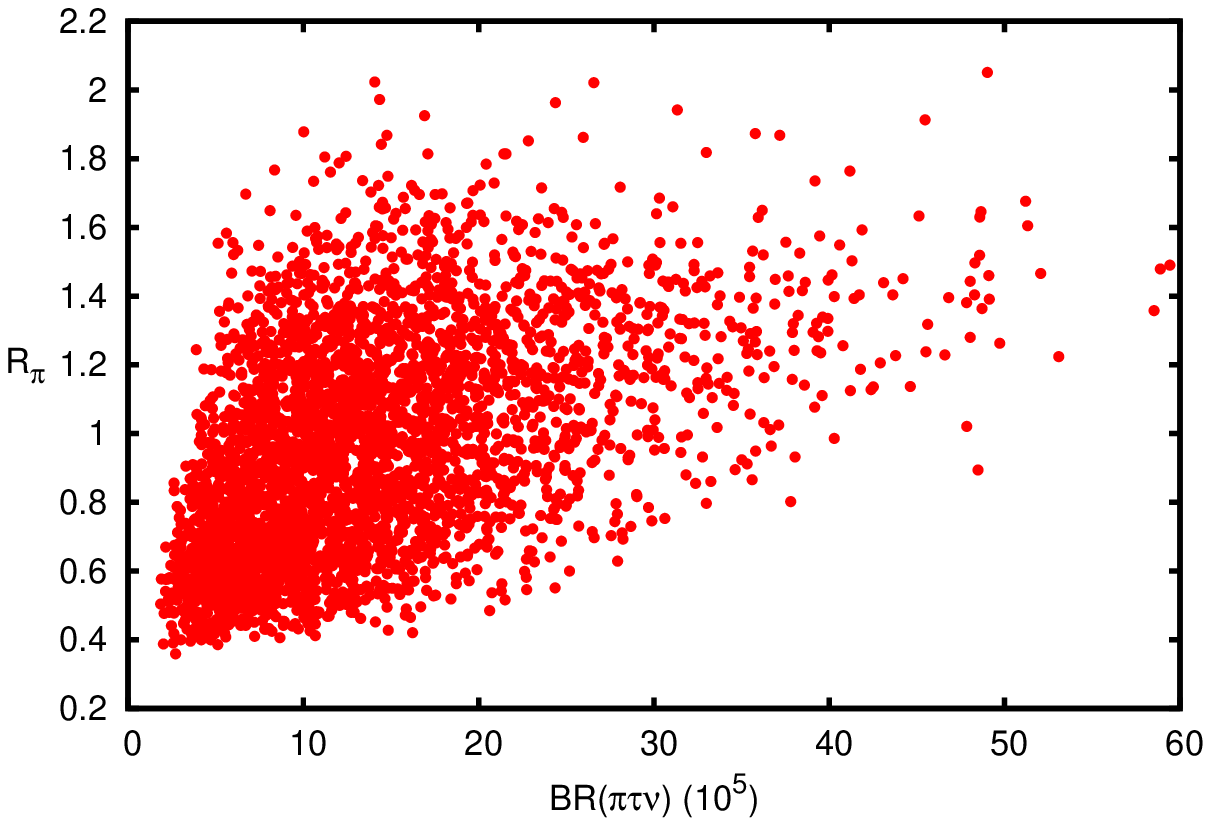}
\end{center}
\caption{Allowed regions of $V_L$ and $V_R$ are shown in the left panel once the $3\sigma$ experimental constraint is imposed. The corresponding ranges in $\mathcal B(B \to \pi\tau\nu)$ and the ratio $R_{\pi}$ in
the presence of these NP couplings are shown in the right panel.}
\label{vlvr_tau}
\end{figure}

We vary $V_L$ and $V_R$ while keeping all other NP couplings to zero. The allowed ranges of $V_L$ and $V_R$ that satisfies $3\sigma$ constraint coming
from  $R_{\pi}^l$, $R_D$, and $R_D^{\ast}$ are shown in the left panel of Fig.~\ref{vlvr_tau}. We see that the experimental values put a severe constraint on the $(V_L, V_R)$
parameter space. 
In the presence of such NP couplings, the $\Gamma(B_q \to \tau\nu)$, $d\Gamma/dq^2(B \to P\,\tau\nu)$, and $d\Gamma/dq^2(B \to V\,\tau\nu)$, where $P$ stands for
pseudoscalar and $V$ stands for vector meson, can be written as
\begin{eqnarray}
\label{eq:vlvr}
\Gamma(B_q \to \tau\nu) &=& \Gamma(B_q \to \tau\nu)|_{\rm SM}\,G_A^2\,,\nonumber \\
\frac{d\Gamma}{dq^2}(B \to P\,\tau\,\nu) &=& \Big[\frac{d\Gamma}{dq^2}(B \to P\,\tau\,\nu)\Big]_{\rm SM}\,G_V^2\,, \nonumber \\ 
\frac{d\Gamma}{dq^2}(B \to V\,\tau\,\nu) &=& \frac{8\,N\,|\overrightarrow{p}_V|}{3}\Bigg\{(\mathcal A_0^2\,G_A^2 + \mathcal A_{||}^2\,G_A^2 + \mathcal A_{\perp}^2\,G_V^2)\,
\Big(1 + \frac{m_{\tau}^2}{2\,q^2}\Big) + \frac{3\,m_{\tau}^2}{2\,q^2}\,\mathcal A_t^2\,G_A^2\Bigg\}\,.
\end{eqnarray}
It is evident that, the value of $\mathcal B(B_c \to \tau\nu)$ varies as $G_A^2$, whereas, $\mathcal B(B \to \pi\tau\nu)$ and the ratio $R_{\pi}$ varies as $G_V^2$ in the 
presence of these NP couplings. The ranges
in $B \to \pi\tau\nu$ branching ratio and the ratio $R_{\pi}$ in the presence of $V_L$ and $V_R$ are shown in the right panel of Fig.~\ref{vlvr_tau}. 
The resulting ranges in $\mathcal B(B_c \to \tau\nu)$, $\mathcal B(B \to \pi\tau\nu)$, and $R_{\pi}$ are
\begin{eqnarray*}
&&\mathcal B(B_c \to \tau\nu) = (1.02,\,3.95)\%\,, \qquad\qquad
\mathcal B(B \to \pi\tau\nu)  = (1.86,\,59.42)\times 10^{-5}\,,\nonumber \\
&&
R_{\pi} = (0.36,\,2.05)\,.
\end{eqnarray*}
We see a significant deviation from the the SM
expectation in such new physics scenario. Measurement of the $\mathcal B(B_c \to \tau\nu)$, $\mathcal B(B \to \pi\tau\nu)$ and the ratio $R_{\pi}$ will put additional constraints on the NP
parameters.
\begin{figure}[htbp]
\begin{center}
\includegraphics[width=5cm,height=4cm]{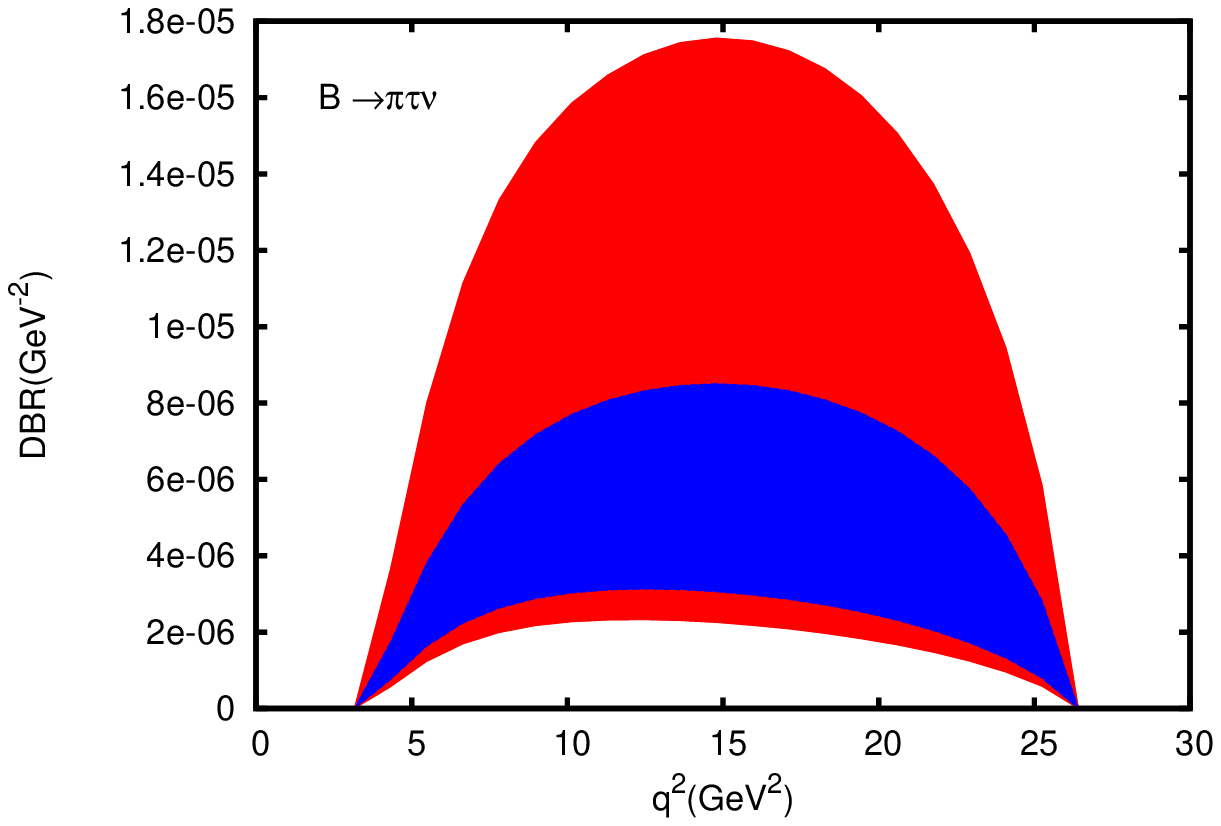}
\includegraphics[width=5cm,height=4cm]{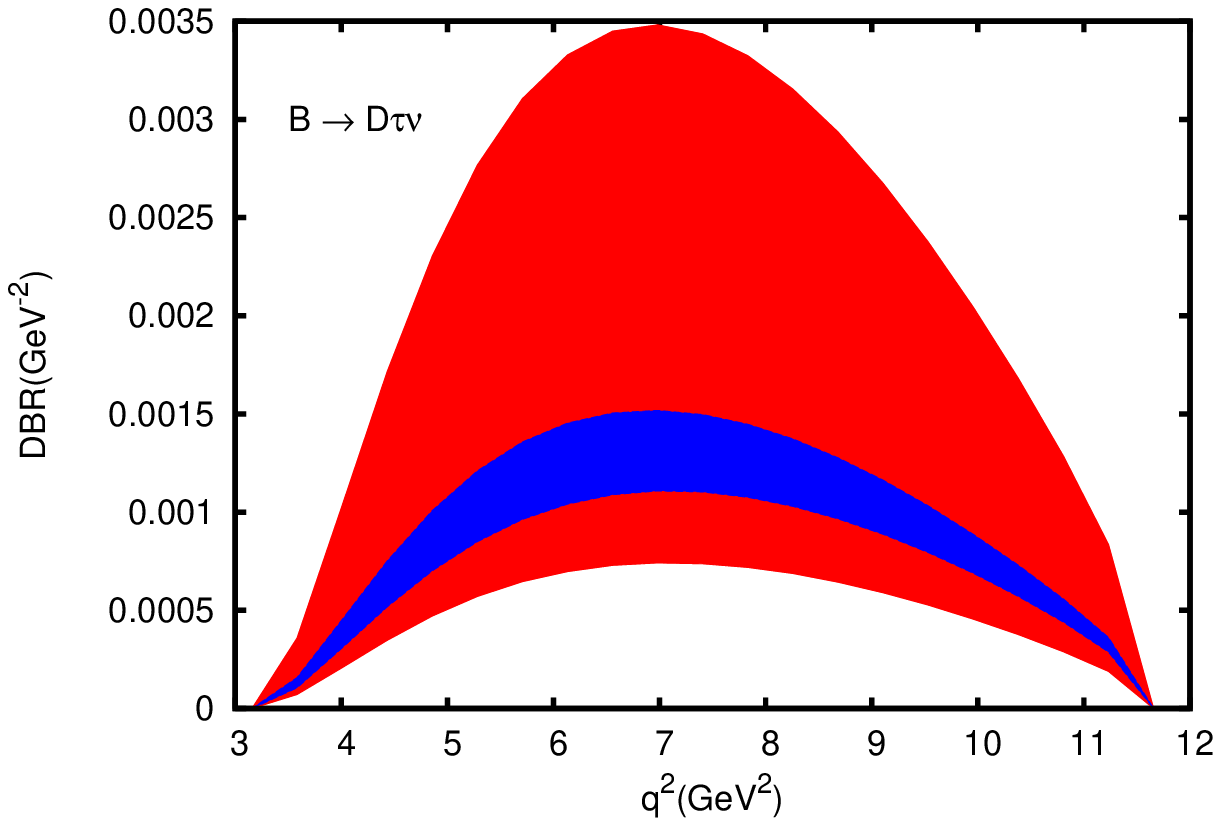}
\includegraphics[width=5cm,height=4cm]{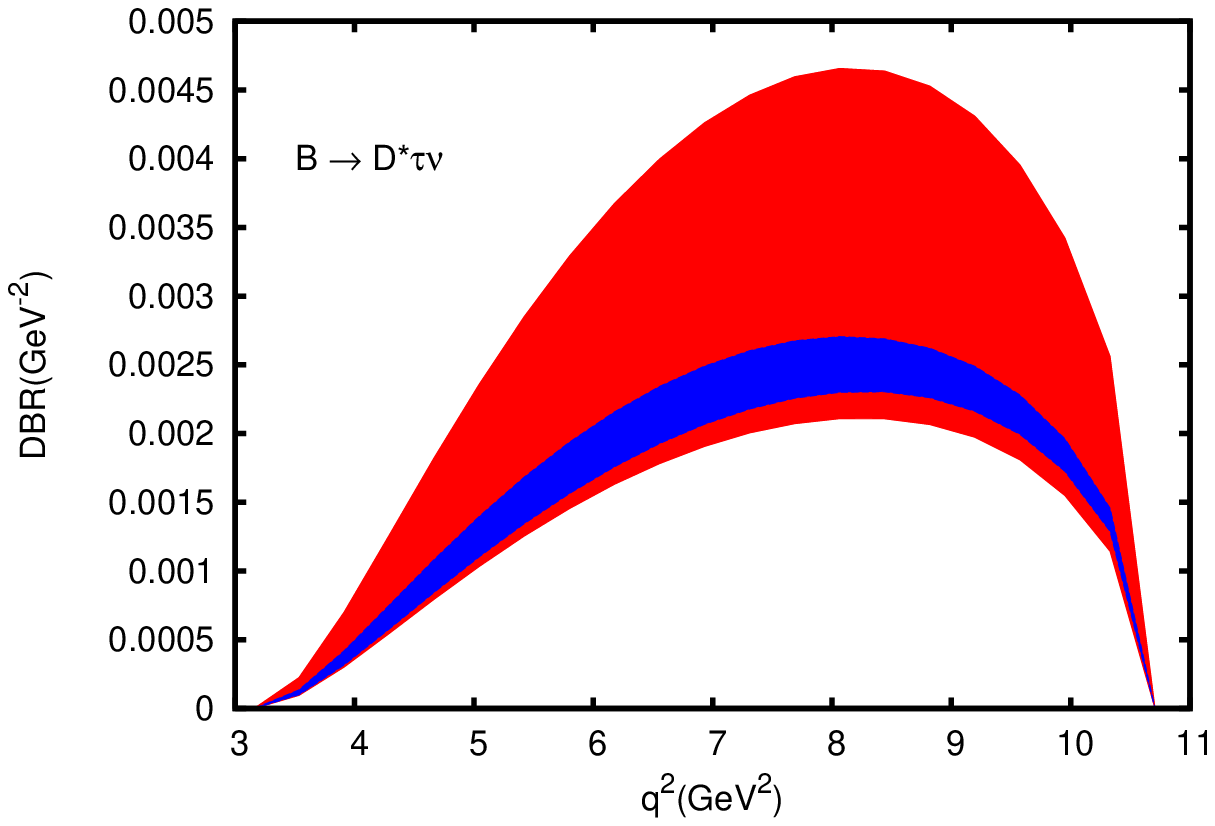}
\includegraphics[width=5cm,height=4cm]{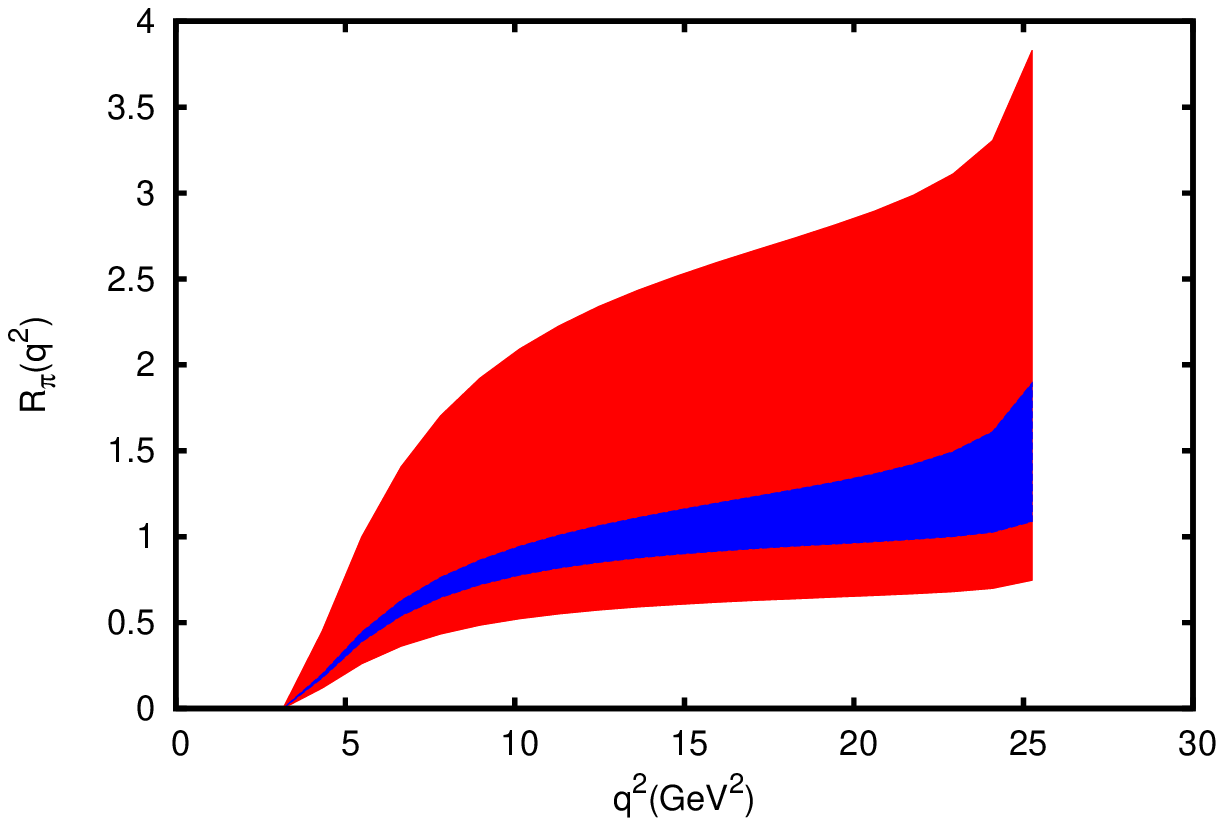}
\includegraphics[width=5cm,height=4cm]{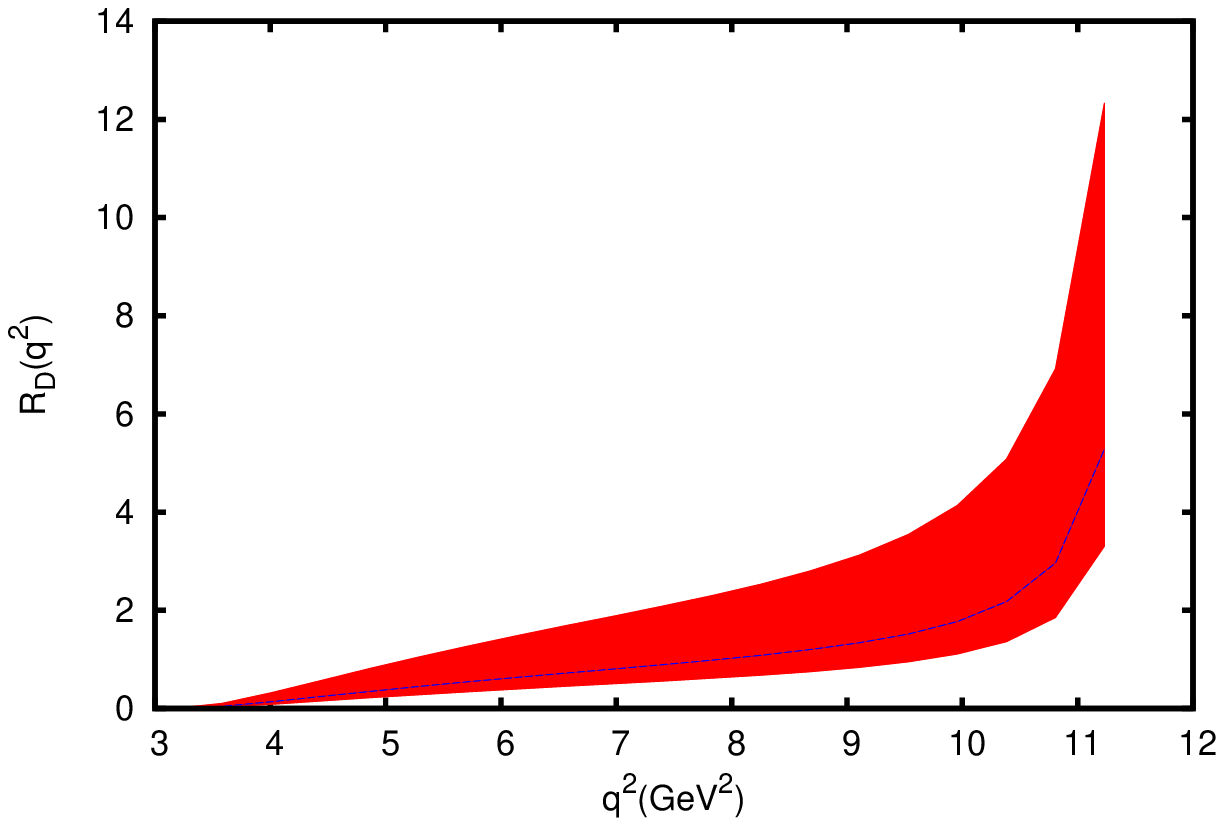}
\includegraphics[width=5cm,height=4cm]{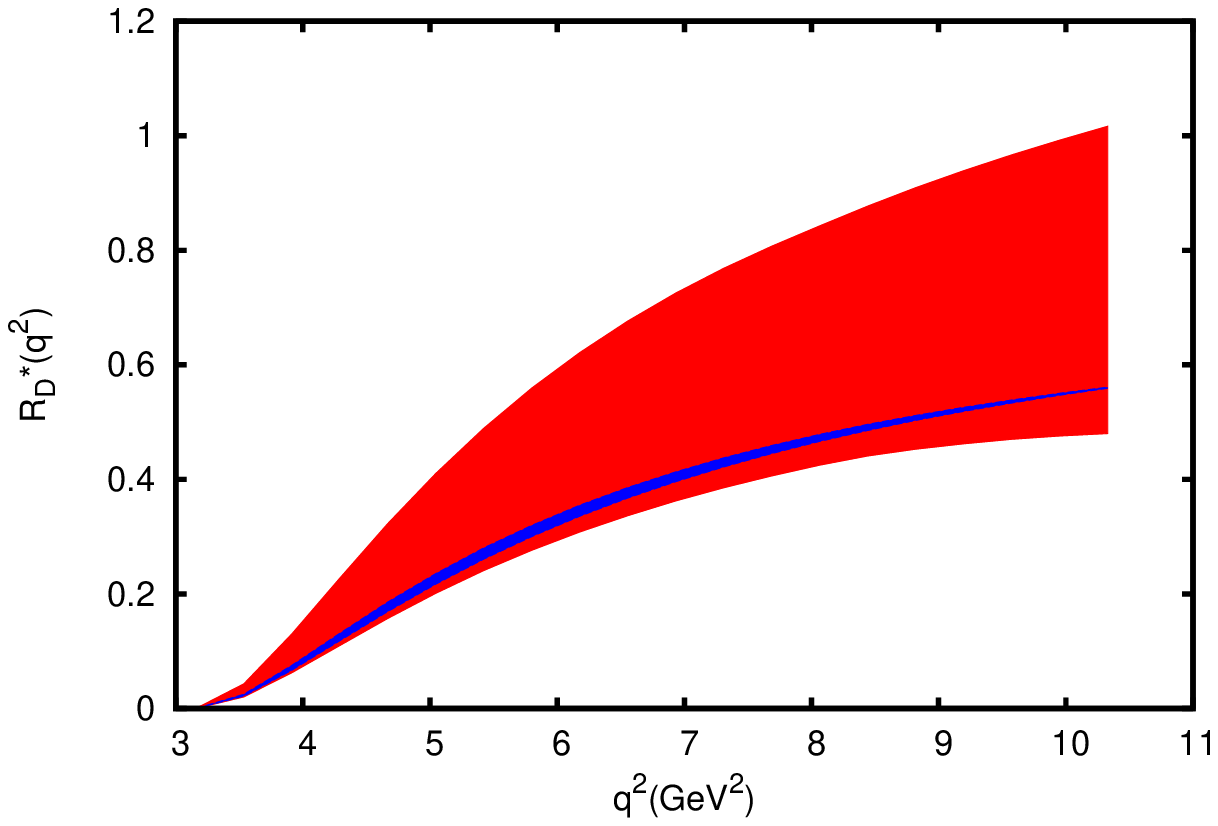}
\includegraphics[width=5cm,height=4cm]{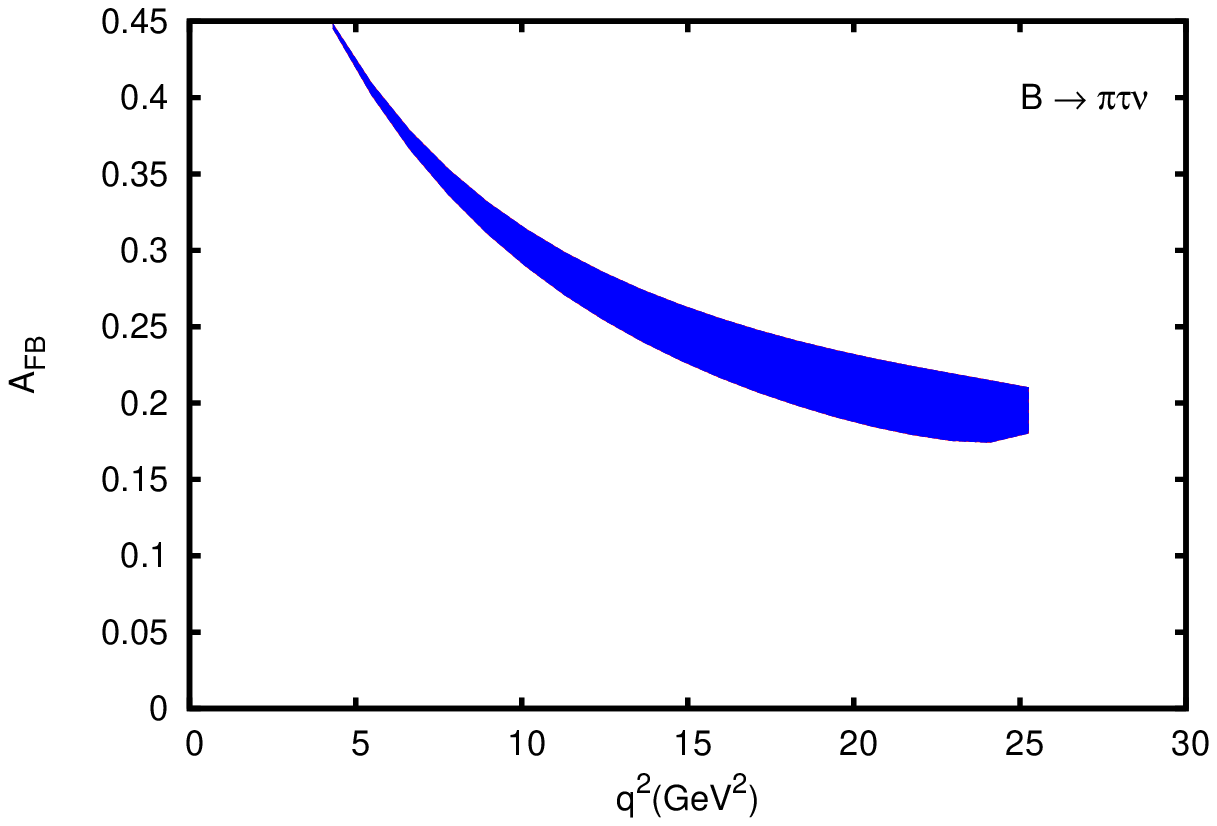}
\includegraphics[width=5cm,height=4cm]{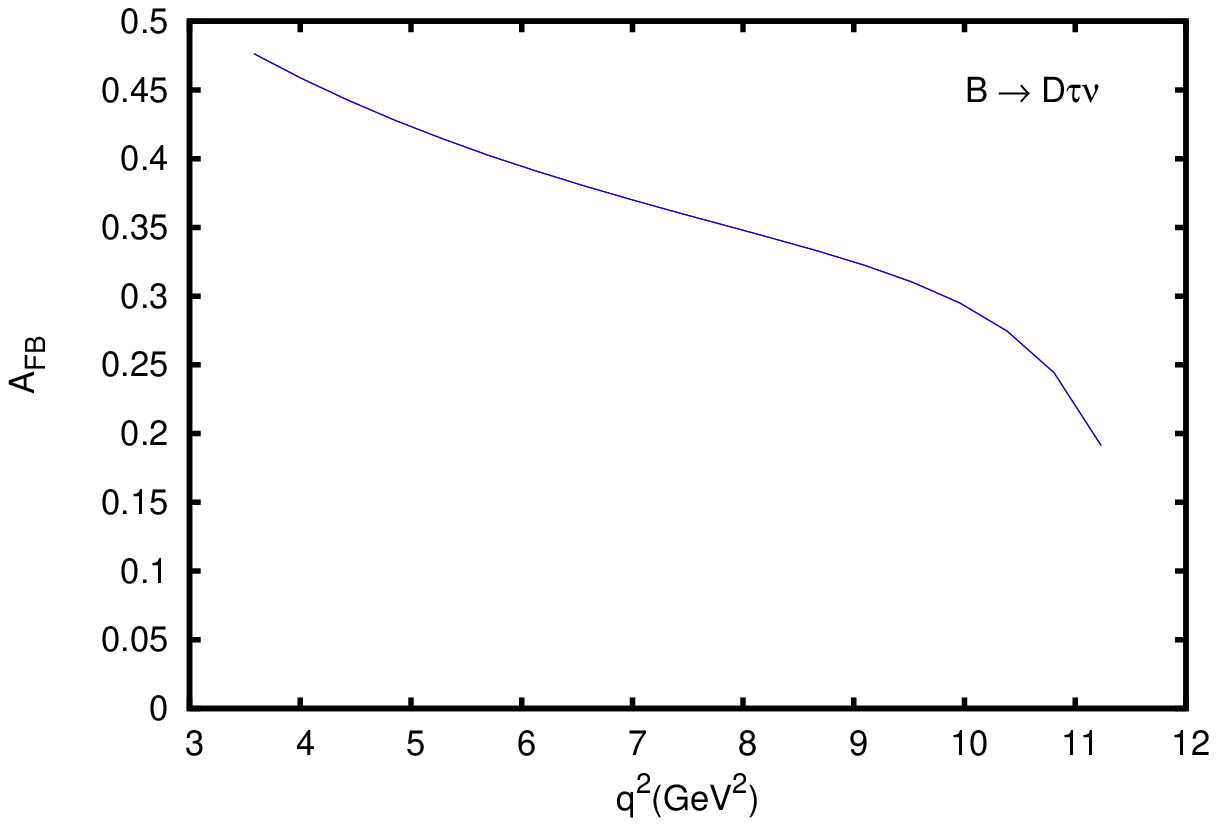}
\includegraphics[width=5cm,height=4cm]{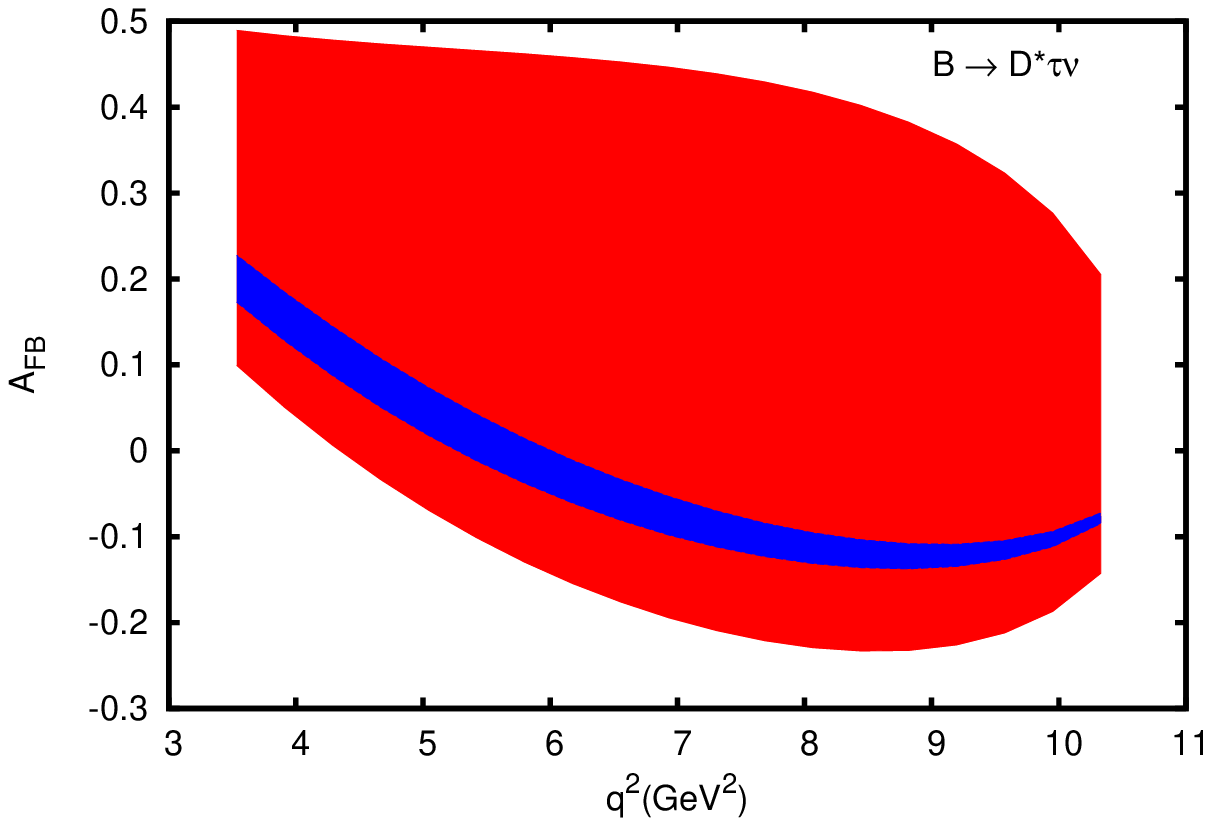}
\end{center}
\caption{Range in DBR$(q^2)$, $R(q^2)$, and the forward backward asymmetry $A_{FB}(q^2)$ for the $B \to \pi\tau\nu$, $B \to D\tau\nu$, and $B \to D^{\ast}\tau\nu$ decay modes. The darker~(blue)
interior region corresponds to the SM prediction, whereas, the lighter~(red), larger region corresponds to the allowed $(V_L,\,V_R)$ NP couplings of Fig.~\ref{vlvr_tau}.}
\label{obs_vlvr_tau}
\end{figure}
We want to see the effects of these NP couplings on various observables that we defined in Sec.~\ref{th}. In Fig.~\ref{obs_vlvr_tau}, we show in blue~(dark) bands the SM
range and show in red~(light) bands the range of each observable once the NP couplings $V_L$ and $V_R$ are switched on. It is clear from Fig.~\ref{obs_vlvr_tau} that,
the differential branching ratios~(DBR) and the ratio of branching ratio get considerable deviations once we include the NP couplings. This is expected and can be understood 
very easily from Eq.~(\ref{eq:vlvr}). In the presence of $V_L$ and $V_R$ alone, the DBR and the ratio for $B \to P\,\tau\,\nu$ decays depends on only $G_V$ coupling and is proportional
to $G_V^2$. 
Whereas, for $B \to V\tau\nu$ decay mode the DBR and the ratio depends on $G_V$ as well as $G_A$ couplings and is proportional to $G_V^2$ and $G_A^2$ as can be seen from Eq.~(\ref{eq:vlvr}). 
We see that the DBR for each decay mode can increase by $100\%$ at the peak of its distribution. Similar conclusions can be made for the ratio of branching ratios as well where we see a $100\%$
increase at the peak of its distribution.
The forward-backward asymmetry, as we expected, does not vary with $V_L$ and $V_R$ for the $B \to \pi\tau\nu$ and the $B \to D\tau\nu$ decay modes. Since it depends on $G_V$ couplings
only, the NP dependency gets canceled in the
ratio as can be seen from Eq.~(\ref{eq:afbplnu}). However, for $B \to D^{\ast}\tau\nu$, the deviation is quite large. Again, it can be very easily understood from Eq.~(\ref{eq:afbvlnu}). 
It is mainly because of the presence of $G_V$ as well as $G_A$ couplings. We see a zero crossing
at $q^2 \approx 6.0\,{\rm GeV^2}$ in the SM for this decay mode.
However, in the presence of such NP, depending on $V_L$ and $V_R$, there may or may not be a zero crossing as is evident from Fig.~\ref{obs_vlvr_tau}.

Again, we want to emphasize the fact that
a pure $G_V$ coupling will contribute to the $B \to P\,\tau\nu$ as well as $B \to V\,\tau\nu$ decay processes, whereas a pure $G_A$ coupling will contribute to the $B \to \tau\nu$ as well as the      
$B \to V\,\tau\,\nu$ decay modes. We do not consider pure $G_V$ and $G_A$ couplings for
our analysis as a pure $G_V$ or a pure $G_A$ type NP coupling will not be able to accommodate all the existing data since current experiments
on $b \to u$ and $b \to c$ semi-(leptonic) decays suggest that there could be new physics in all the three decay modes. Hence, if NP is present in $R_{\pi}^l$, $R_D$, and $R_{D^{\ast}}$, one
can rule out the possibility of having a pure $G_V$ or a pure $G_A$ type of NP couplings.

\subsection{Scenario B}
\begin{figure}[htbp]
\begin{center}
\includegraphics[width=8cm,height=5cm]{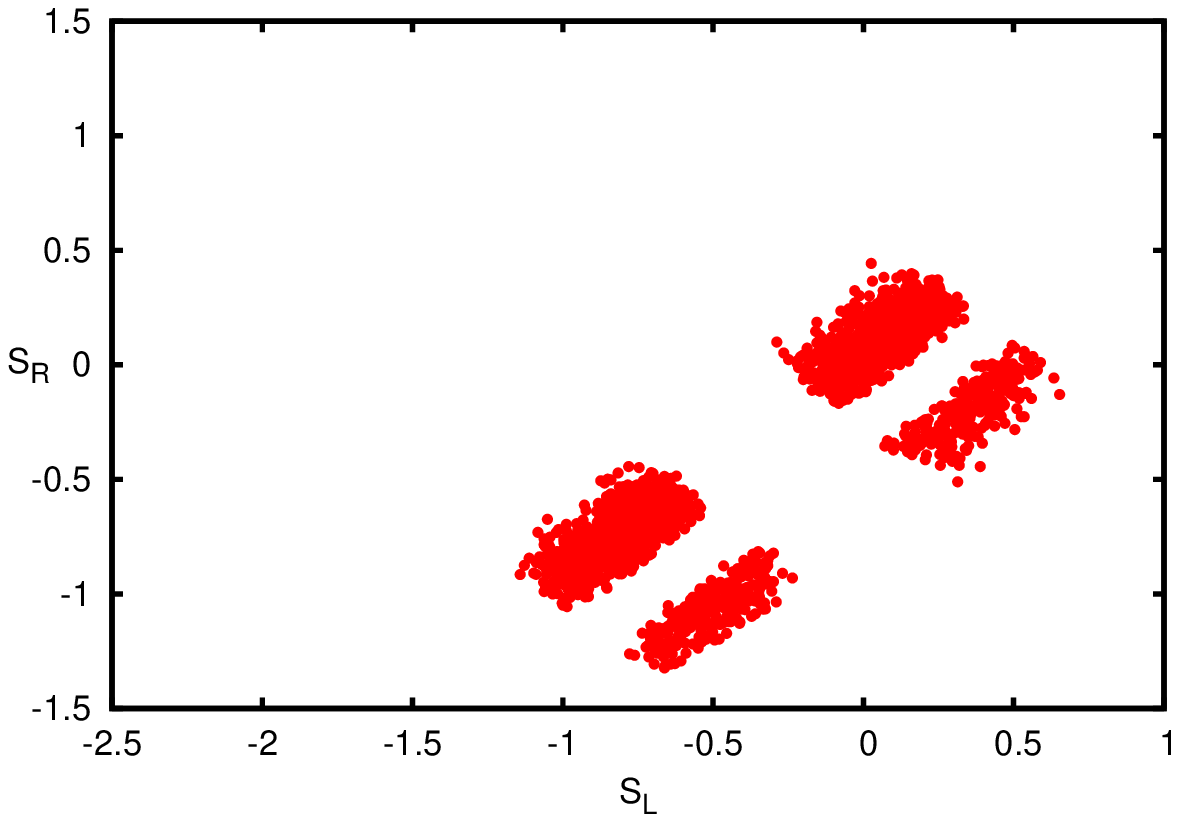}
\includegraphics[width=8cm,height=5cm]{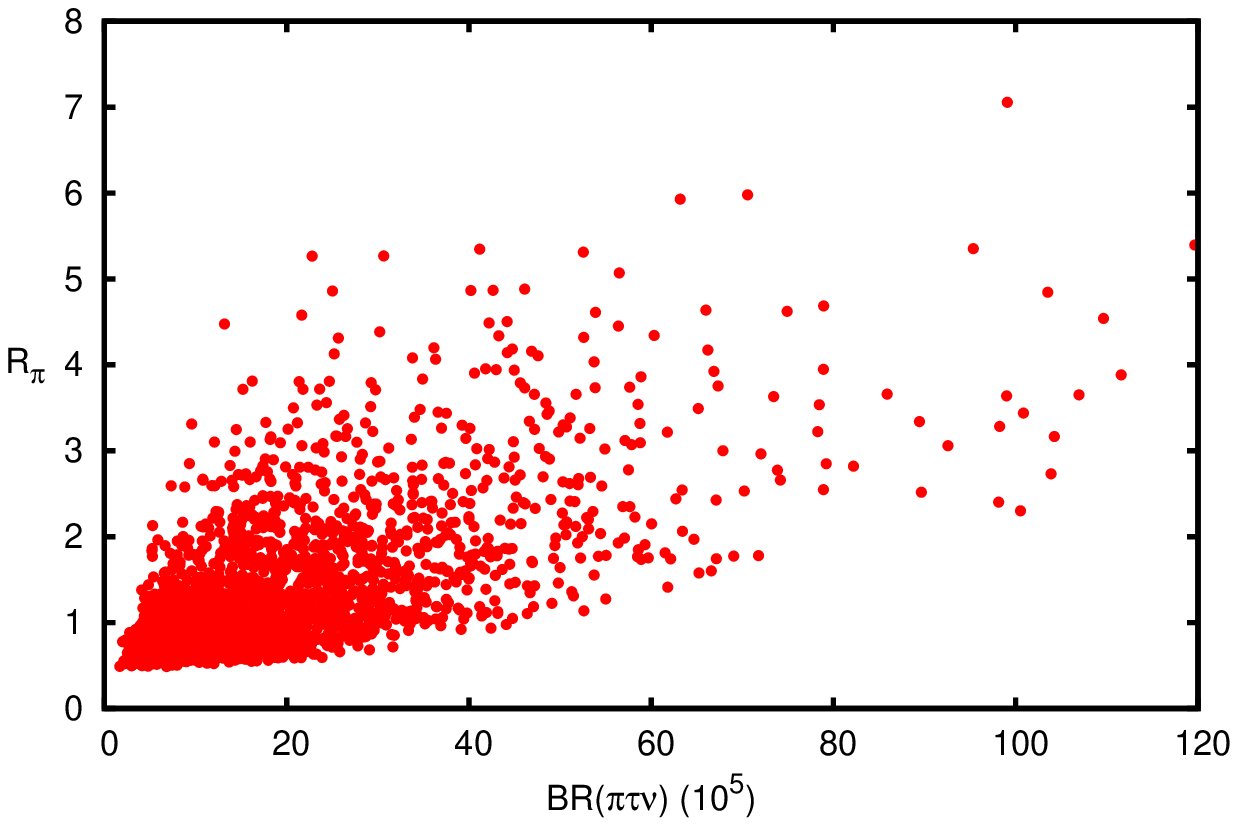}
\end{center}
\caption{Allowed ranges of $(S_L,\,S_R)$ is shown in the left panel once the experimental constraint is imposed. The right panel shows the ranges of $B \to \pi\tau\nu$ branching
fractions and the ratio $R_{\pi}$ with these NP couplings.}
\label{slsr_tau}
\end{figure}

Here we consider nonzero $S_L$ and $S_R$ couplings and keep all other NP couplings to zero. 
The allowed ranges of $S_L$ and $S_R$ that satisfy the $3\sigma$ experimental constraints are shown in the left panel of Fig.~\ref{slsr_tau}. 
In the presence of $S_L$ and $S_R$, the $\Gamma(B_q \to \tau\nu)$, $d\Gamma/dq^2(B \to P\,\tau\nu)$, and $d\Gamma/dq^2(B \to V\,\tau\nu)$ can be written as
\begin{eqnarray}
\label{eq:slsr}
\Gamma(B_q \to \tau\nu) &=& \Gamma(B_q \to \tau\nu)|_{\rm SM}\,\Big[1 - \frac{m_B^2}{m_{\tau}\,(m_b + m_q)}\,G_P\Big]^2\,,\nonumber \\
\frac{d\Gamma}{dq^2}(B \to P\,\tau\,\nu) &=& \frac{8\,N\,|\overrightarrow{p}_P|}{3}\Bigg\{H_0^2\Big(1 + \frac{m_{\tau}^2}{2\,q^2}\Big) + \frac{3\,m_{\tau}^2}{2\,q^2}\,H_t^2 + 
\frac{3}{2}\Big(H_S^2\,G_S^2 + \frac{2\,m_{\tau}}{\sqrt{q^2}}\,H_t\,H_S\,G_S\Big)\Bigg\}\,, \nonumber \\ 
\frac{d\Gamma}{dq^2}(B \to V\,\tau\,\nu) &=& \frac{8\,N\,|\overrightarrow{p}_V|}{3}\,\Bigg\{(\mathcal A_0^2 + \mathcal A_{||}^2 + \mathcal A_{\perp}^2)\Big(1 + \frac{m_{\tau}^2}{2\,q^2}\Big) +
\frac{3\,m_{\tau}^2}{2\,q^2}\,\mathcal A_t^2 \nonumber \\
&&+ \frac{3}{2}\Big(\mathcal A_P^2\,G_P^2 + \frac{2\,m_{\tau}}{\sqrt{q^2}}\,\mathcal A_t\,\mathcal A_P\,G_P\Big)\Bigg\}
\end{eqnarray}

We see that $B \to \tau\nu$ and $B \to D^{\ast}\tau\nu$ depend on pure $G_P$ coupling, whereas, 
$B \to \pi\tau\nu$ and $B \to D\tau\nu$ depend on pure $G_S$ coupling. Hence, we do not consider pure $G_P$ and pure $G_S$ NP couplings for our analysis as these will not simultaneously 
explain all the existing data. 
\begin{figure}[htbp]
\begin{center}
\includegraphics[width=5cm,height=4cm]{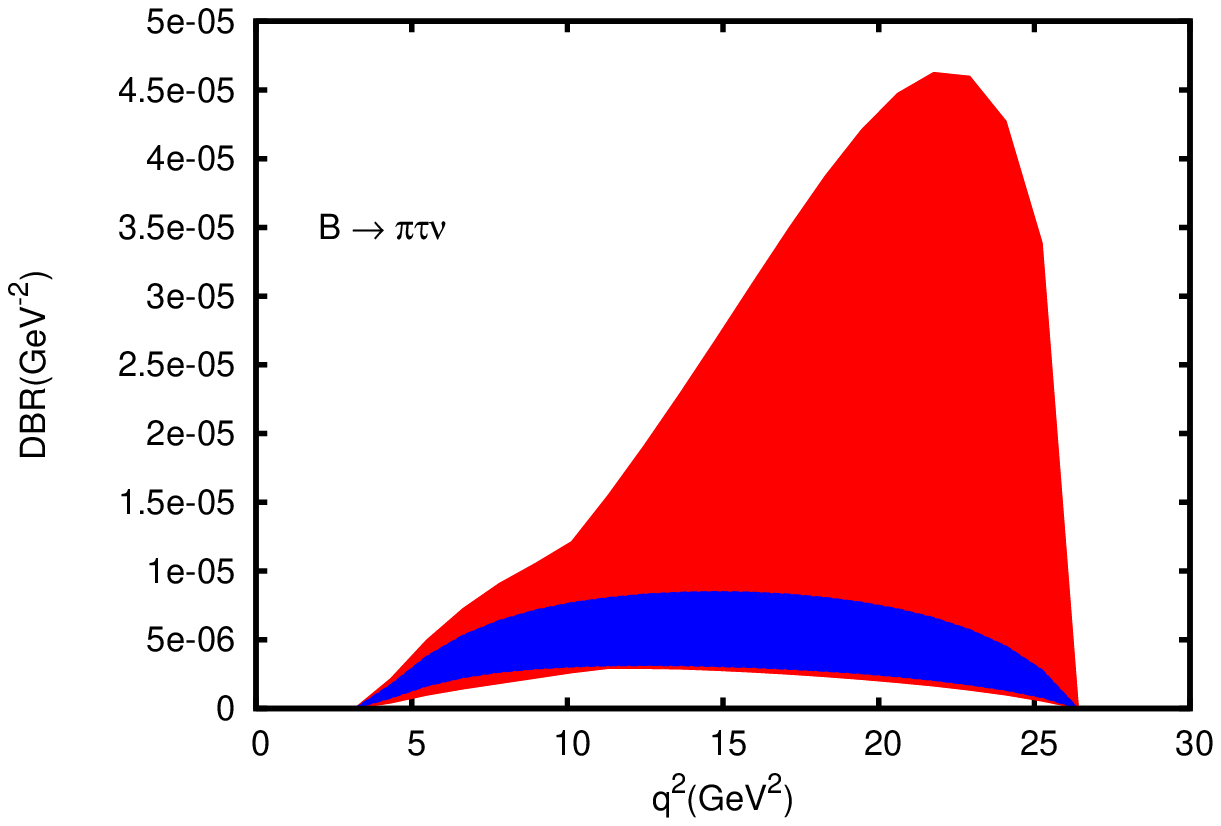}
\includegraphics[width=5cm,height=4cm]{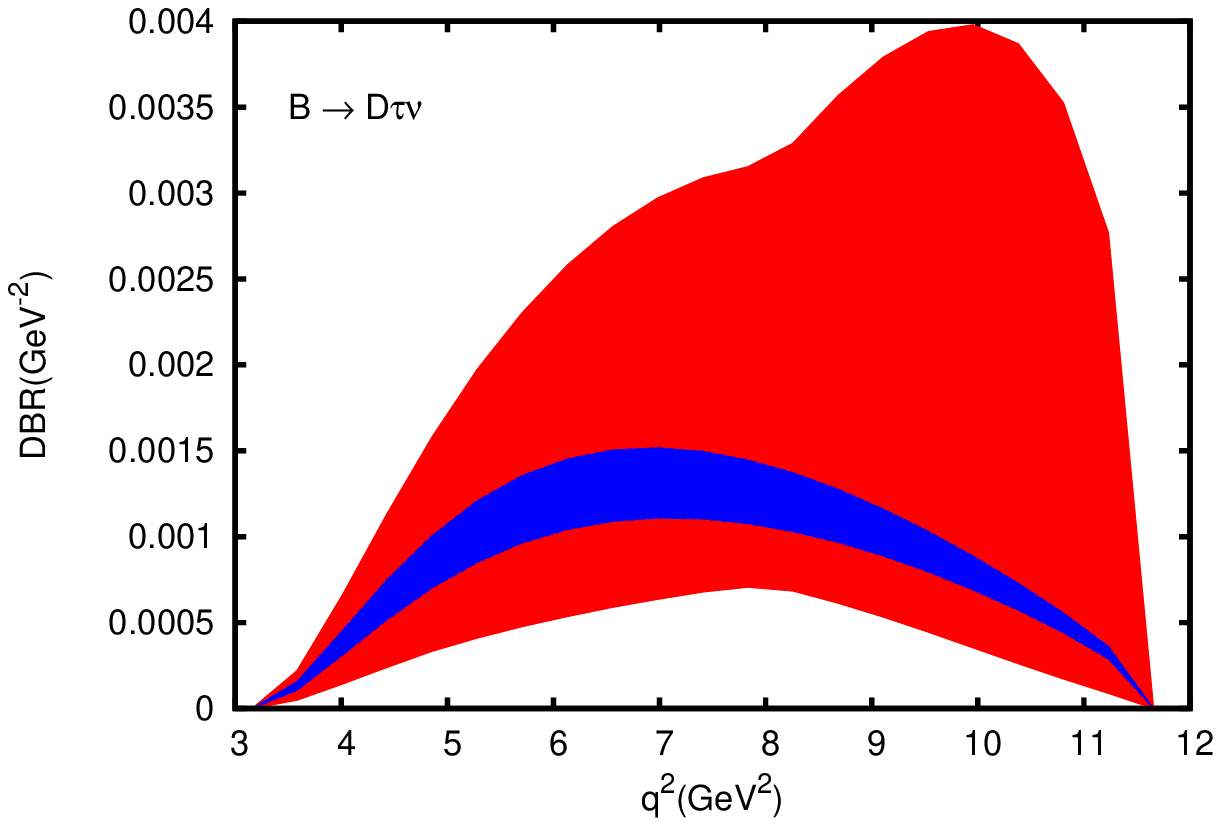}
\includegraphics[width=5cm,height=4cm]{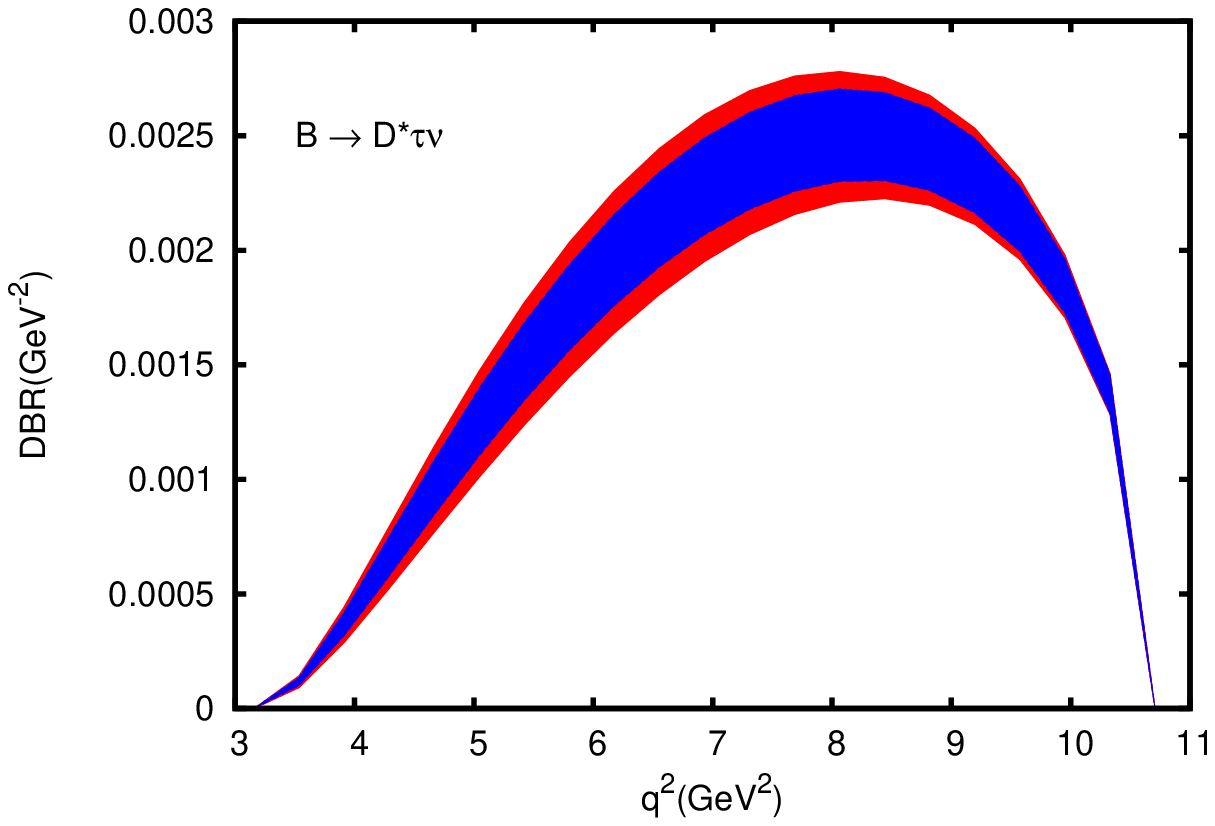}
\includegraphics[width=5cm,height=4cm]{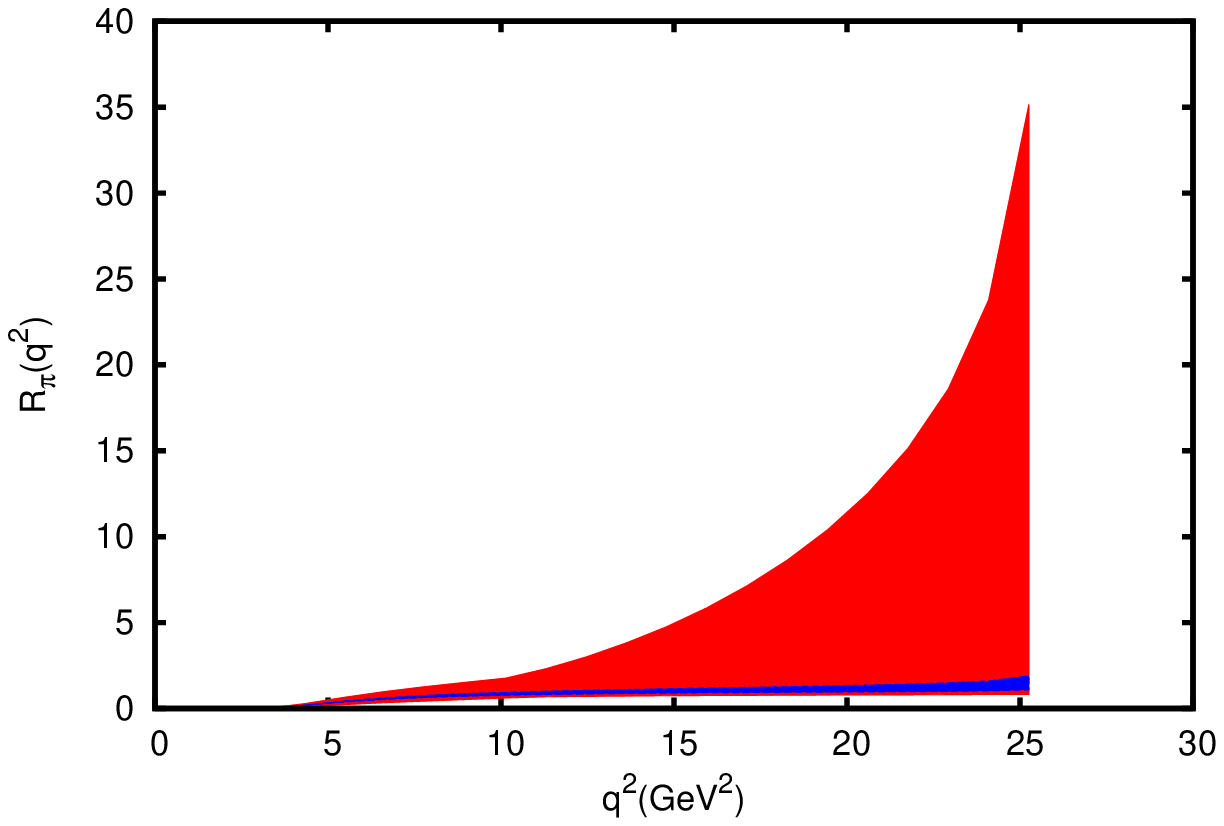}
\includegraphics[width=5cm,height=4cm]{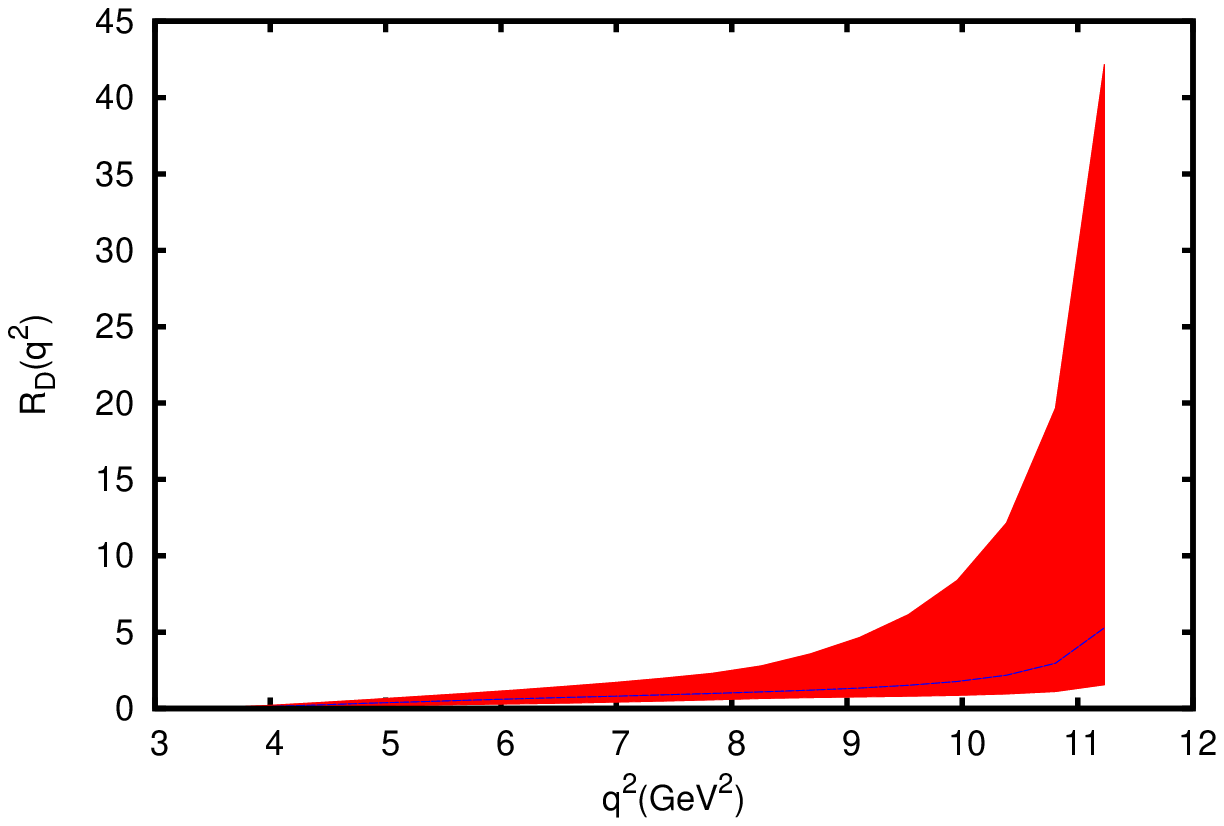}
\includegraphics[width=5cm,height=4cm]{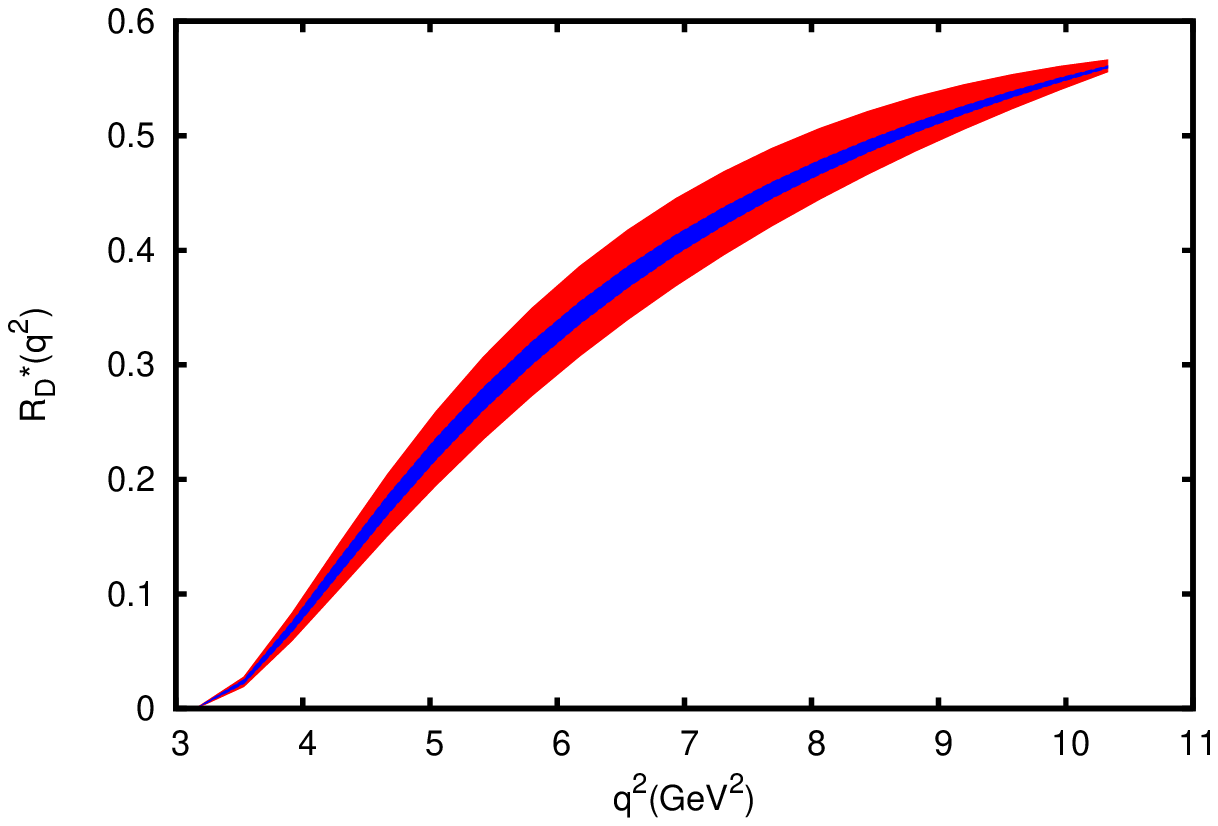}
\includegraphics[width=5cm,height=4cm]{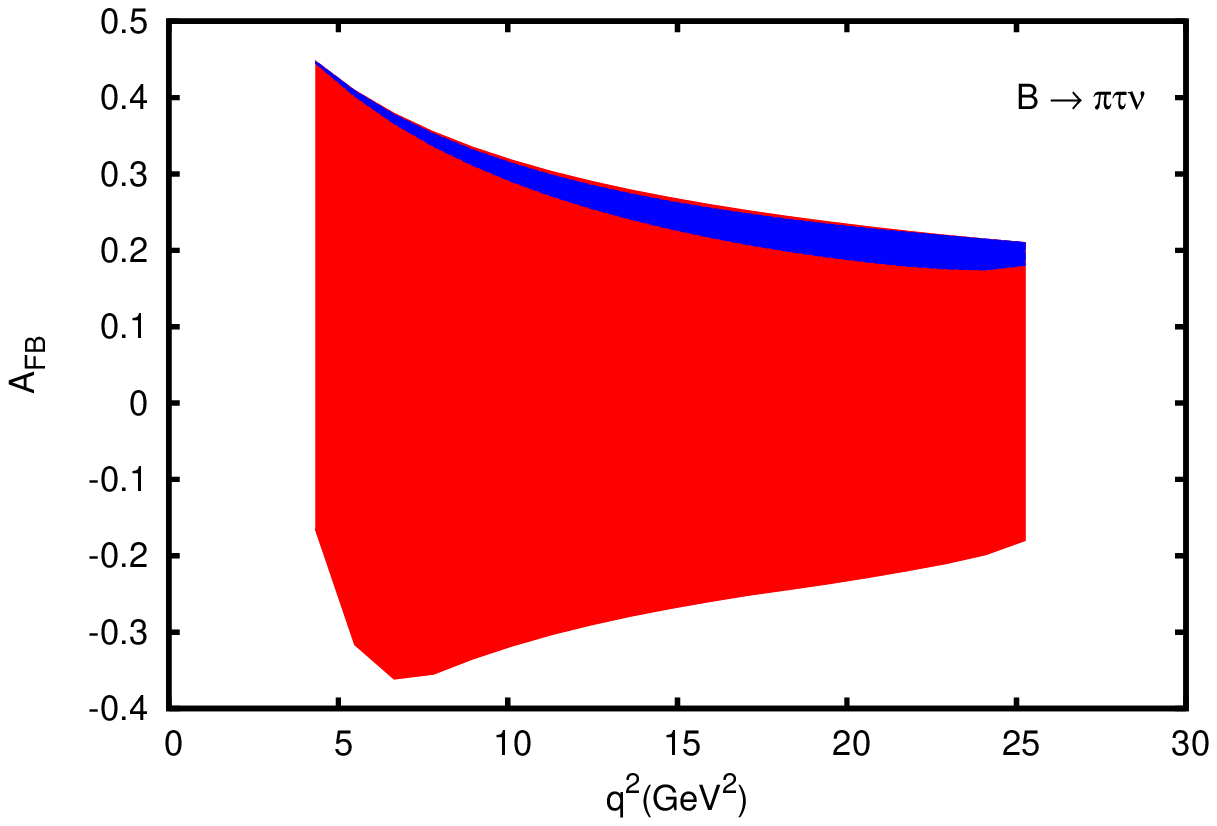}
\includegraphics[width=5cm,height=4cm]{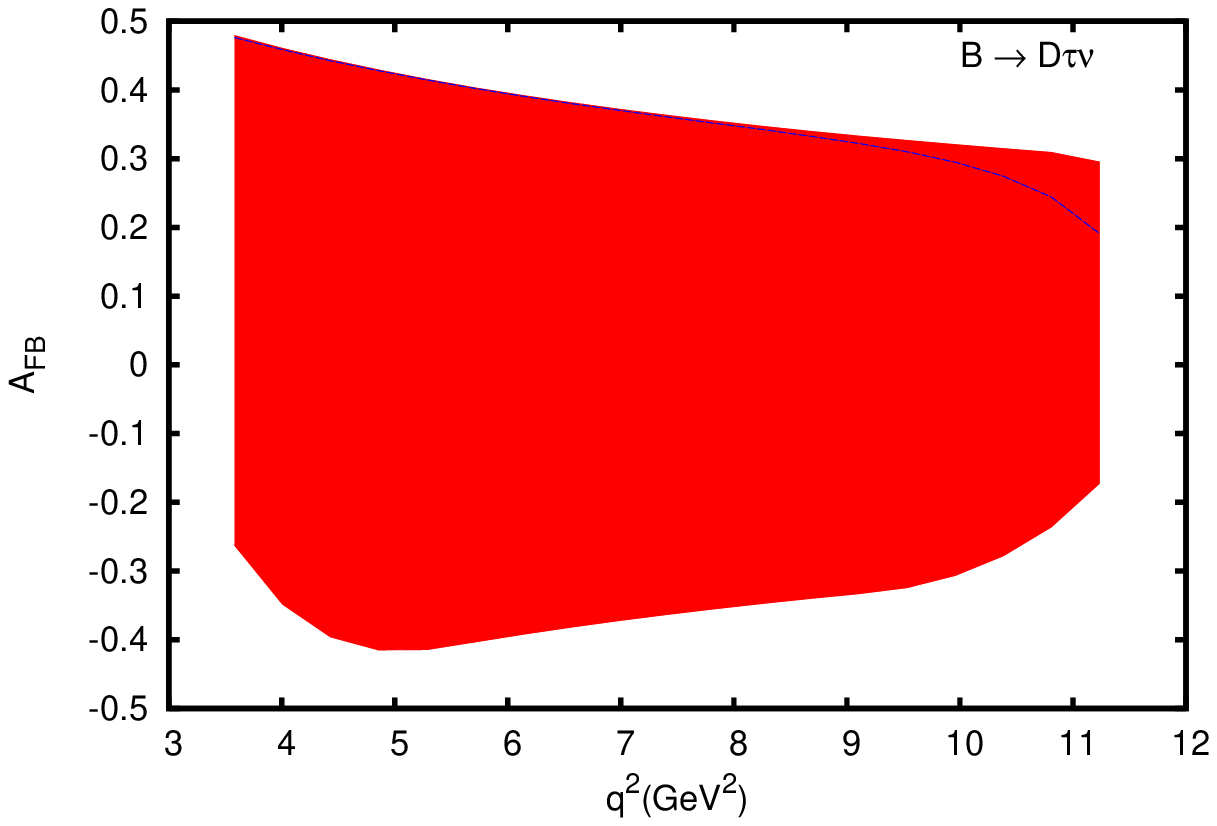}
\includegraphics[width=5cm,height=4cm]{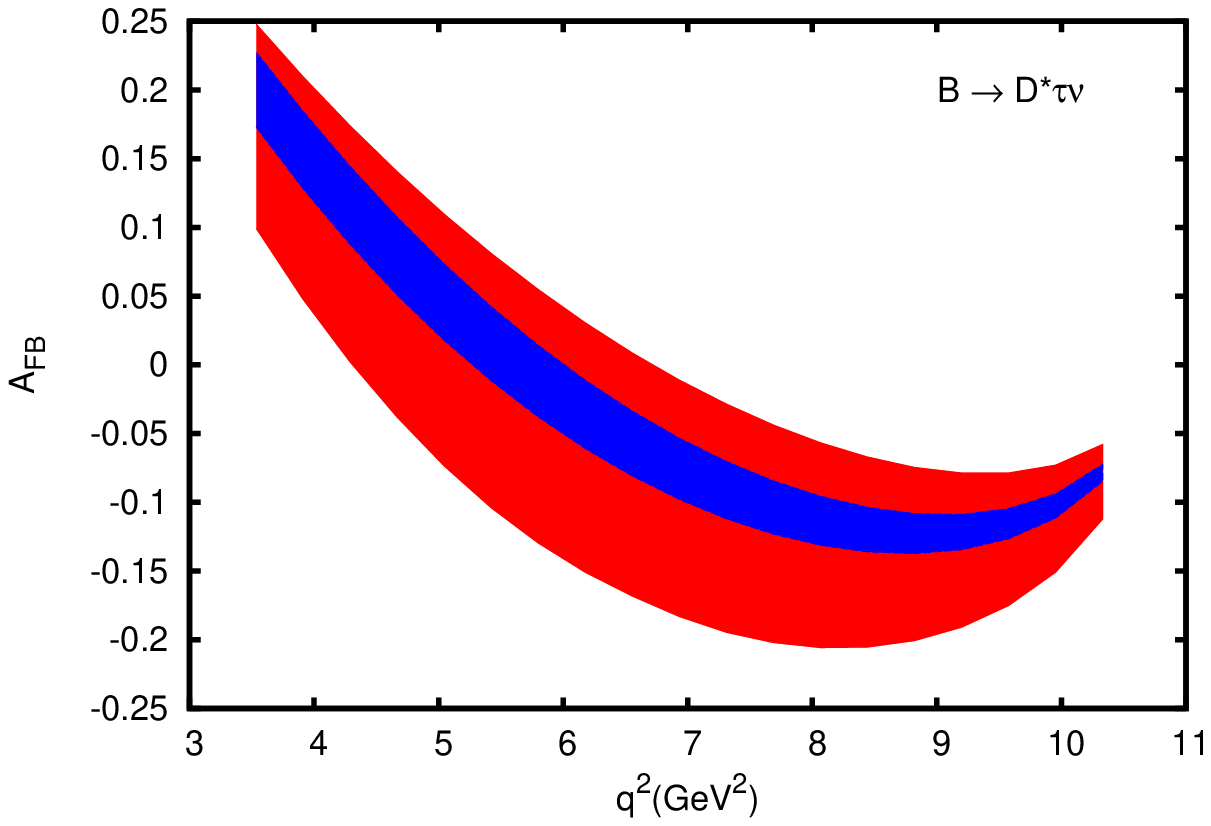}
\end{center}
\caption{Range in DBR$(q^2)$, $R(q^2)$, and the forward-backward asymmetry $A_{FB}(q^2)$ for the $B \to \pi\tau\nu$, $B \to D\tau\nu$, and $B \to D^{\ast}\tau\nu$ decay modes. The darker~(blue) 
interior region corresponds to the SM prediction, whereas, the lighter~(red), larger region corresponds to the allowed $(S_L,\,S_R)$ NP couplings of Fig.~\ref{slsr_tau}.}
\label{obs_slsr_tau}
\end{figure}
The effects of these NP couplings on the $\mathcal B(B \to \pi\tau\nu)$ and the ratio $R_{\pi}$ is shown in the right panel of Fig.~\ref{slsr_tau}. In the presence of such NP, the $3\sigma$ allowed ranges of 
the branching ratio of $B_c \to \tau\nu$, $B \to \pi\tau\nu$, and the ratio $R_{\pi}$ of the branching ratios of $B \to \pi\tau\nu$ to the corresponding $B \to \pi\,l\,\nu$ are
\begin{eqnarray*}
&&\mathcal B(B_c \to \tau\nu) = (0.21,\,13.66)\%\,, \qquad\qquad
\mathcal B(B \to \pi\tau\nu)  = (1.69,\,119.66)\times 10^{-5}\,,\nonumber \\
&&
R_{\pi} = (0.49,\,7.06)\,.
\end{eqnarray*}
We see that the $\mathcal B(B_c \to \tau\nu)$, $\mathcal B(B \to \pi\tau\nu)$, and the ratio $R_{\pi}$ are quite sensitive to the $S_L$ and $S_R$ NP couplings. The deviation from the SM is quite large
once these NP couplings are switched on.

We now wish to see how different observables behave with $S_L$ and $S_R$.
The corresponding DBR, the ratio $R(q^2)$, and the forward-backward asymmetries $A_{FB}(q^2)$ as a function of $q^2$ are shown in Fig.~\ref{obs_slsr_tau}. We see that deviation from the SM is much larger in the case
of $B \to \pi\tau\nu$ and $B \to D\tau\nu$ decay modes than the $B \to D^{\ast}\tau\nu$ decay mode. We see that the variation is quite similar in $B \to \pi\tau\nu$ and $B \to D\tau\nu$ decay modes. It is
expected as both the decay modes depend on the NP couplings through $G_S$,
whereas the $B \to D^{\ast}\tau\nu$ depends on the NP couplings through $G_P$ and hence the variation is quite different from the $B \to \pi\tau\nu$ and $B \to D\tau\nu$ decay modes. 
Again, the peak of the distribution of differential branching ratio for the $B \to \pi\tau\nu$ and $B \to D\tau\nu$ can shift to a higher $q^2$ region once the NP couplings are introduced.

Again in the SM, as mentioned earlier, we see a zero crossing in the forward-backward asymmetry for the $B \to D^{\ast}\tau\nu$ decay mode. Moreover, we observe no such zero crossing in case of 
$B \to \pi\tau\nu$ and $B \to D\tau\nu$ decay modes.
However, once the NP couplings $S_L$ and $S_R$ are switched on, we see a zero crossing for the $B \to \pi\tau\nu$ as well as the $B \to D\tau\nu$ decay modes. Depending on the value of the NP couplings, 
there may be a zero crossing or there could be a total change of sign of the $A_{FB}$ parameter as can be
seen from Fig.~\ref{obs_slsr_tau}. Thus, we see that, the forward-backward asymmetry in the case of $B \to \pi\tau\nu$ and $B \to D\tau\nu$ is very sensitive to the $S_L$ and $S_R$ couplings.
In the case of $B \to D^{\ast}\tau\nu$ decay mode, however, the sensitivity is much smaller than the $B \to \pi\tau\nu$ and $B \to D\tau\nu$ modes. It is worth mentioning that, depending on the 
value of the NP couplings, there can be a zero crossing for the $B \to D^{\ast}\tau\nu$ decay process which is marginally different from the SM, as is evident from Fig.~\ref{obs_slsr_tau}.

\subsection{Scenario C}
\begin{figure}[htbp]
\begin{center}
\includegraphics[width=8cm,height=5cm]{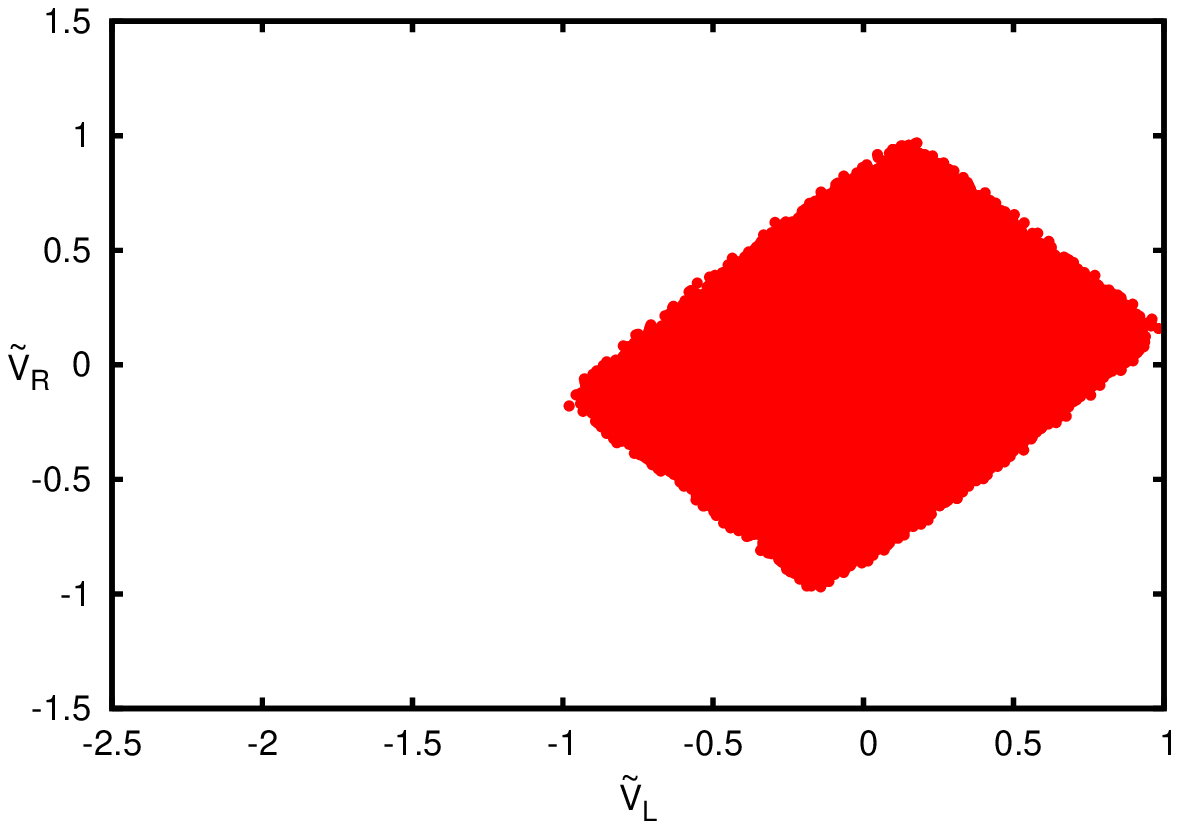}
\includegraphics[width=8cm,height=5cm]{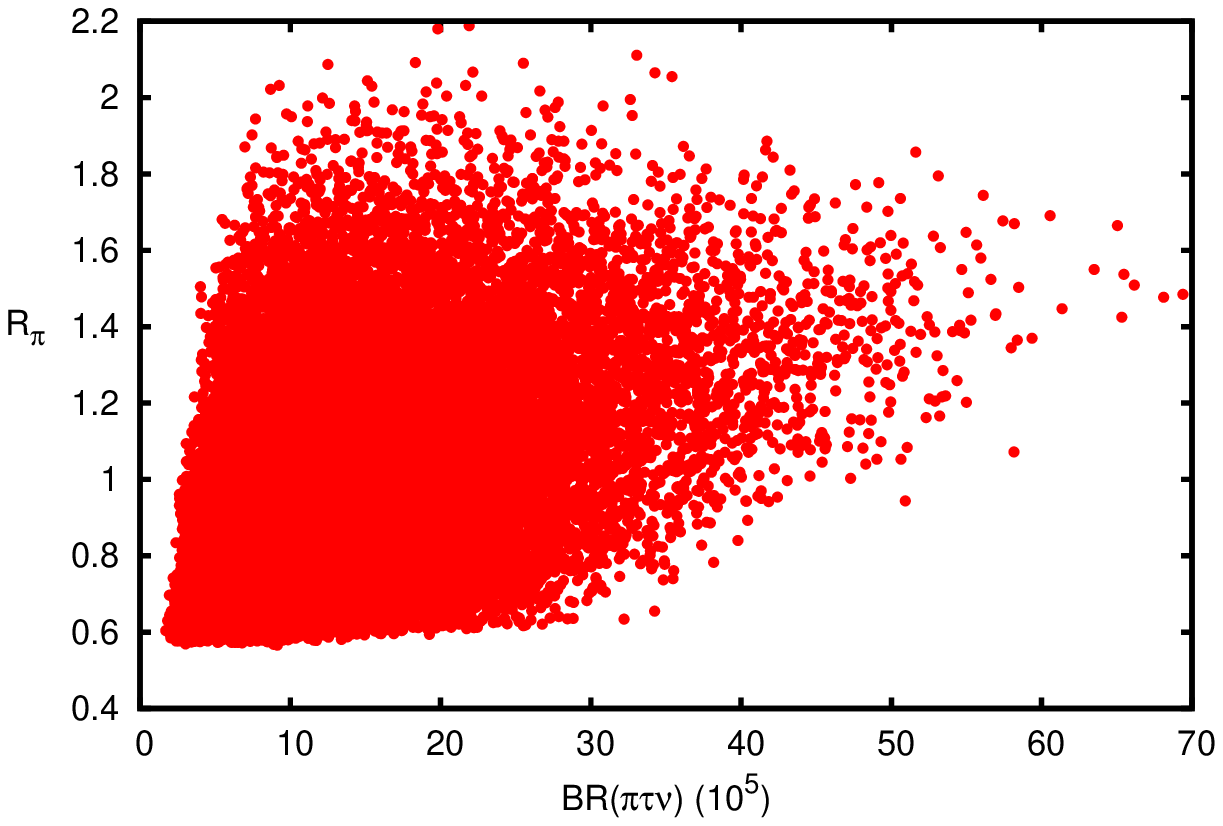}
\end{center}
\caption{Range in $\widetilde{V}_L$ and $\widetilde{V}_R$ is shown in the left panel once the $3\sigma$ experimental constraint is imposed. The resulting range in the $\mathcal B(B \to \pi\tau\nu)$ and
$R_{\pi}$ is shown in the right panel with these NP couplings.}
\label{vltvrt_tau}
\end{figure}

We set all the other NP couplings to zero while varying $\widetilde{V}_L$ and $\widetilde{V}_R$. These couplings are related to the right-handed neutrino interactions. As already mentioned in Sec.~\ref{th}, the
decay rate depends quadratically on these NP couplings. The linear term that comes from the interference between the SM and the NP is negligible due to the mass of the neutrino. The allowed ranges of
$\widetilde{V}_L$ and $\widetilde{V}_R$ are shown in the left panel of Fig.~\ref{vltvrt_tau}. It is evident that the parameter space is much less restricted than Scenario A~$(V_{L,\,R} \ne 0)$ and 
Scenario B~$(S_{L,\,R} \ne 0)$. 
\begin{figure}[htbp]
\begin{center}
\includegraphics[width=5cm,height=4cm]{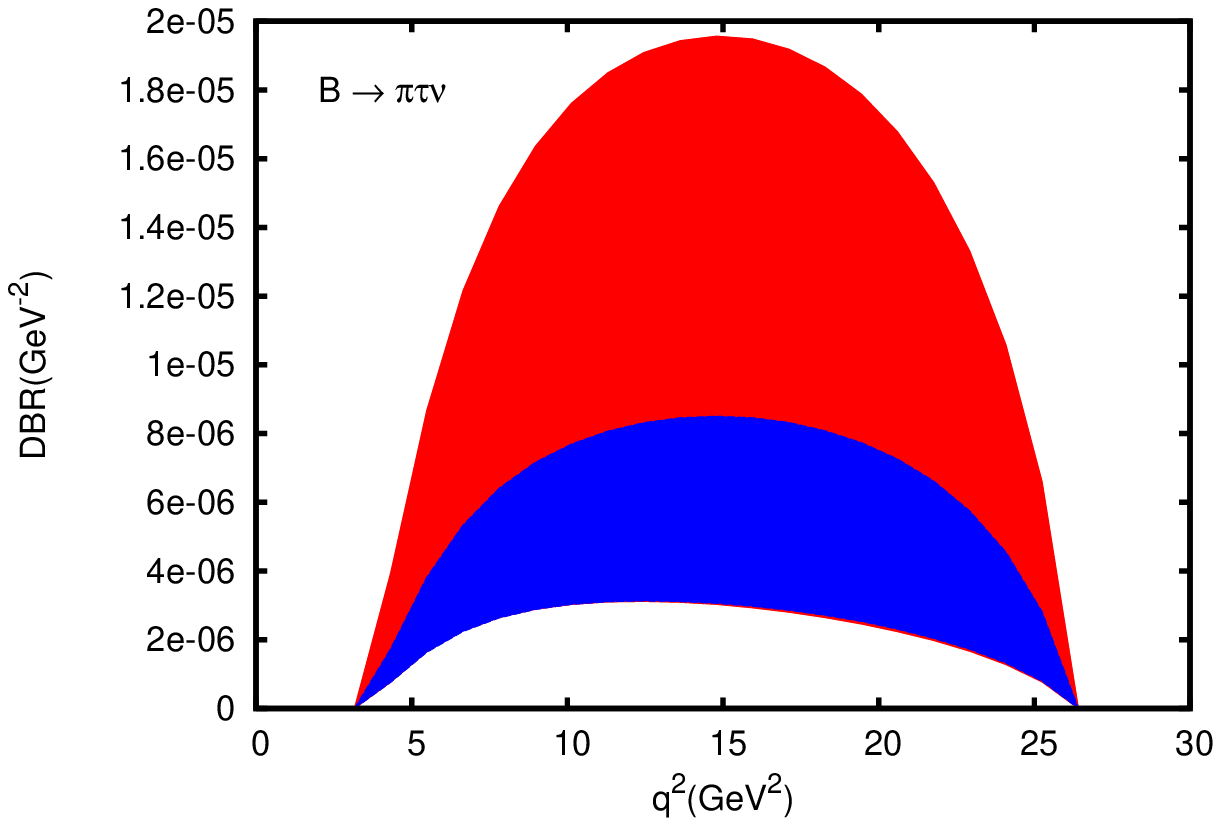}
\includegraphics[width=5cm,height=4cm]{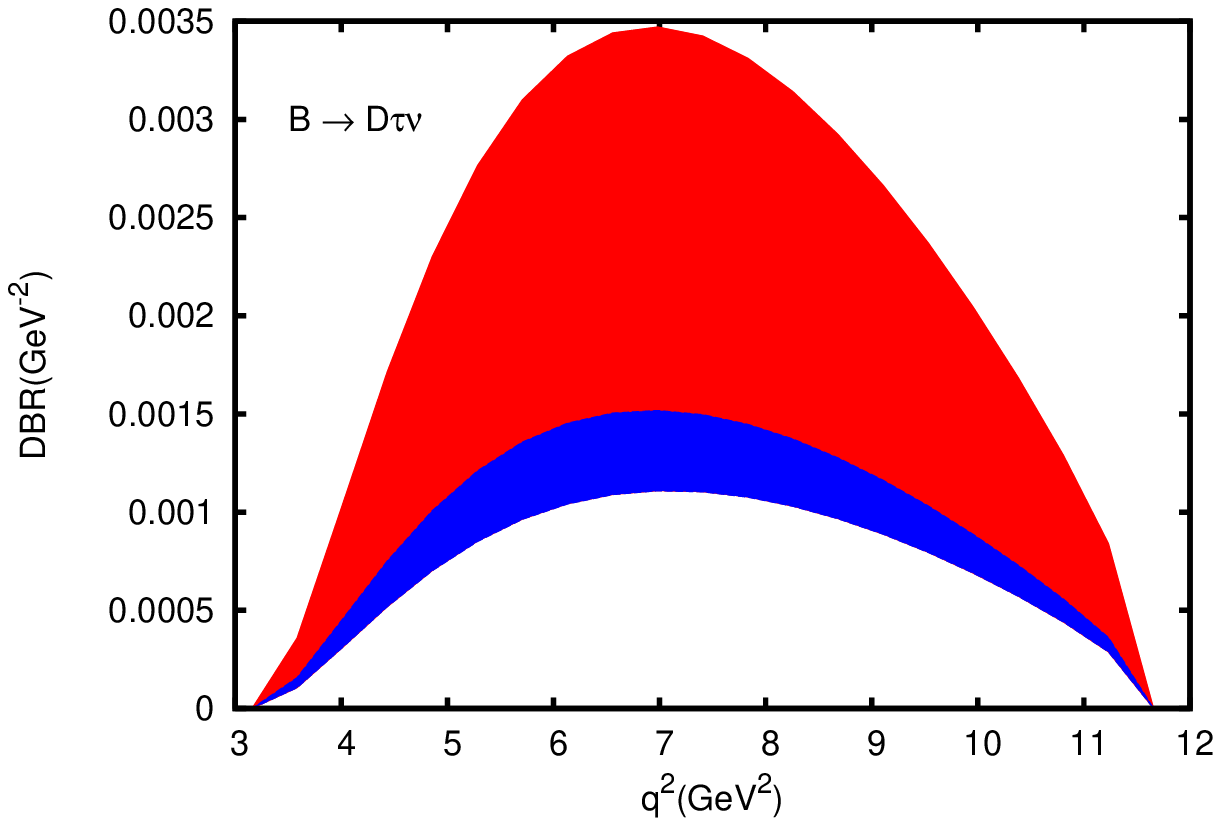}
\includegraphics[width=5cm,height=4cm]{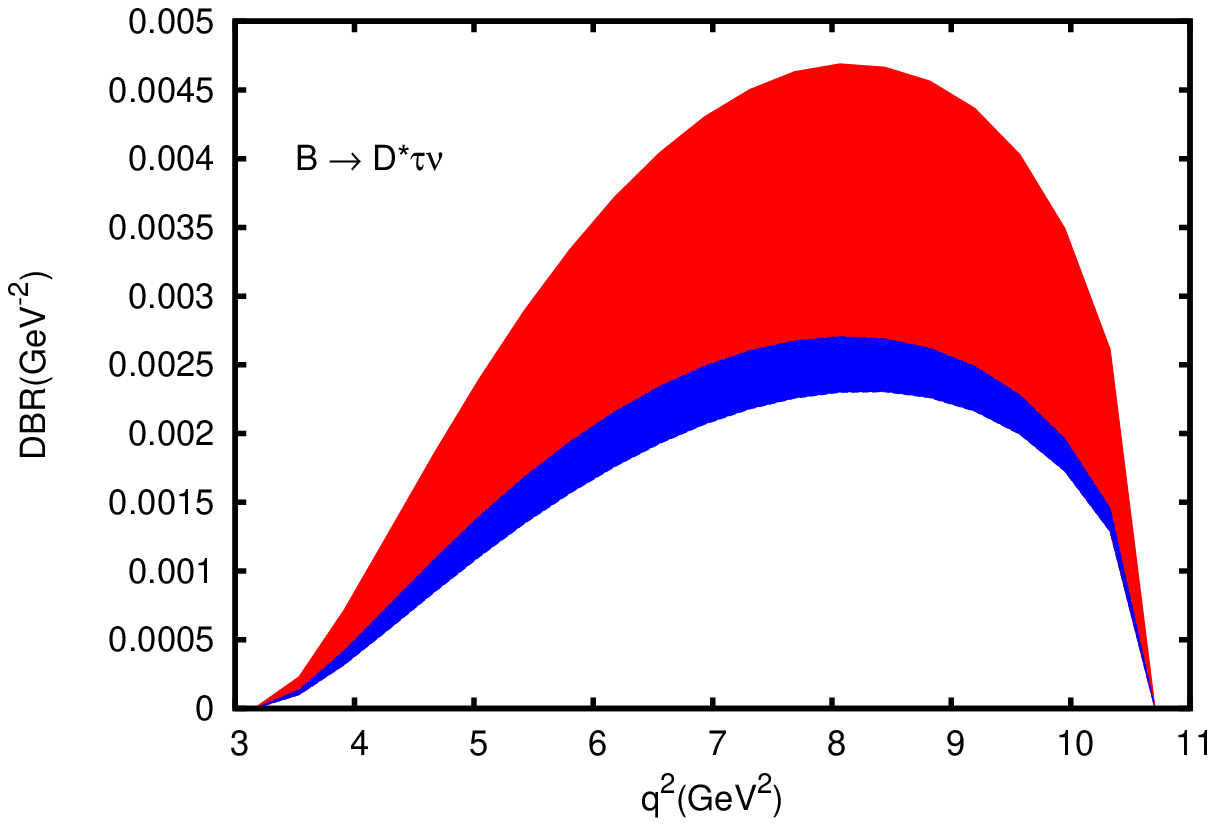}
\includegraphics[width=5cm,height=4cm]{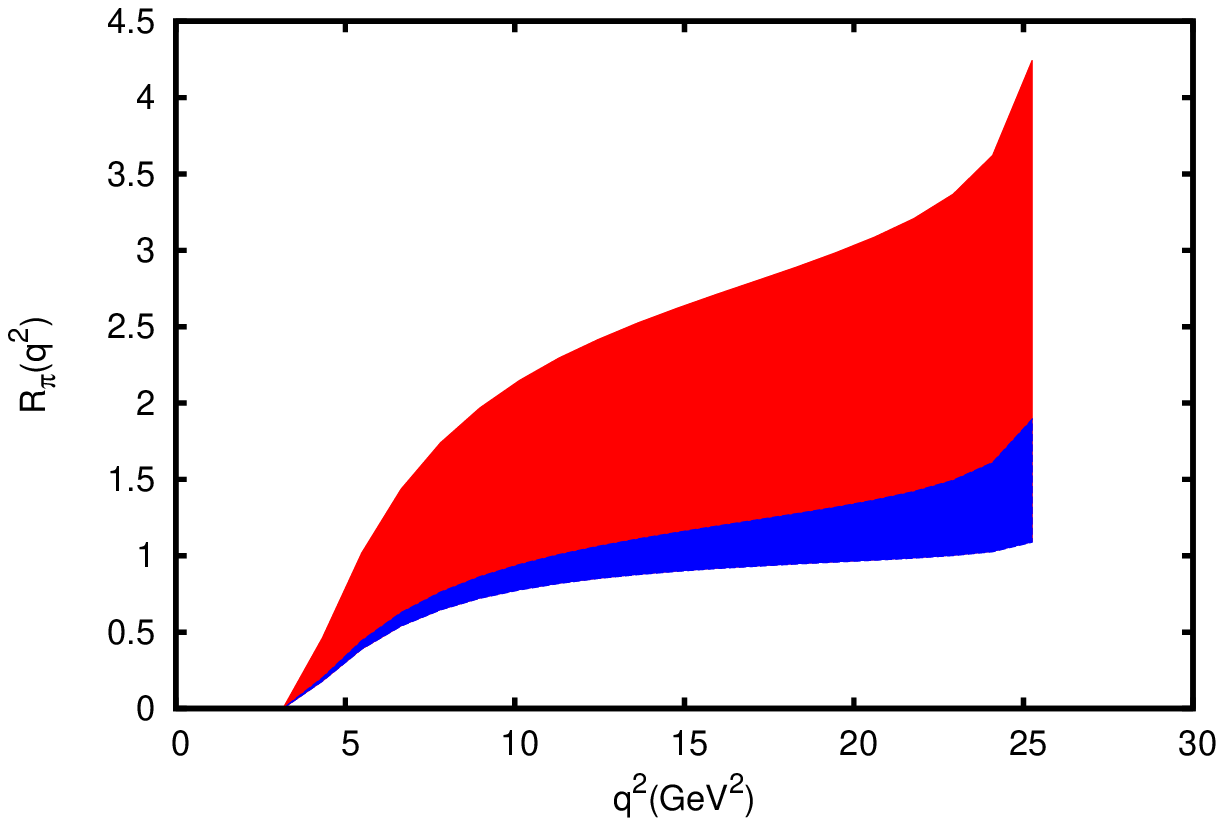}
\includegraphics[width=5cm,height=4cm]{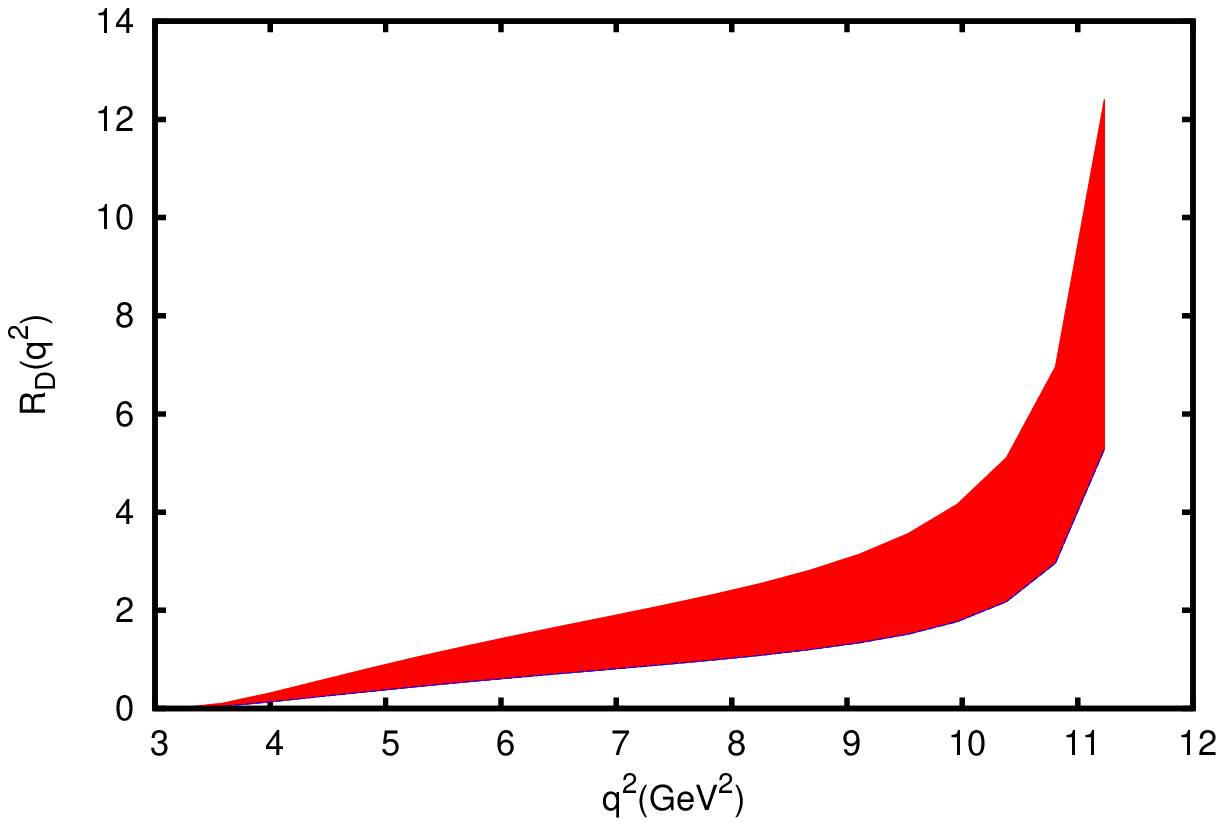}
\includegraphics[width=5cm,height=4cm]{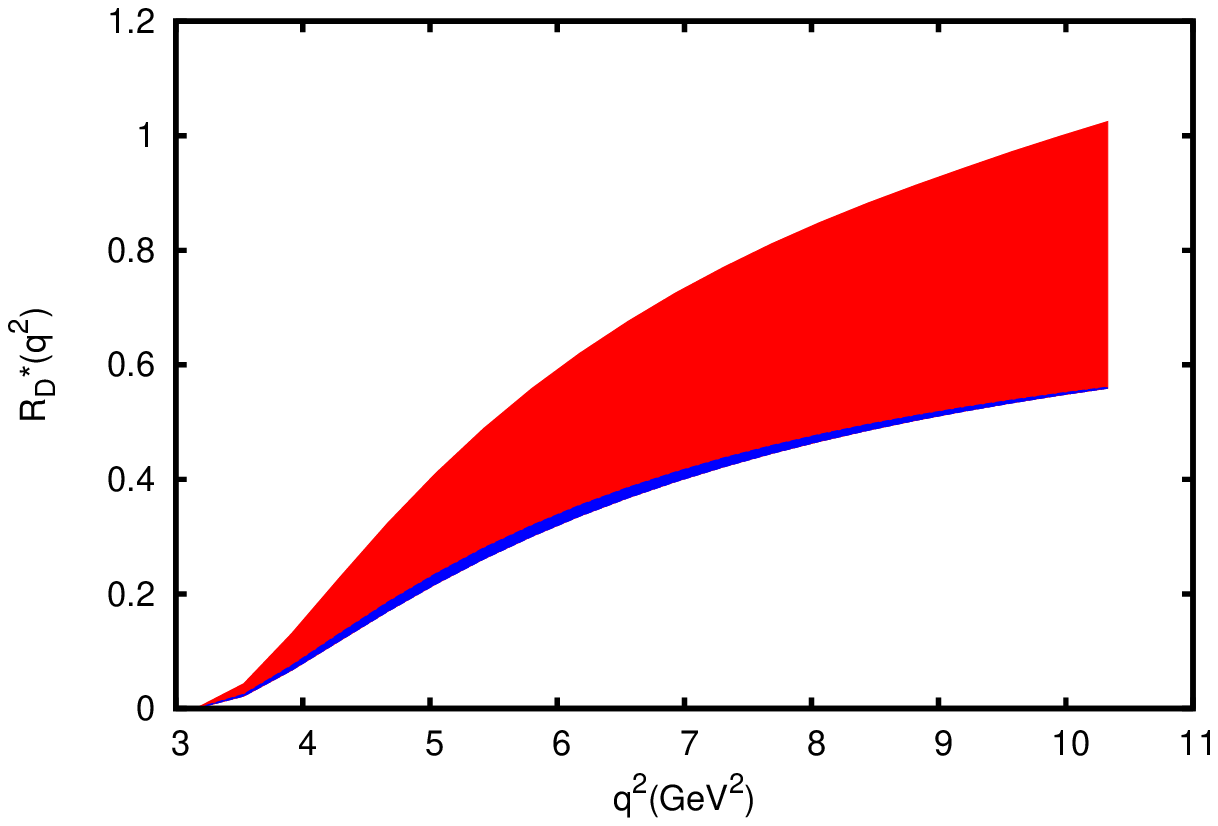}
\includegraphics[width=5cm,height=4cm]{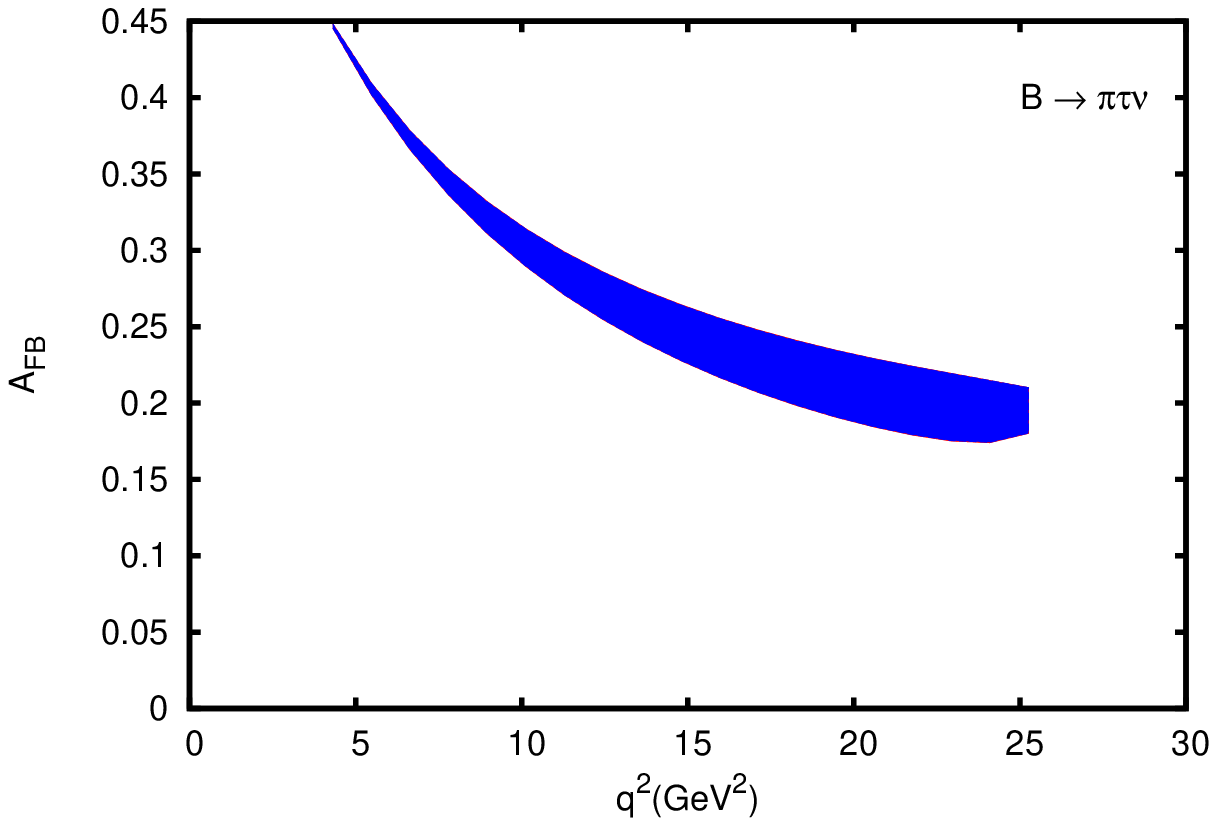}
\includegraphics[width=5cm,height=4cm]{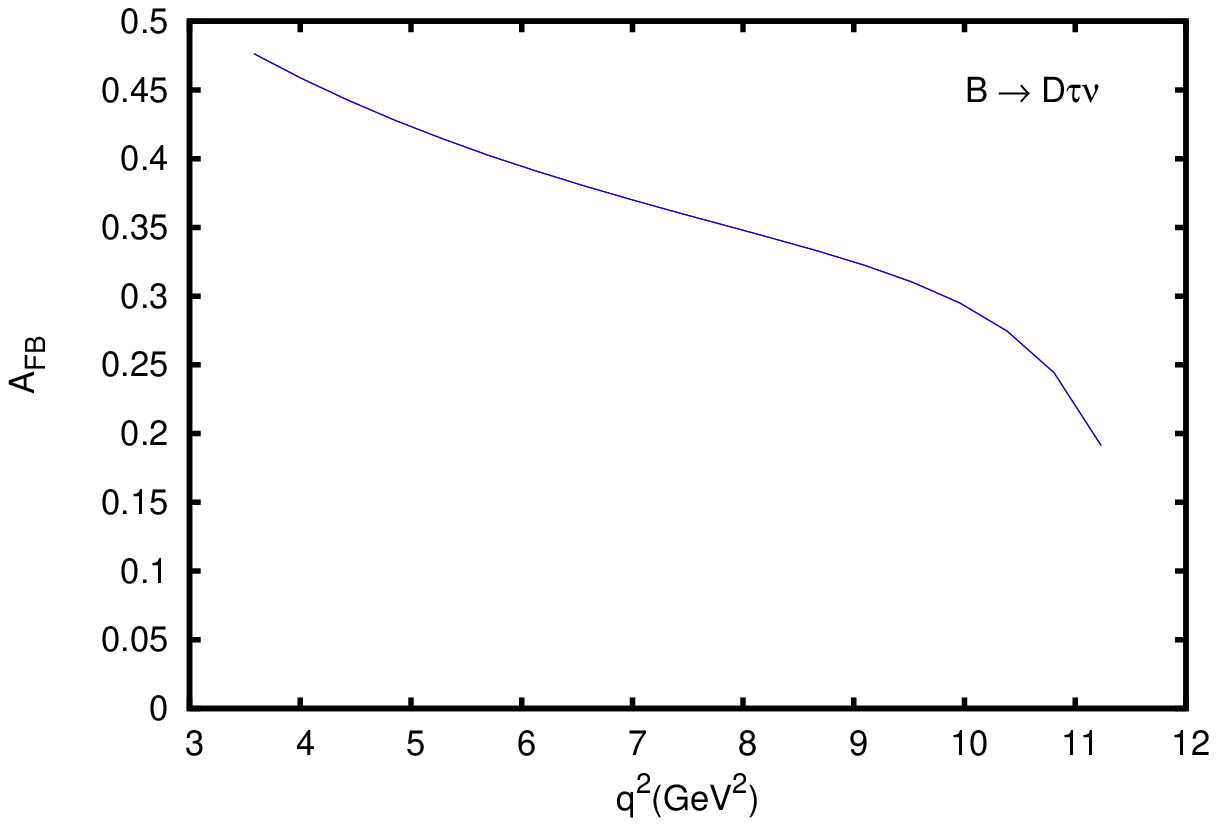}
\includegraphics[width=5cm,height=4cm]{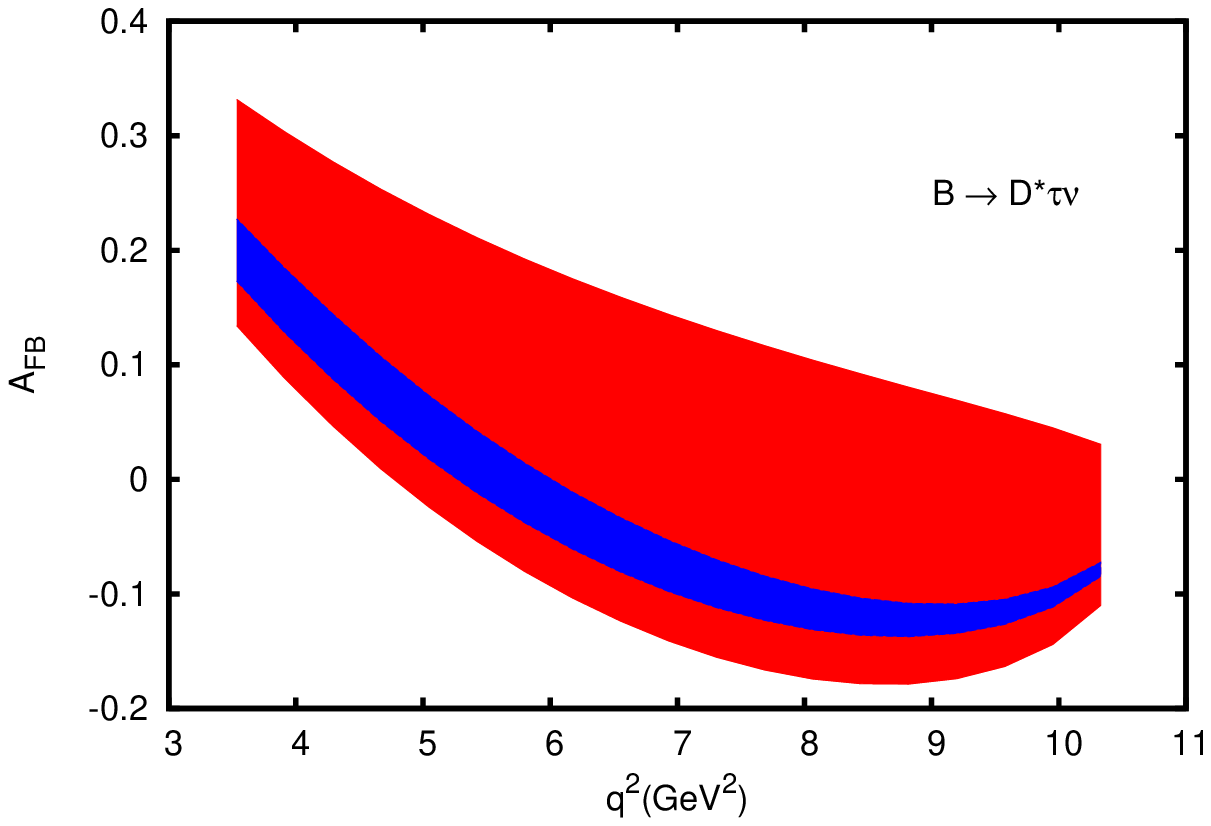}
\end{center}
\caption{Range in DBR$(q^2)$, $R(q^2)$, and $A_{FB}(q^2)$ for the $B \to \pi\tau\nu$, $B \to D\tau\nu$, and the $B \to D^{\ast}\tau\nu$ decay modes. The dark~(blue) band corresponds to
the SM range, whereas, the light~(red) band corresponds to the NP couplings $(\widetilde{V}_L,\,\widetilde{V}_R)$ that are shown in the left panel of Fig.~\ref{vltvrt_tau}.}
\label{obs_vltvrt_tau}
\end{figure}

In the presence of such NP couplings, the $\Gamma(B_q \to \tau\nu)$, $d\Gamma/dq^2(B \to P\,\tau\nu)$, and $d\Gamma/dq^2(B \to V\,\tau\nu)$, where $P$ stands for
pseudoscalar and $V$ stands for vector meson, can be written as
\begin{eqnarray}
\label{eq:vltvrt}
\Gamma(B_q \to \tau\nu) &=& \Gamma(B_q \to \tau\nu)|_{\rm SM}\,\Big(1 + \widetilde{G}_A^2\Big)\,,\nonumber \\
\frac{d\Gamma}{dq^2}(B \to P\,\tau\,\nu) &=& \Big(\frac{d\Gamma}{dq^2}(B \to P\,\tau\,\nu)\Big)_{\rm SM}\,\Big(1 + \widetilde{G}_V^2\Big)\,, \nonumber \\ 
\frac{d\Gamma}{dq^2}(B \to V\,\tau\,\nu) &=& \frac{8\,N\,|\overrightarrow{p}_V|}{3}\Bigg\{\Big[\mathcal A_0^2\,(1 + \widetilde{G}_A^2) + \mathcal A_{||}^2\,(1 + \widetilde{G}_A^2) + 
\mathcal A_{\perp}^2\,(1 + \widetilde{G}_V^2)\Big]\,
\Big(1 + \frac{m_{\tau}^2}{2\,q^2}\Big) \nonumber \\
&&+ 
\frac{3\,m_{\tau}^2}{2\,q^2}\,\mathcal A_t^2\,(1 + \widetilde{G}_A^2)\Bigg\}
\end{eqnarray}
It is evident from Eq.~(\ref{eq:vltvrt}) that the $B \to \tau\nu$ decay branching ratio depends on the NP couplings through $\widetilde{G}_A^2$ term
and the $B \to D^{\ast}\tau\nu$ branching ratio depend on $\widetilde{V}_L$ and $\widetilde{V}_R$ couplings through $\widetilde{G}_A^2$ as well as $\widetilde{G}_V^2$ term, whereas the $B \to \pi\tau\nu$ and $B \to D\tau\nu$
branching ratios depend on these couplings through $\widetilde{G}_V^2$ term.
The corresponding $3\sigma$ allowed ranges of $\mathcal B(B \to \pi\tau\nu)$ and the ratio $R_{\pi}$ is shown in the right panel of Fig.~\ref{vltvrt_tau}. The ranges are
\begin{eqnarray*}
&&\mathcal B(B_c \to \tau\nu) = (1.09,\,4.13)\%\,, \qquad\qquad
\mathcal B(B \to \pi\tau\nu)  = (1.71,\,69.39)\times 10^{-5}\,,\nonumber \\
&&
R_{\pi} = (0.57,\,2.19)\,,
\end{eqnarray*}
and are quite similar to Scenario A. Again, a significant deviation from the SM prediction is expected in such NP scenario.

The allowed ranges of all the different observables with these NP couplings are shown in Fig.~\ref{obs_vltvrt_tau}.
We see that the differential branching ratio, the ratio of branching ratio, and the forward backward asymmetry parameters vary quite significantly with the inclusion of the NP couplings.
The $q^2$ distribution looks quite similar to what we obtain for Scenario A. Although, the differential branching ratio and the ratio of branching ratios are quite sensitive to 
$\widetilde{V}_L$ and $\widetilde{V}_R$, the forward-backward asymmetry for the $B \to \pi\tau\nu$ and $B \to D\tau\nu$ does not depend on the
NP couplings at all. However, for the $B \to D^{\ast}\tau\nu$ decay mode, all the three observables are very sensitive to these right-handed neutrino couplings. Again, depending on these
NP couplings, there may be a zero crossing in the $q^2$ distribution of the $A_{FB}$ parameter which can be quite different from the SM prediction.
%%%%%%%%%%%%%55
\subsection{Scenario D}
\begin{figure}[htbp]
\begin{center}
\includegraphics[width=8cm,height=5cm]{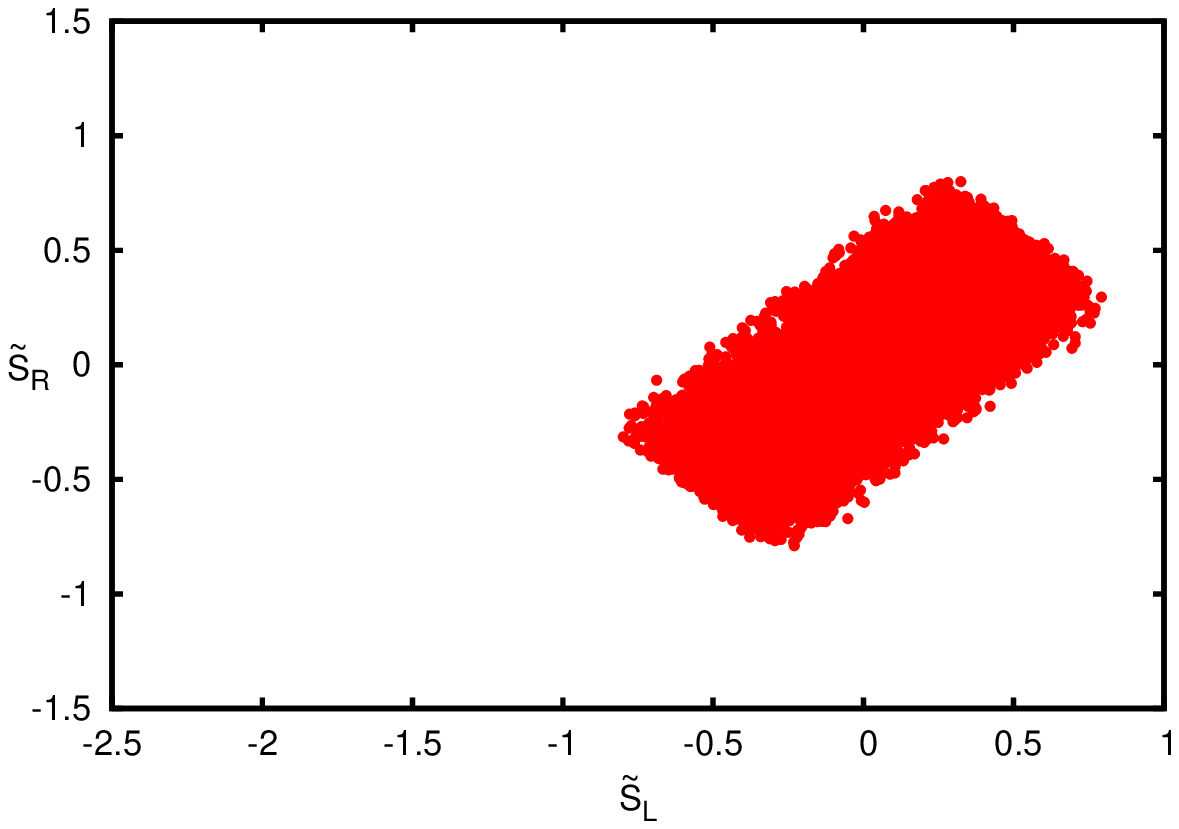}
\includegraphics[width=8cm,height=5cm]{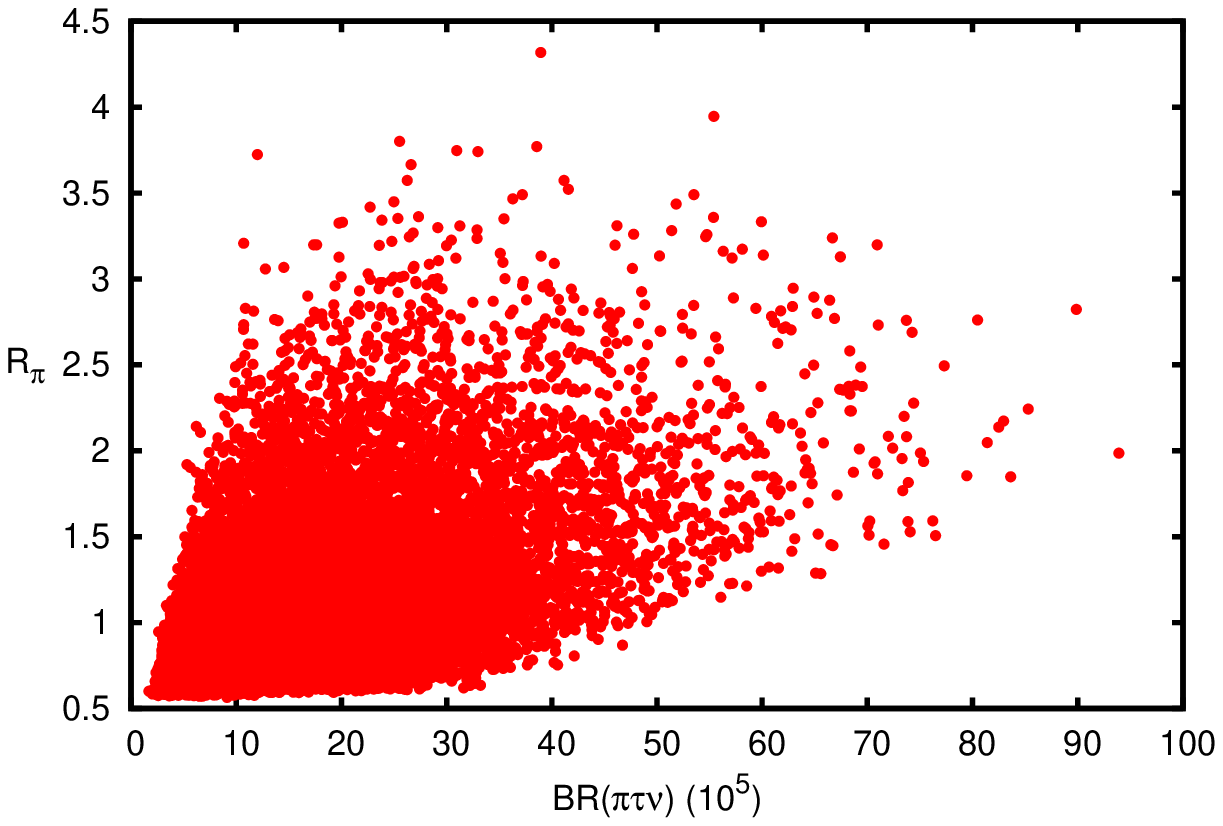}
\end{center}
\caption{Left panel shows the allowed range in $\widetilde{S}_L$ and $\widetilde{S}_R$ with the $3\sigma$ experimental constraint imposed. The resulting range in $B \to \pi\tau\nu$ branching ratio
and the ratio $R_{\pi}$ is shown in the right panel once the NP $\widetilde{S}_L$ and $\widetilde{S}_R$ are included.} 
\label{sltsrt_tau}
\end{figure}

We include the new physics effects coming from the $\widetilde{S}_L$ and $\widetilde{S}_R$ alone while keeping all the other NP couplings to zero. We impose the experimental constraint coming from
the measured data of $R_{\pi}^l$, $R_D$, and $R_{D^{\ast}}$ and the resulting allowed ranges of $\widetilde{S}_L$ and $\widetilde{S}_R$ are shown in the left panel of Fig.~\ref{sltsrt_tau}.
Similar to $\widetilde{V}_L$ and $\widetilde{V}_R$, these couplings also arise due to the right-handed neutrino interactions. The decay rate depends on these NP couplings quadratically and hence the
parameter space is less constrained.
\begin{figure}[htbp]
\begin{center}
\includegraphics[width=5cm,height=4cm]{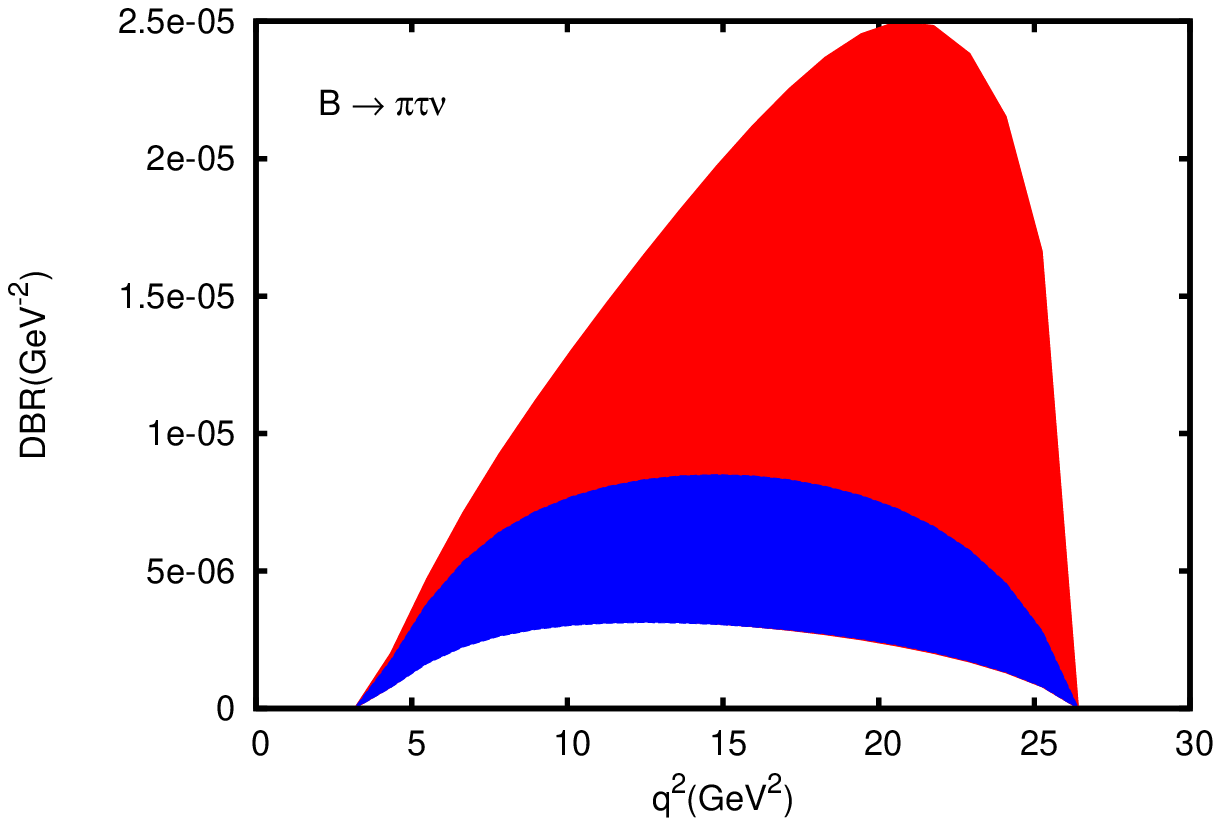}
\includegraphics[width=5cm,height=4cm]{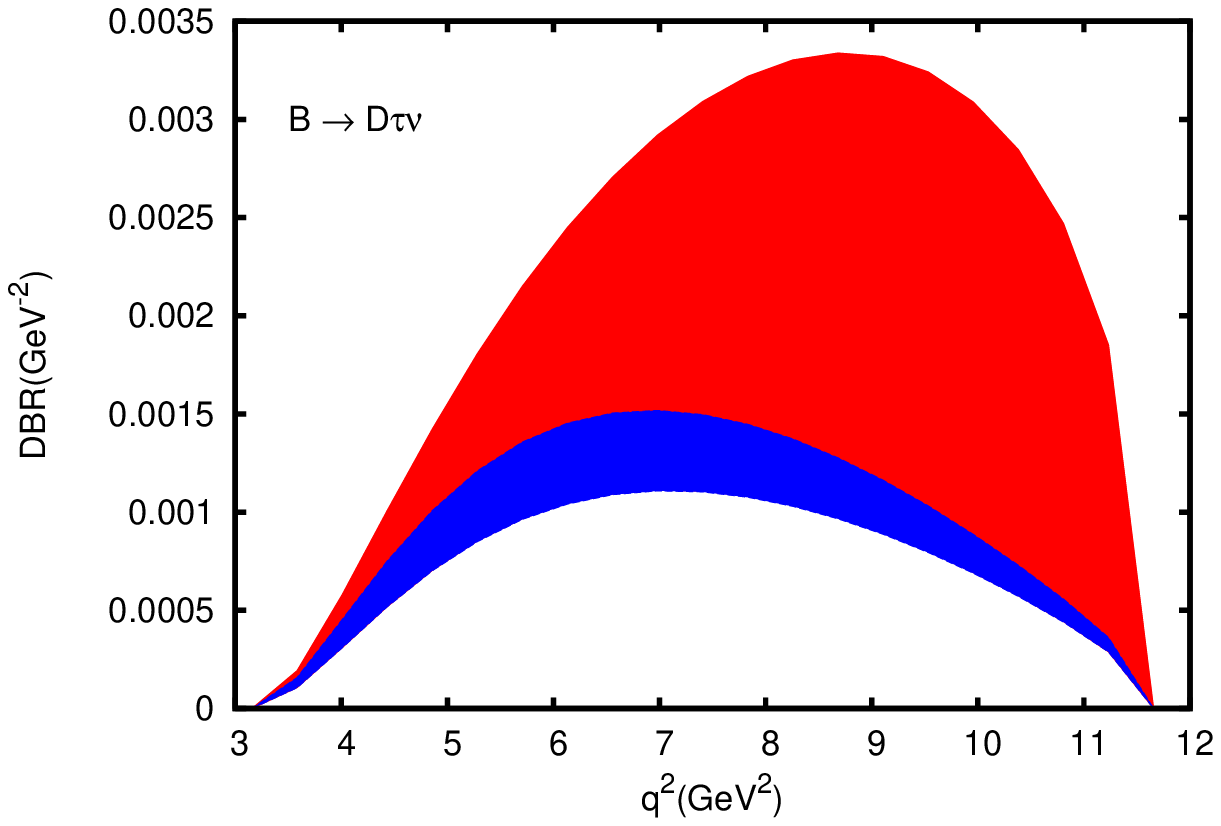}
\includegraphics[width=5cm,height=4cm]{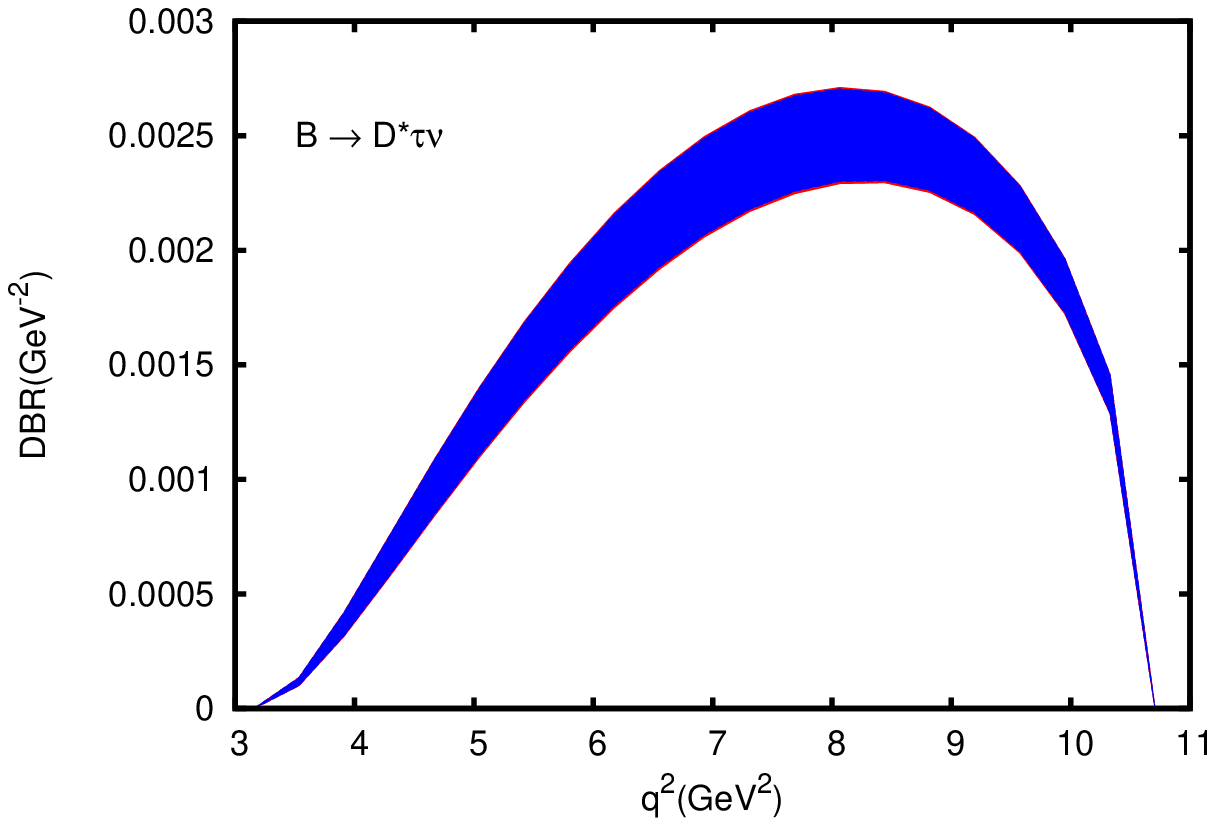}
\includegraphics[width=5cm,height=4cm]{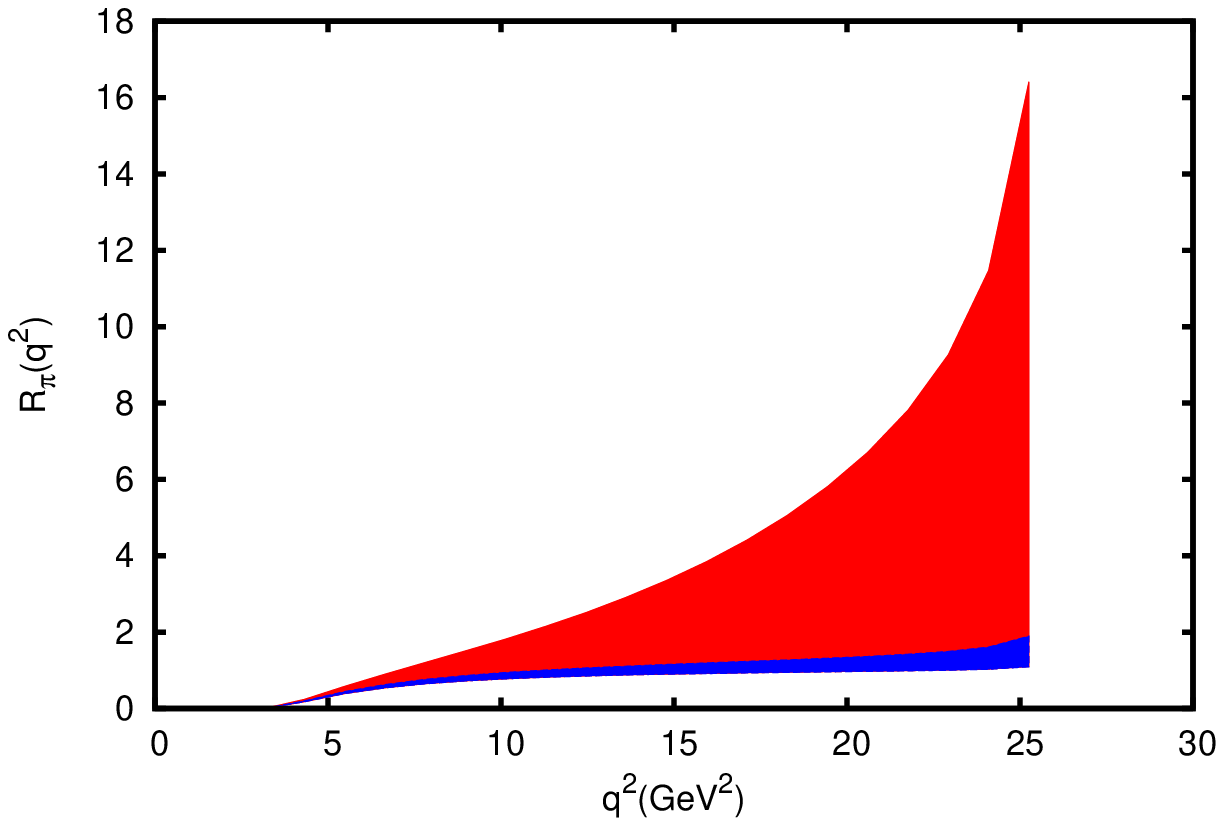}
\includegraphics[width=5cm,height=4cm]{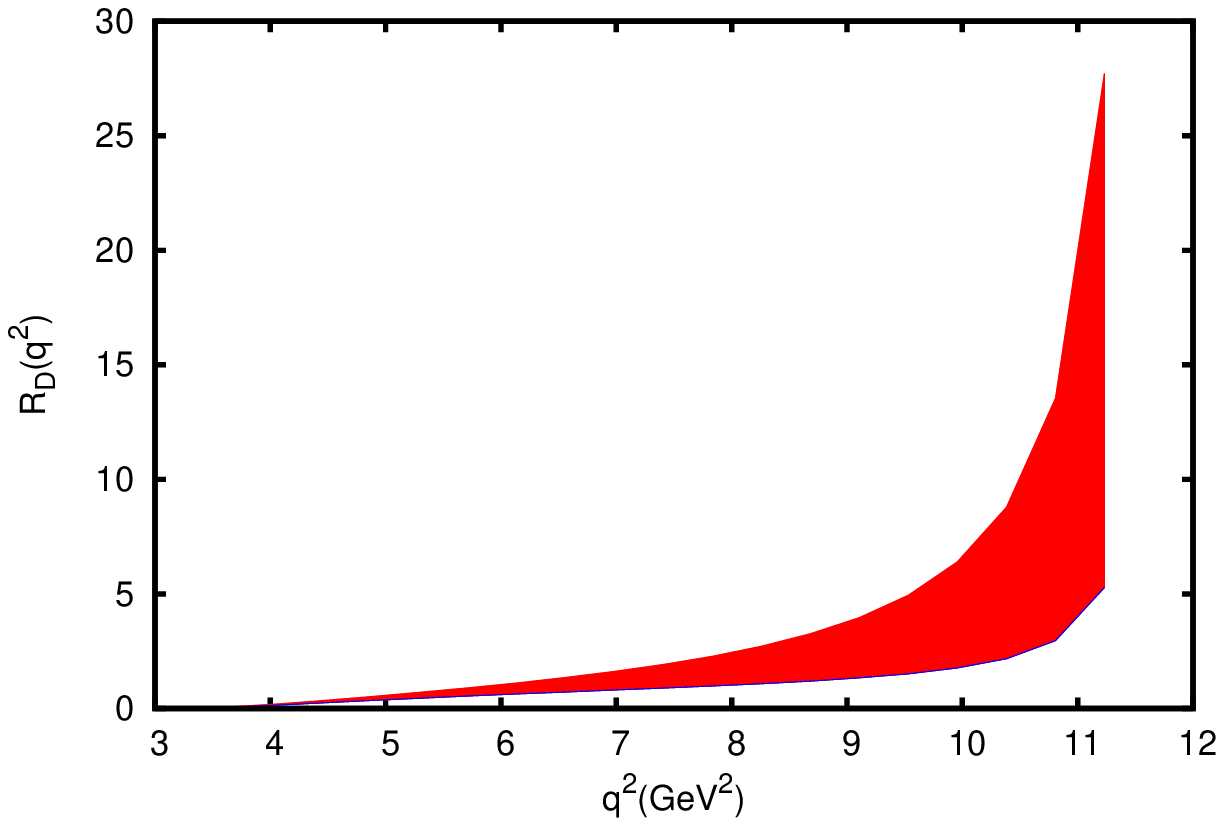}
\includegraphics[width=5cm,height=4cm]{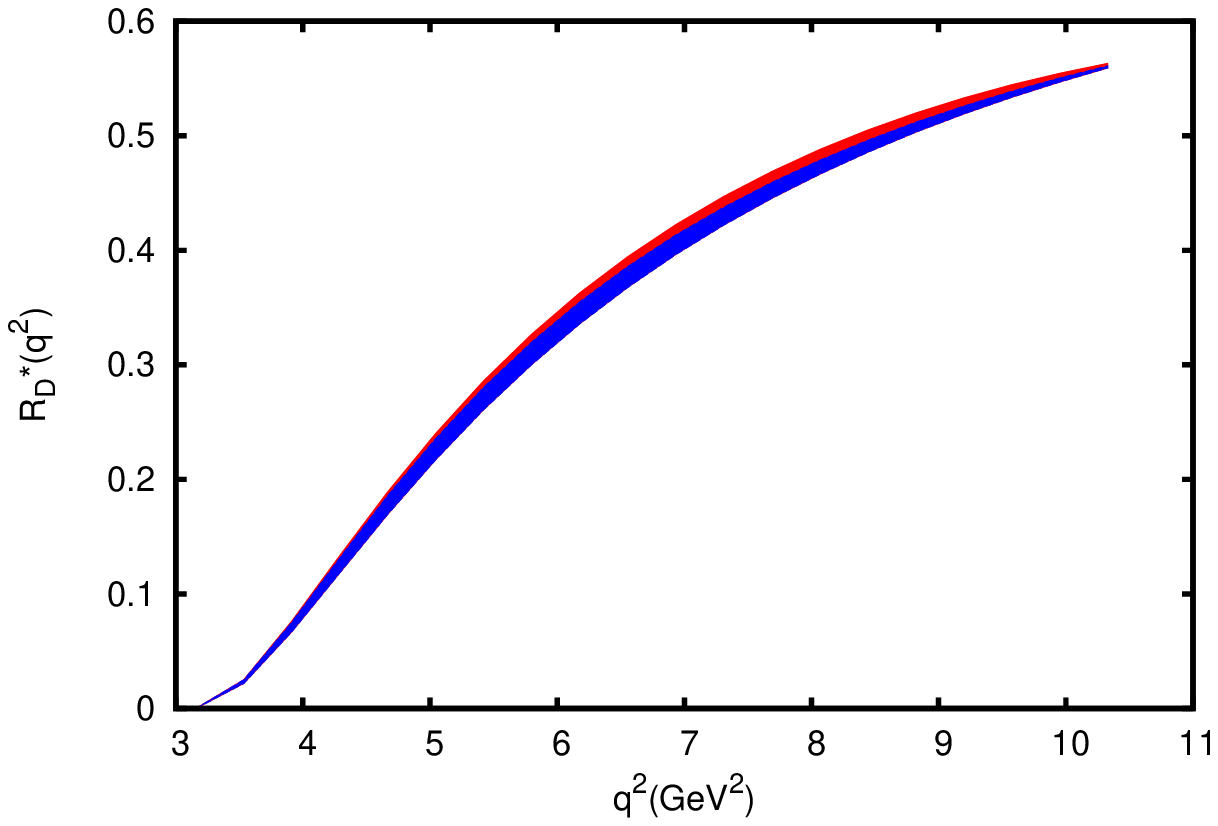}
\includegraphics[width=5cm,height=4cm]{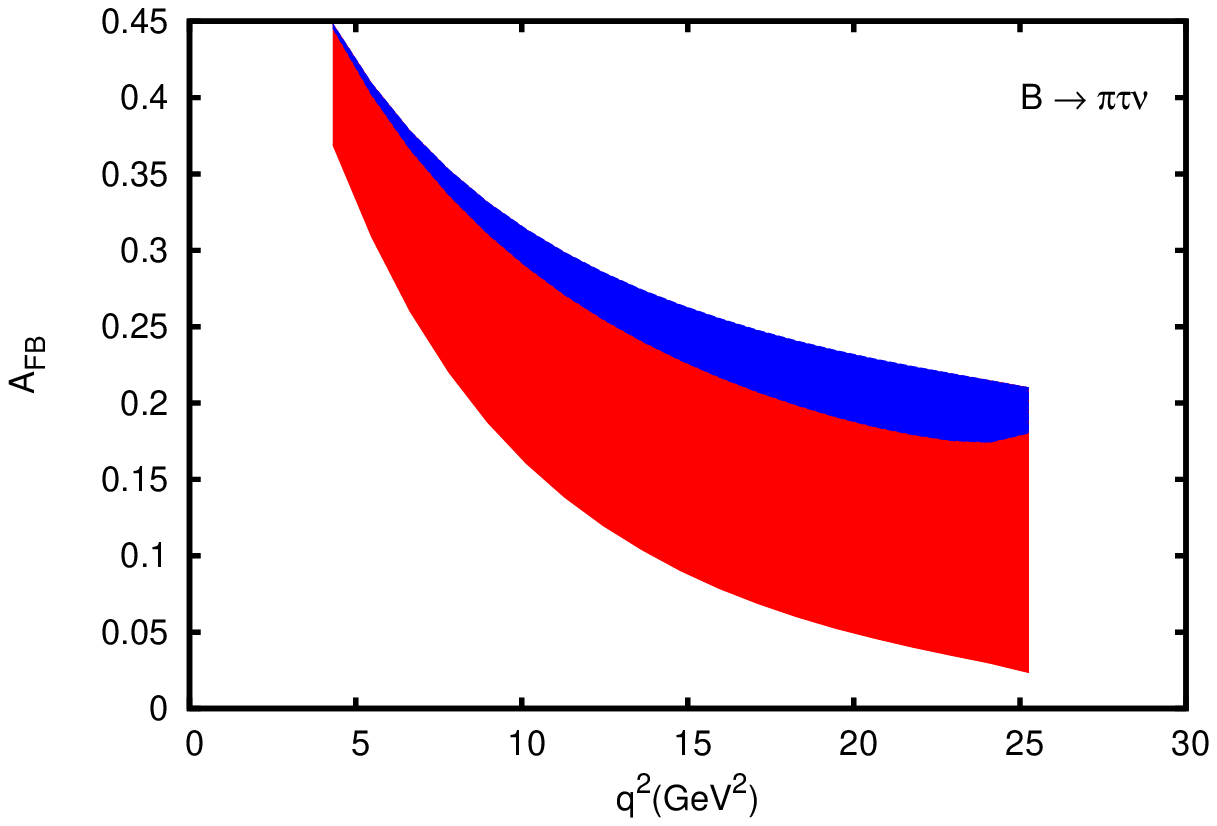}
\includegraphics[width=5cm,height=4cm]{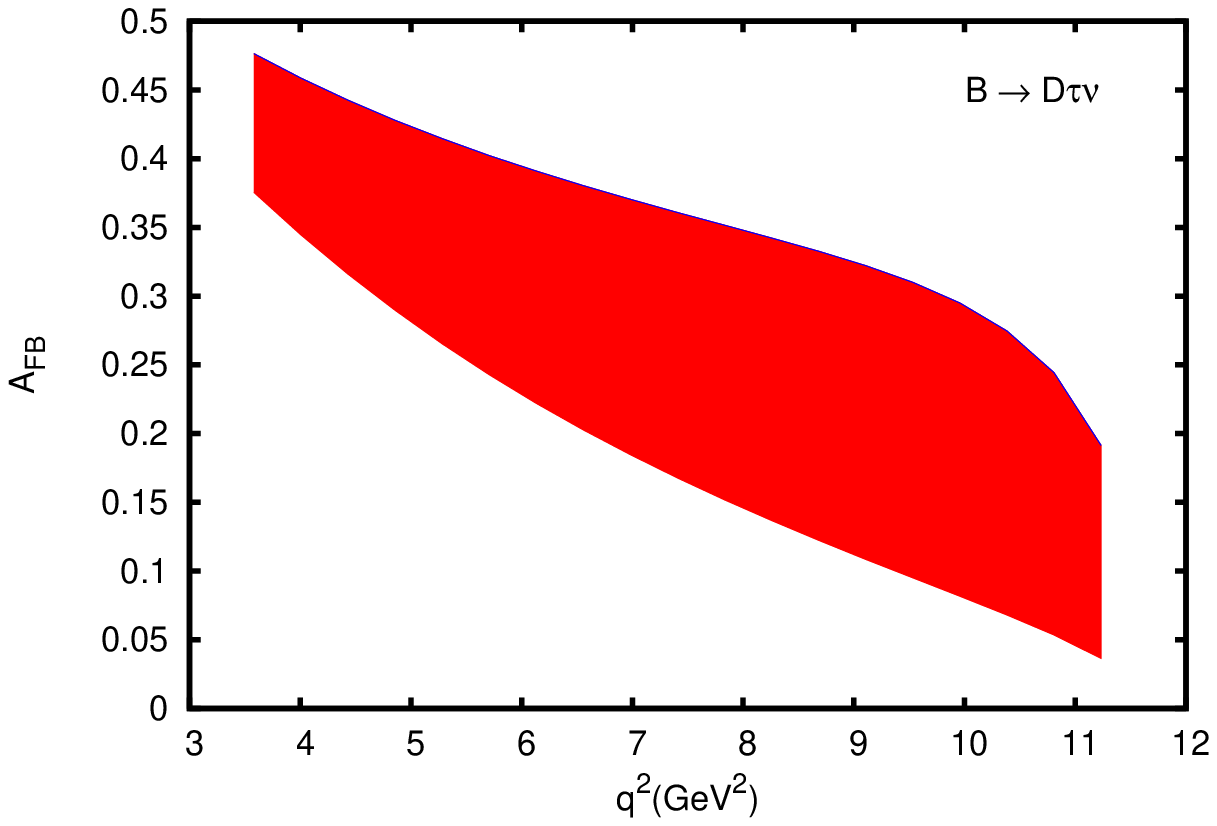}
\includegraphics[width=5cm,height=4cm]{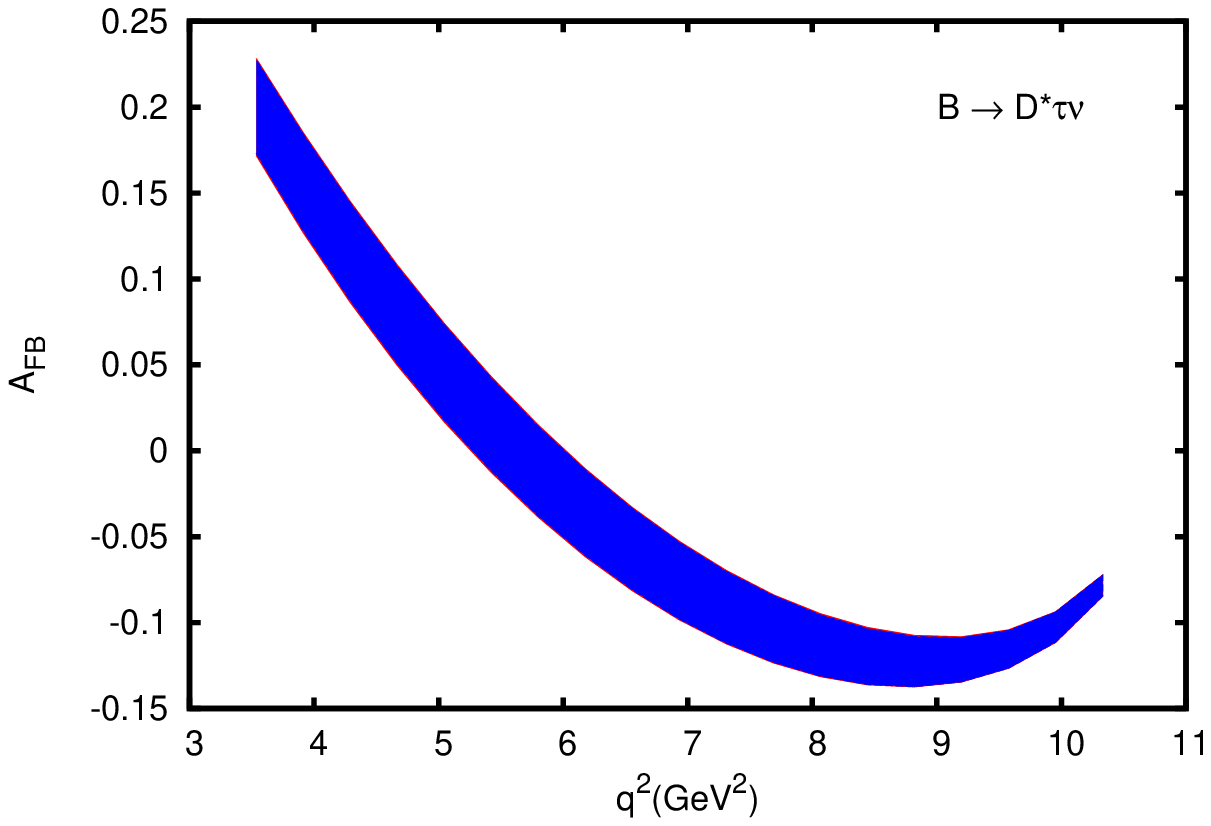}
\end{center}
\caption{Range in various observables such as DBR$(q^2)$, $R(q^2)$, and $A_{FB}(q^2)$ for the $B \to \pi\tau\nu$, $B \to D\tau\nu$, and the $B \to D^{\ast}\tau\nu$ decays. The allowed range in
each observable is shown in light~(red) band once the NP couplings $(\widetilde{S}_L,\,\widetilde{S}_R)$ are varied within the allowed ranges as shown in the left panel of Fig.~\ref{sltsrt_tau}.
The corresponding SM prediction is shown in dark~(blue) band. }
\label{obs_sltsrt_tau}
\end{figure}
In the presence of $\widetilde{S}_L$ and $\widetilde{S}_R$, the $\Gamma(B_q \to \tau\nu)$, $d\Gamma/dq^2(B \to P\,\tau\nu)$,
and $d\Gamma/dq^2(B \to V\,\tau\nu)$ can be written as
\begin{eqnarray}
\label{eq:sltsrt}
\Gamma(B_q \to \tau\nu) &=& \Gamma(B_q \to \tau\nu)|_{\rm SM}\,\Big[1 + \frac{m_B^4}{m_{\tau}^2\,(m_b + m_q)^2}\,\widetilde{G}_P^2\Big]\,,\nonumber \\
\frac{d\Gamma}{dq^2}(B \to P\,\tau\,\nu) &=& \frac{8\,N\,|\overrightarrow{p}_P|}{3}\Bigg\{H_0^2\Big(1 + \frac{m_{\tau}^2}{2\,q^2}\Big) + \frac{3\,m_{\tau}^2}{2\,q^2}\,H_t^2 + 
\frac{3}{2}\,H_S^2\,\widetilde{G}_S^2 \Bigg\}\,, \nonumber \\ 
\frac{d\Gamma}{dq^2}(B \to V\,\tau\,\nu) &=& \frac{8\,N\,|\overrightarrow{p}_V|}{3}\,\Bigg\{(\mathcal A_0^2 + \mathcal A_{||}^2 + \mathcal A_{\perp}^2)\Big(1 + \frac{m_{\tau}^2}{2\,q^2}\Big) +
\frac{3\,m_{\tau}^2}{2\,q^2}\,\mathcal A_t^2 + \frac{3}{2}\mathcal A_P^2\,\widetilde{G}_P^2\Bigg\}\,.
\end{eqnarray}
The $3\sigma$ allowed ranges of the $B \to \pi\tau\nu$ branching ratio and the ratio $R_{\pi}$ are shown in the right panel of Fig.~\ref{sltsrt_tau}. The ranges of
$\mathcal B(B_c \to \tau\nu)$, $\mathcal B(B \to \pi\tau\nu)$, and $R_{\pi}$ are
\begin{eqnarray*}
&&\mathcal B(B_c \to \tau\nu) = (1.11,\,16.71)\%\,, \qquad\qquad
\mathcal B(B \to \pi\tau\nu)  = (1.70,\,93.90)\times 10^{-5}\,,\nonumber \\
&&
R_{\pi} = (0.56,\,4.32)\,.
\end{eqnarray*}

The effect of these NP couplings on various observables are quite similar to the scenario where only the $S_L$ and $S_R$ are nonzero. The allowed ranges of all the observables are 
shown in Fig.~\ref{obs_sltsrt_tau}.
The differential branching ratio, the ratio of branching 
ratios, and the forward-backward aymmetry parameters deviate quite significantly from the SM prediction for the $B \to \pi\tau\nu$ and $B \to D\tau\nu$ decay modes, whereas there is no or very little
deviation of these observables from the SM value in case of $B \to D^{\ast}\tau\nu$ decay process. We see that the $B \to \tau\nu$ and $B \to D^{\ast}\tau\nu$ decay branching ratios depend on these
NP couplings through $\widetilde{G}_P^2$ terms, but, the $B \to \pi\tau\nu$ and $B \to D\tau\nu$ decay branching fractions depend on these NP couplings through $\widetilde{G}_S^2$ terms. Hence, we
see similar behavior for the $B \to \pi\tau\nu$ and $B \to D\tau\nu$ decay modes. However, as expected, the variation in the $B \to D^{\ast}\tau\nu$ decay mode is quite different from the $B \to \pi\tau\nu$ and 
the $B \to D\tau\nu$ decay modes.
Again, we see that the peak of the distribution of $B \to \pi\tau\nu$ and $B \to D\tau\nu$ decay branching ratios shift toward a large $q^2$ region.
Although, the effects of these right-handed couplings are quite similar to its left-handed counterpart, there are some differences. We do not see any zero crossing in 
the $q^2$ distribution of the $A_{FB}$ parameter for the $B \to \pi\tau\nu$ and $B \to D\tau\nu$ decay modes.
%%%%%%%

\section{Conclusion}
\label{con}
$B$ decay measurements have been providing us a lot of useful information regarding the nature of new physics.
Several recent measurements in the rare processes have put severe constraints on the NP parameters.
Precision measurements in $B$ meson decays have been a great platform for indirect evidences of beyond the
standard model physics. The recent measurements of the ratio of the branching ratio $R_D$ of $B \to D\,\tau\,\nu$ to that of $B \to D\,l\,\nu$ and $R_D^{\ast}$ of $B \to D^{\ast}\,\tau\,\nu$ to that of $B \to D^{\ast}\,l\,\nu$ 
differ from the standard model expectation at the $3.4\sigma$ level. It is still not conclusive enough that new physics is indeed present in this $b\to c\,\tau\,\nu$ processes. More precise measurements will reveal the nature
of the new physics. Similar new physics effects have been observed in $b \to u\,\tau\,\nu$ processes as well. The measurement of the branching ratio of $B \to \tau\nu$ and the ratio $R_{\pi}^l$ of the branching ratio 
of $B \to \tau\nu$ to $B \to \pi\,l\,\nu$ decays differ from the standard model expectation at more than the $2.5\sigma$ level. A lot of phenomenological studies have been done in order to explain all these discrepancies. 
In this paper, we consider an effective Lagrangian for the $b \to q\,l\,\nu$ transition processes in the presence of NP, where $q = u,\,c$, and perform a combined analysis of $B \to \tau\nu$, $B \to D\tau\nu$ and $B \to D^{\ast}\tau\nu$
decay processes. Our work differs significantly from others as we include the right-handed neutrino couplings. We assume that new physics is present only in the third generation leptons. 
We look at four different new physics scenarios. The results of our analysis are as follows.
 
We assume new physics in the third generation lepton only and see the effect of each new physics couplings on various observables. We first find the allowed ranges of each NP coupling using a $3\sigma$ constraint
coming from the most recent data of $R_{\pi}^l$, $R_D$, and $R_{D^{\ast}}$. 
For nonzero $V_L$ and $V_R$ couplings, the differential branching
ratio and the ratio of branching ratios are quite sensitive to these NP couplings for each decay mode. However, the forward-backward asymmetry for the $B \to \pi\tau\nu$ and $B \to D\tau\nu$ is not sensitive to these 
couplings at all. The forward-backward asymmetry is quite sensitive to these NP couplings for $B \to D^{\ast}\tau\nu$ decays and the deviation from
the standard model prediction can be quite significant depending on the value of $V_L$ and $V_R$. Although, we see a zero crossing in the $q^2$ distribution, it may or may not be there depending on the NP couplings.
Again, even if we see a zero crossing, it can deviate quite significantly from the standard model prediction.

In the case of $S_L$ and $S_R$ couplings, all the observables such as the differential branching ratio, ratio of branching ratios, and the forward-backward asymmetry are quite sensitive to the NP couplings for the
$B \to \pi\tau\nu$ and $B \to D\tau\nu$ decays. However, the sensitivity is somewhat reduced for the $B \to D^{\ast}\tau\nu$ decay mode. Although, in the standard model, there is no zero crossing in the forward-backward asymmetry parameter for the $B \to \pi\tau\nu$ and $B \to D\tau\nu$ decay modes, however, depending on the value of $S_L$ and $S_R$, one might see a zero crossing for both the decay modes.
For the $B \to D^{\ast}\tau\nu$ mode, the zero crossing can be similar or marginally different from the standard model one.

For the right-handed neutrino couplings $(\widetilde{V}_L,\,\widetilde{V}_R)$ and $(\widetilde{S}_L,\,\widetilde{S}_R)$, the effects are quite similar to its left-handed counterpart $(V_L,\,V_R)$ and $(S_L,\, S_R)$. 
However, the sensitivity is somewhat reduced. 

Although current experimental results are pointing towards the third generation leptons for possible new physics, there could be, in principle, new physics in the first two generations as well.
If there is NP in all generation leptons, then it might be possible to identify it by measuring the forward-backward asymmetry for $B \to \pi\,l\,\nu$, $B \to D\,l\,\nu$, and
$B \to D^{\ast}\,l\,\nu$ decay modes, where $l$ could be either an electron or a muon. It will provide useful information regarding the NP couplings $(S_L,\,S_R)$ and $(\widetilde{S}_L,\,\widetilde{S}_R)$. 
Similarly, measurement of the branching ratio of $B_c \to \tau\nu$ and $B \to \pi\tau\nu$ and the ratio $R_{\pi}$ will put additional constraints on the nature of NP couplings. Retaining our current
approach, we could also sharpen our estimates once improved measurements of various branching ratios and the ratio of branching ratios become available. At the same time, reducing the theoretical uncertainties in various
form factors and decay constants will also improve our estimates in future.
%%%%%%%%%%%%%%%%%%%%%%%%%%%
\vskip 1.0cm
\acknowledgements 
R.~D. and A.~K.~G. would like to thank BRNS, Government of India for financial support.
%\pagebreak
\appendix
\label{app}
\section{Kinematics and Helicity Amplitudes}
\label{kha}
We use the helicity method of Refs.~\cite{Korner,Kadeer} to calculate the different helicity amplitudes for
a $B$ meson decaying to pseudoscalar(vector) meson along with a charged lepton and an antineutrino in the final state.
We know that the amplitude square of the decay $B\to P(V)\,l\,\nu$ can be factorised into leptonic $(L_{\mu\nu})$ and hadronic $(H_{\mu\nu})$
tensors. That is
\begin{eqnarray}
|\mathcal M(B\to P(V)\,l\,\nu)|^2 &=& |\langle P(V)\,l\,\nu |\mathcal L_{\rm eff}|B\rangle|^2 = L_{\mu\nu}H^{\mu\nu}\,.
\end{eqnarray}

The leptonic and hadronic tensor product $L_{\mu\nu}\,H^{\mu\nu}$  depends on the polar angle $\cos\theta_l$, where $\theta_l$ is the angle between the $P~(V)$ meson three momentum vector and
the lepton three momentum vector in the $q^2$ rest frame, and can be worked out using the completeness relation of the polarization four vectors $\epsilon(t,\pm,0)$, i.e,
\begin{eqnarray}
\sum_{m,\,m^{\prime} = t,\pm,0}\,\epsilon^{\mu}(m)\,\epsilon^{\ast\,\nu}(m^{\prime})\,g_{m\,m^{\prime}} = g^{\mu\nu}\,,
\end{eqnarray}
where $g_{m\,m^{\prime}} = {\rm diag}(+,\,-,\,-,\,-)$. Using this approach, one can factorize $L_{\mu\nu}\,H^{\mu\nu}$ in terms of two Lorentz invariant quantities such that
\begin{eqnarray}
L_{\mu\nu}\,H^{\mu\nu} &=& L^{\mu^\prime\nu^\prime}\,g_{\mu^\prime\mu}\,g_{\nu^\prime\nu}H^{\mu\nu} =
\sum_{m,m^\prime,n,n^\prime}\,L^{\mu^\prime\nu^\prime}\,\epsilon_{\mu^\prime}(m)\,\epsilon^*_\mu(m^\prime)\,g_{mm^\prime}\,\epsilon^*_{\nu^\prime}(n)\,\epsilon_\nu(n^\prime)\,g_{nn^\prime}\,H^{\mu\nu}\,  \nonumber\\
&&=
\sum_{m,m^\prime,n,n^\prime}\,\Big(L^{\mu^\prime\nu^\prime}\,\epsilon_{\mu^\prime}(m)\,\epsilon^*_{\nu^\prime}(n)\,\Big)\Big(H^{\mu\nu}\,\epsilon^*_\mu(m^\prime)\,\epsilon_\nu(n^\prime)\Big)\,g_{mm^\prime}\,g_{nn^\prime}\,  \nonumber\\
&&=\sum_{m,\,m^{\prime},\,n,\,n^{\prime}}\,L(m,\,n)\,H(m^{\prime},\,n^{\prime})\,g_{m\,m^{\prime}}\,g_{n\,n^{\prime}}\,,
\end{eqnarray}
where $L(m,\,n)$ and $H(m^{\prime},\,n^{\prime})$ can now be evaluated in different Lorentz frames. We evaluate $L(m,\,n)$ in the $l-\nu$ center-of-mass frame, i.e, in the $q^2$ rest frame
and $H(m^{\prime},\,n^{\prime})$ in the $B$ meson rest frame.

In the $B$ meson rest frame, the helicity basis $\epsilon$ is taken to be
\begin{eqnarray} 
&&\epsilon(0)=\frac{1}{\sqrt{q^2}}(|p_{M}|,0,0,-q_0)\,,\qquad\qquad
\epsilon(\pm)=\pm \frac{1}{\sqrt 2}(0,\pm 1,-i,0)\,,\qquad\qquad \nonumber\\
&&\epsilon(t)=\frac{1}{\sqrt{q^2}}(q_0,0,0,-|p_{M}|)\,,   
\end{eqnarray}
where $ q_0=(m_B^2-m_{M}^2+q^2)/\,2m_B\, $ and $q = p_B-p_{M}$ is the momentum transfer, respectively. Here $m_M$ and $p_M$ denote the mass and the four momentum of the final state pseudoscalar(vector) meson $M$, respectively.
Again, we have $ |p_{M}|=\lambda^{1/2}(m_B^2,m_{M}^2,q^2)/{2m_B} $. In the $B$ meson rest frame, the $B$ and $M$ meson four momenta $p_B$ and $p_{M}$ are
\begin{eqnarray}
&&p_B=(m_B,0,0,0)\,,\qquad\qquad
p_{M}=(E_M,0,0,|\vec{p}_{M}|)\,,
\end{eqnarray}
where the $ E_{M}=(m_B^2 + m_{M}^2-q^2)/\,2m_B$. 
For a vector meson in the final state, the polarization four vectors obey the following orthonormality condition
\begin{eqnarray}
&& \epsilon^*_\alpha(m)\,\epsilon^\alpha(m^\prime)=-\,\delta_{mm^\prime}\,
\end{eqnarray} 
and the completeness relation
\begin{eqnarray}
\sum_{m,m^\prime}\,\epsilon_\alpha(m)\,\epsilon_\beta(m^\prime)\,\delta_{mm^\prime}=-g_{\alpha\beta}+\frac{(p_V)_\alpha(p_V)_\beta}{m_V^2}\,.
\end{eqnarray}

The leptonic tensor $L(m,\,n)$ is evaluated in the $l - \nu_l$ center-of-mass frame, i.e, in the $q^2$ rest frame. In this frame, the helicity basis $\epsilon$ is taken to be
\begin{eqnarray} 
&&\epsilon(0)=(0,0,0,-1)\,,\qquad\qquad
\epsilon(\pm)=\pm \frac{1}{\sqrt 2}(0,\pm 1,-i,0)\,,\qquad\qquad
\epsilon(t)=(1,0,0,0)
\end{eqnarray}

In the $q^2$ rest frame, the four momenta of the lepton and the antineutrino pair can be written as 
\begin{eqnarray}
&&p^{\mu}_l=(E_l,\,|p_l|\,\sin\theta_l,\,0,\,-|p_l|\,\cos\theta_l)\,,\nonumber \\
&&p^{\mu}_{\nu}=(|p_l|,\,-|p_l|\,\sin\theta_l,\,0,\,|p_l|\,\cos \theta_l)\,,
\end{eqnarray} 
where the lepton energy $E_l=(q^2 + m_l^2)/2 \sqrt{q^2}$ and the magnitude of its three momenta is $|p_l|=(q^2- m_l^2)/2 \sqrt {q^2}$.

\section{B to $ \pi$ Form Factors}
\label{ffpi}

For the $B \to \pi$ transition form factors, there are two nonperturbative methods for calculating
the $B \to \pi$ form factors: light-cone sum rules~(LCSR) and lattice QCD~(LQCD).
QCD light-cone sum rules with pion distribution amplitudes 
allow one to calculate the $B \to \pi$ form factors at small and intermediate momentum transfers $0 \le q^2 \le q^2_{\rm max}$,
where $q^2_{\rm max}$ varies from $12$ to $16\,{\rm GeV}^2$~\cite{bpiff}. 
The most recent lattice QCD computations with three dynamical flavors predict these form factors at $q^2 \ge 16\,{\rm GeV}^2$ , in the upper part of
the semileptonic region $0 \le q^2 \le (m_B - m_{\pi})^2$ , with an accuracy reaching
$10\%$. There are also recent results available in the quenched approximation on a fine
lattice~\cite{lattice}. Very recently, in Ref.~\cite{Khodjamirian}, the author uses the sum rule results for the form factors as an input for a z-series parametrization that yield the
$q^2$ shape in the whole semileptonic region of $B \to \pi\,l\,\nu$. The relevant formulae for $F_{+}(q^2)$ and $F_0(q^2)$ pertinent for our discussion, taken from Ref.~\cite{Khodjamirian},
are
\begin{eqnarray}
&&F_{+}(q^2) = \frac{F_{+}(0)}{\Big(1-\frac{q^2}{m_B^2}\Big)}\Bigg\{1 + \sum_{k=1}^{N-1}\,b_k\,\Big(z(q^2,t_0)^k - z(0,t_0)^k -
(-1)^{N-k}\,\frac{k}{N}\,\Big[z(q^2,t_0)^N - z(0,t_0)^N\Big]\Big)\Bigg\} \nonumber \\
&&F_0(q^2) = F_0(0)\,\Bigg\{1 + \sum_{k=1}^N\,b_k^0\,\Big(z(q^2,t_0)^k - z(0,t_0)^k\Big)\Bigg\}
\end{eqnarray}
where by default $F_{+}(0) = F_0(0)$ %= 0.281^{+0.018}_{-0.022}$. 
and
\begin{eqnarray}
&&z(q^2,t_0) = \frac{\sqrt{(m_B + m_{\pi})^2 - q^2} - \sqrt{(m_B + m_{\pi})^2 - t_0}}{\sqrt{(m_B + m_{\pi})^2 - q^2} + \sqrt{(m_B + m_{\pi})^2 - t_0}}
\end{eqnarray}
where the auxiliary parameter $t_0$ is defined as $t_0 = (m_B + m_{\pi})^2 - 2\sqrt{m_B\,m_{\pi}}\,\sqrt{(m_B + m_{\pi})^2 - q^2_{\rm min}}$.
The central values of $F_{+}(0) = F_0(0)$ and the slope parameters $b_1$ and $b^0_1$ are
\begin{eqnarray}
&&F_0(0) = F_{+}(0) = 0.281\pm 0.028\,,\qquad\qquad
b_1 = -1.62\pm 0.70\,,\qquad\qquad
b_1^0 = -3.98\pm 0.97\,.
\end{eqnarray}
For the uncertainties, we add the various errors reported in Ref.~\cite{Khodjamirian} in quadrature.
%%%%%%%%%%%%%%%%%
\section{$B \to D,\,D^{\ast}$ form Factors using HQET}
\label{ffdd}
In the heavy quark effective theory one can write the hadronic matrix elements of current between two hadrons in inverse powers of heavy quark mass and the hadronic form factor
in a reduced single universal form, which is a function of the kinematic variable $ v_B.v_{P(V)} $, where $ v_B $ and $ v_{P(V)}$ are the four velocity of the $B$ meson and the pseudoscalar~(vector) meson,
respectively. The weak vector and axial vector currents are parametrized as~\cite{Falk} 
\begin{eqnarray}
\langle \,D(v^{\prime}) | \bar c\,\gamma_\mu\,b | B(v) \rangle &=&\sqrt{m_B\,m_D}\,\Big[h_+ (\omega)(v+v^\prime)_\mu+h_- (\omega)(v-v^\prime)_\mu \Big] \,, \nonumber\\
\langle D^*(v^{\prime},\epsilon^\prime) | \bar c\,\gamma_\mu\,b | B(v) \rangle &=&i\sqrt{m_B\,m_D}\,h_V(\omega)\,\varepsilon_{\mu\nu\alpha\beta}\, \epsilon^{\prime\ast\nu}\, v^{\prime\alpha}v^\beta\,, \nonumber\\
\langle D^*(v^{\prime},\epsilon^\prime) | \bar c\,\gamma_\mu\,\gamma_5\,b | B(v) \rangle &=& \sqrt{m_B\,m_D}\,\Big[\,h_{A_1}(\omega)\,(\omega+1)\,\epsilon_\mu^{\prime\ast} - h_{A_2}(\omega) 
\epsilon^{\prime\ast}\cdot v \, v_\mu\, \nonumber\\
&& - h_{A_3}(\omega)\,\epsilon^{\prime\ast}.v\,v_\mu^\prime\,\Big]\,,
\end{eqnarray}
where the kinemetic variable $\omega = v_B.v_{(D,\,D^{\ast})} = (m_B^2 + m_{(D,\,D^{\ast})}^2-q^2)/\,2\,m_B\,m_{(D,\,D^{\ast})}$.
Now, for the $B \to D$ form factors $F_+(q^2)$ and $ F_0(q^2)$, we obtain
\begin{eqnarray}
&&F_+(q^2)=\frac{V_1(\omega)}{r_D}\,, \qquad\qquad
F_0(q^2)=\frac{(1+\omega)\,r_D}{2}\,S_1(\omega)\,,
\end{eqnarray} 
where $ V_1(\omega)$ and $S_1(\omega)$, taken from Ref.~\cite{Caprini}, are
\begin{eqnarray}
&&V_1(\omega)=\Big[\,h_+(\omega)-\frac{(1-r)}{(1+r)}\,h_-(\omega)\,\Big]\,,      \nonumber\\
&&S_1(\omega)=\Big[\,h_+(\omega)-\frac{(1+r)(\omega-1)}{(1-r)(\omega+1)}\,h_-(\omega)\,\Big]\,,
\end{eqnarray} 
and 
\begin{eqnarray} 
&&r_D=\frac{2\sqrt{m_B\,m_D}}{(m_B+m_D)}\,, \qquad\qquad
r=\frac{m_D}{m_B} .
\end{eqnarray}
We follow Ref.~\cite{Sakaki} and parametrized $V_1(\omega)$ in terms of $\rho_1$ and $z$ parameters as
\begin{eqnarray}
&&V_1(\omega)=V_1(1)\,\Big[1 - 8\,\rho_1^2\,z + (51\,\rho_1^2 - 10)\,z^2 - (252\,\rho_1^2 - 84)\,z^3\,\Big]\,,
\end{eqnarray}
where $ z=(\sqrt{\omega+1}-\sqrt{2})/(\sqrt{\omega+1}+\sqrt{2})$.
The numerical value of $V_1(1)$ and $\rho_1^2$ are~\cite{Aubert:2009ac} 
\begin{eqnarray}
&& V_1(1)|V_{cb}|=(43.0\pm1.9\pm1.4)\times 10^{-3} , \nonumber\\
&& \rho_1^2=1.20\pm 0.09\pm 0.04. 
\end{eqnarray}
The form factor $S_1(\omega)$ has the following parametrization~\cite{Sakaki}.
\begin{eqnarray}
&& S_1(\omega)=1.0036[1-0.0068(\omega-1)+0.0017(\omega-1)^2-0.0013(\omega-1)^3]V_1(\omega).
\end{eqnarray}

We now concentrate on the $B\to V$ i.e. $ B\to D^{\ast}$ form factor in the HQET~\cite{Fajfer} by defining the universal form factor
$h_{A_1} $ which can be related to $A_0(q^2),\,A_1(q^2),\,A_2(q^2),\,{\rm and}\,V(q^2)$ as 
\begin{eqnarray}
&& A_1 (q^2)= r_{D^*}\frac{\omega+1}{2} h_{A_1}(\omega) \,, \nonumber\\
&& A_0(q^2)=\frac{R_0(\omega)}{r_{D^*}} h_{A_1}(\omega)\,,\nonumber\\
&& A_2(q^2)=\frac{R_2(\omega)}{r_{D^*}} h_{A_1}(\omega)\,,\nonumber\\
&& V_0(q^2)=\frac{R_1(\omega)}{r_{D^*}} h_{A_1}(\omega)\,
\end{eqnarray}
where $r_{D^*}=2\sqrt{m_B\,m_{D^*}}/(m_B+m_{D^*}) $. The $\omega$ dependence of the form factors in the limit of heavy quark can be written as~\cite{Caprini, Fajfer}
\begin{eqnarray}
&& h_{A_1}(\omega)=h_{A_1}(1)[1-8\,\rho^2z+(53\rho^2-15)z^2-(231\rho^2-91)z^3]\,,\nonumber\\
&& R_1(\omega)=R_1(1)-0.12(\omega-1)+0.05(\omega-1)^2\,,\nonumber\\
&& R_2(\omega)=R_2(1)+0.11(\omega-1)-0.06(\omega-1)^2\,,\nonumber\\
&& R_0(\omega)=R_0(1)-0.11(\omega-1)+0.01(\omega-1)^2\,,
\end{eqnarray}
where, we use the following numerical values of the free parameters from Refs.~\cite{Dungel:2010uk, Fajfer} for our numerical analysis. That is
\begin{eqnarray}
&&h_{A_1}(1)\,|V_{cb}| = (34.6 \pm 0.2 \pm 1.0)\times 10^{-3}\,, \nonumber \\
&&\rho_1^2 = 1.214 \pm 0.034 \pm 0.009 \,, \nonumber \\
&&R_1(1) = 1.401 \pm 0.034 \pm 0.018 \,, \nonumber \\
&&R_2(1) = 0.864 \pm 0.024 \pm 0.008 \,, \nonumber \\
&&R_0(1) = 1.14 \pm 0.114
\end{eqnarray}

%%%%%%%%%%%%%%%%%% REFERENCES %%%%%%%%%%%%%%%%%%%%%%%%%%%%

%%%%%%%%%%%%%%%%%%%%%%%%%%
\end{document}